\documentclass[11pt]{article}
\usepackage{jheppub}
\usepackage[T1]{fontenc}
\usepackage{axodraw2,pix}
\usepackage{bm}
\usepackage{bbold} 

\graphicspath{{./}{./figures/}}

\newcommand{\beq}{\begin{equation}}
\newcommand{\eeq}{\end{equation}}
\newcommand{\bea}{\begin{eqnarray}}
\newcommand{\eea}{\end{eqnarray}}
\newcommand{\bi}{\begin{itemize}}
\newcommand{\ei}{\end{itemize}}

\newcommand{\mc}[1]{\mathcal{#1}}
\newcommand{\nn}{\nonumber}  
\newcommand\MSbar{$\overline{\rm MS}$}
\newcommand{\hsq}{\phi^\dagger \phi}
\newcommand{\bub}{\rmii{c}}

\newcommand{\freq}{\omega_\bub}

\newcommand{\nocontentsline}[3]{}
\newcommand{\tocless}[2]{\bgroup\let\addcontentsline=\nocontentsline#1{#2}\egroup}

\newcommand{\rmi}[1]{{\mbox{\scriptsize #1}}}
\newcommand{\rmii}[1]{{\mbox{\tiny\rm{#1}}}}
\renewcommand{\vec}[1]{{\bf #1}}
\newcommand{\bmu}{\bar\mu}
\newcommand{\gammaE}{{\gamma_\rmii{E}}}
\newcommand{\T}{\rmii{$T$}}

\newcommand{\Nf}{N_{\rm f}}
\newcommand{\Nc}{N_{\rm c}}
\newcommand{\Tc}{T_{\rm c}}

\newcommand{\Tp}{T_p}
\newcommand{\Hp}{H_p}
\newcommand{\Hs}{H_{\ast}}

\newcommand{\mD}{m_\rmii{D}}
\newcommand{\mh}{\mu_{h}}

\newcommand{\g}{g}
\newcommand{\gp}{g'}
\newcommand{\gY}{g_{Y}}
\newcommand{\gs}{g_\rmi{s}}
\newcommand{\alphas}{\alpha_\rmi{s}}

\newcommand{\nB}{n_\rmii{B}}

\newcommand{\Tint}[1]{{\hbox{$\sum$}\!\!\!\!\!\!\!\int\,}_{\!\!\!\!\raise-0.9ex\hbox{$\scriptstyle{#1}$}}}
\newcommand{\Tinti}[1]{{{\Sigma}\!\!\!\!\raise0.3ex\hbox{$\int$}_\rmii{${#1}$}}}
\newcommand{\Tintip}[1]{{{\Sigma'}\!\!\!\!\!\raise0.3ex\hbox{$\int$}_\rmii{${#1}$}}}

\newcommand{\bsl}[1]{\,\slash\!\!\!\!{#1}\,}

\title{Theoretical uncertainties for cosmological first-order phase transitions}
\author[a]{Djuna Croon,}
\author[b]{Oliver Gould,}
\author[b,c]{Philipp Schicho,}
\author[c]{Tuomas V. I. Tenkanen,}
\author[a,d]{and Graham White}

\affiliation[a]{TRIUMF, 4004 Wesbrook Mall, Vancouver, BC V6T 2A3, Canada}
\affiliation[b]{Helsinki Institute of Physics, University of Helsinki, FI-00014, Finland}
\affiliation[c]{Albert Einstein Center for Fundamental Physics, Institute for Theoretical Physics, University of Bern, Sidlerstrasse 5, CH-3012 Bern, Switzerland}
\affiliation[d]{Kavli IPMU (WPI), UTIAS, The University of Tokyo, Kashiwa, Chiba 277-8583, Japan}

\emailAdd{dcroon@triumf.ca}
\emailAdd{oliver.gould@helsinki.fi}
\emailAdd{philipp.schicho@helsinki.fi}
\emailAdd{tenkanen@itp.unibe.ch}
\emailAdd{graham.white@ipmu.jp}
\date{\today}
\abstract{
We critically examine the magnitude of theoretical uncertainties in perturbative calculations of first-order phase transitions, using the Standard Model effective field theory as our guide.
In the usual daisy-resummed approach, we find large uncertainties due to renormalisation scale dependence, which amount to two to three orders-of-magnitude uncertainty in
the peak gravitational wave amplitude, relevant to experiments such as LISA.
Alternatively, utilising dimensional reduction in a more sophisticated perturbative approach
drastically reduces this scale dependence,
pushing it to higher orders.
Further, this approach resolves other thorny problems with daisy resummation: it
is gauge invariant which is explicitly demonstrated for the Standard Model, and avoids
an uncontrolled derivative expansion
in the bubble nucleation rate.
}

\preprint{HIP-2020-26/TH}

\begin{document}
\maketitle

\section{Introduction}
\label{sec:introduction}

A first-order phase transition in the early universe gives rise to a stochastic gravitational wave background (SGWB) which may be observable today.%
\footnote{
    For recent reviews, see Refs.~\cite{Weir:2017wfa,Caprini:2018mtu,Mazumdar:2018dfl,Senaha:2020mop}.
}
As a consequence, upcoming gravitational wave experiments open a new window into particle physics phenomenology in the early universe.
The frequency window 
of space-based interferometer experiments, such as LISA, may probe the nature of the electroweak phase transition which typically produce a gravitational wave background peaking in the mHz range~\cite{Caprini:2015zlo,Caprini:2019egz}. A first-order electroweak phase transition is motivated in particular by the baryon asymmetry of the universe (BAU),
as it provides the necessary departure from equilibrium~\cite{Morrissey:2012db,White:2016nbo}. Although state of the art calculations of the Standard Model (SM) indicate a crossover transition \cite{Kajantie:1995kf,Kajantie:1996mn,Kajantie:1996qd,Csikor:1998eu,DOnofrio:2015gop}, it is straightforward to extend it by new scalar fields \cite{Gil:2012ya,Carena:2012np,Profumo:2014opa,Kozaczuk:2014kva,Vaskonen:2016yiu,Dorsch:2016nrg,Chiang:2017nmu,Beniwal:2018hyi,Bruggisser:2018mrt,Athron:2019teq,Kainulainen:2019kyp}
or effective operators~\cite{Grojean:2004xa,Delaunay:2007wb,Chala:2018ari}
to catalyse a strong first-order electroweak phase transition (EWPT).%
\footnote{
    Some more exotic possibilities are
    a multistep transition~\cite{Patel:2012pi,Patel:2013zla,Blinov:2015sna},
    monopoles in the early universe~\cite{Arunasalam:2017eyu},
    modified couplings in the early universe~\cite{Baldes:2016rqn,Ellis:2019flb},
    utilising the QCD transition~\cite{vonHarling:2017yew,Ipek:2018lhm,Croon:2019ugf} and
    utilising vector-like fermions~\cite{Angelescu:2018dkk}.
}
Aside from the electroweak transition,
hidden sectors can have dark transitions \cite{Schwaller:2015tja,Croon:2018erz,Croon:2018kqn,Hall:2019ank,Croon:2019rqu,Croon:2019iuh}, which could provide a unique window on a dark sector, which only interacts gravitationally.

In order to accurately predict the SGWB in any of these models, one requires a robust mapping between the observables and Lagrangian parameters. This problem can be partitioned in two~\cite{Croon:2018erz,Caprini:2019egz}:
the mapping between Lagrangian parameters and thermal parameters; and
the mapping between thermal parameters and the SGWB, in particular its peak frequency and amplitude. Only if both these mappings are well understood, can the SGWB serve as complimentary to other probes, such as collider experiments \cite{Curtin:2014jma,Kotwal:2016tex,Ramsey-Musolf:2019lsf} and direct detection experiments \cite{Assamagan:2016azc,Baldes:2018emh}. In some cases, the reach of a detector such as LISA may even improve collider constraints on effective operators compared to collider upgrades~\cite{Chala:2018ari}.
However, to reach such precision, the theoretical uncertainties associated with the SGWB predictions need to be under control.
With these motivations in mind, this paper will study the theoretical uncertainties for different methods of calculating the thermal parameters of a phase transition. 

As has long been recognised, at high temperatures the long-wavelength modes of bosons become strongly coupled~\cite{Linde:1980ts}. This thwarts the usual perturbative expansion.
(At least) two different approaches to resumming perturbation theory have been developed to ameliorate this problem at high-$T$:
daisy resummation and dimensional reduction.
We compare and contrast these two approaches, estimating the numerical magnitude of theoretical uncertainties.

It was realised early on~\cite{Dolan:1973qd,Kirzhnits:1976ts} that the so called {\em daisy} diagrams cause the largest infrared contributions, and should be resummed. Concrete resummation methods were developed and utilised to two-loop order.
Dubbed
the Parwani~\cite{Parwani:1991gq} and
the Arnold-Espinosa~\cite{Arnold:1992rz}
resummations, these approaches differ in details though are methodologically similar.
The Parwani resummation method allows for a smooth transition to the correct low-$T$ behaviour, but generates unphysical linear terms in the potential which shifts the symmetric minimum away from the origin~\cite{Cline:2011mm,Laine:2017hdk}.
The Arnold-Espinosa resummation method avoids unphysical linear terms,
though by only screening the IR bosonic modes, it fails at sufficiently low temperatures.
Various attempts to go beyond these leading contributions, and to resum so called {\em super-daisy} diagrams were made
(see for example Ref.~\cite{Curtin:2016urg} and references therein),
though we do not consider these approaches here.
Similarly, we defer comparisons to hard thermal loop perturbation theory to future investigations~\cite{Braaten:1991gm}.

The idea of dimensional reduction is clearest in the Matsubara formalism.
Therein, the equilibrium properties of 4-dimensional QFTs at nonzero temperature, $T$, are described by fields living on $\mathbb{R}^3\times S^{1}_{\beta}$
with the radius of the circle equal to $\beta=1/T$.
Phenomena on length scales much longer than $1/T$ do not see the compact direction, and hence should be describable by a purely 3-dimensional effective field theory.
As a concrete method for resummation of perturbation theory, this idea dates back to the 1980s~\cite{Ginsparg:1980ef,Appelquist:1981vg,Nadkarni:1982kb,Landsman:1989be}.
The 1990s revived this approach~\cite{Farakos:1994kx,Braaten:1995cm,Braaten:1995jr,Kajantie:1995dw},
developed a simpler method to derive the effective coupling constants of the 3d theory and 
laid down a generic recipe. An important further development was the use of lattice Monte-Carlo simulations to study the 3d effective theory~\cite{Farakos:1994xh,Kajantie:1995kf,Kajantie:1996qd}, which led to the discovery that the electroweak transition in the Standard Model is a crossover~\cite{Kajantie:1996mn}. In this work we do not discuss lattice simulations in detail, being mostly interested in theoretical uncertainties of perturbation theory.

A first-order EWPT requires physics beyond the SM (BSM), in particular new Higgs interactions. If the particles responsible for these new interactions are significantly heavier than the electroweak scale, it should be possible to integrate them out at $T=0$. The resulting effective theory contains all possible operators of the SM, with higher-dimensional operators suppressed by the heavy scale of these new interactions. This is the SM effective field theory (SMEFT)~\cite{Buchmuller:1985jz,Grzadkowski:2010es}.
When truncated at dimension-six and keeping to 3rd generation fermions, it contains an additional 60 independent operators.
However, as we are only interested in the electroweak phase transition, which takes place due to symmetry breaking in the Higgs sector, we will restrict ourselves to considerations of the single effective operator
\beq
\label{eq:O6}
\mc{O}_6 = \frac{1}{M^2}\bigl(\phi^\dagger \phi\bigr)^3
\;,
\eeq
where $\phi$ is the SM Higgs doublet.%
\footnote{Note that SMEFT is frequently used in this way in the literature \cite{Grojean:2004xa,Delaunay:2007wb,Chala:2018ari,Ellis:2019flb,Damgaard:2015con,deVries:2017ncy,Balazs:2016yvi,deVries:2018tgs,Phong:2020ybr}} 

Such a dimension-six operator may imply a potential barrier between symmetric and broken phases even before considering loop corrections. In particular, a first-order phase transition can be triggered by a relatively small dimension-six coefficient and negative quartic coupling.
However, the additional operator $\mc{O}_{6}$ is merely one of many in a complete basis for the SMEFT.
Derivative couplings with kinetic and gauge covariant terms (cf.~Ref.~\cite{Buchmuller:1985jz,Grzadkowski:2010es,deBlas:2014mba,Marzocca:2020jze}) make themselves felt at higher energies and can additionally affect the EWPT.
Since $\mc{O}_{6}$ is arguably the dominant higher-dimensional operator in composite Higgs scenarios~\cite{Delaunay:2007wb}, and
expected to dominate in scenarios with extended scalar sectors and large portal couplings,
we refrain from considering the full SMEFT basis.
This does not change the qualitative scope of our analysis.

The phase transition in this effective field theory was studied previously, in Refs.~\cite{Grojean:2004xa,Bodeker:2004ws,Delaunay:2007wb,Cai:2017tmh,Chala:2018ari} (see also Ref.~\cite{Huber:2007vva}).
These works adopted the daisy-resummed approach with and without a high-temperature approximation. It was found that as the cut-off scale $M$ is lowered, the EWPT strengthens, and both the critical temperature and nucleation temperature decrease. At sufficiently small $M\lesssim 650$~GeV, the transition is strong enough to be observable at near-future gravitational wave experiments~\cite{Cai:2017tmh,Chala:2018ari}.
For $M\lesssim 550$~GeV, the bubble nucleation rate never exceeds the Hubble rate and the phase transition never completes.

The strength of the transition is dictated by the single tunable parameter, $M$.
This makes a convenient model to study theoretical uncertainties, the focus of this article. It is also significant to note that as one varies $M$, the thermodynamics of the phase transition in this SMEFT qualitatively reproduces that of single-step transitions in scalar extensions of the SM, such as the two-Higgs doublet model (see e.g.\ Figs.~(5) and (7) in Ref.~\cite{Caprini:2019egz}). However, we will not be concerned with the validity of the SMEFT, or the minimal truncation that we consider, for describing any particular extension of the SM.

To date, many studies have examined the reliability of perturbation theory for first-order phase transitions, recent examples include Refs.~\cite{Patel:2011th,Wainwright:2011qy,Wainwright:2012zn,Curtin:2016urg,Chiang:2017nmu,Chiang:2017zbz,Jain:2017sqm,Laine:2017hdk,Cai:2017tmh,Chiang:2018gsn,Prokopec:2018tnq,Gould:2019qek,Carena:2019une,Kainulainen:2019kyp,Senaha:2020mop}.
Previously
the daisy-resummed and
dimensionally-reduced approaches were compared in Refs.~\cite{Cline:1997bm,Gould:2019qek,Kainulainen:2019kyp},
which also include comparisons to lattice simulations.
In this paper, we go beyond such previous studies by carrying out a comprehensive study of a wide range of different theoretical uncertainties relevant for these two approaches, focusing on the implications for the gravitational wave spectrum.

\begin{figure}[t]
    \centering
    \includegraphics[width=.75\textwidth]{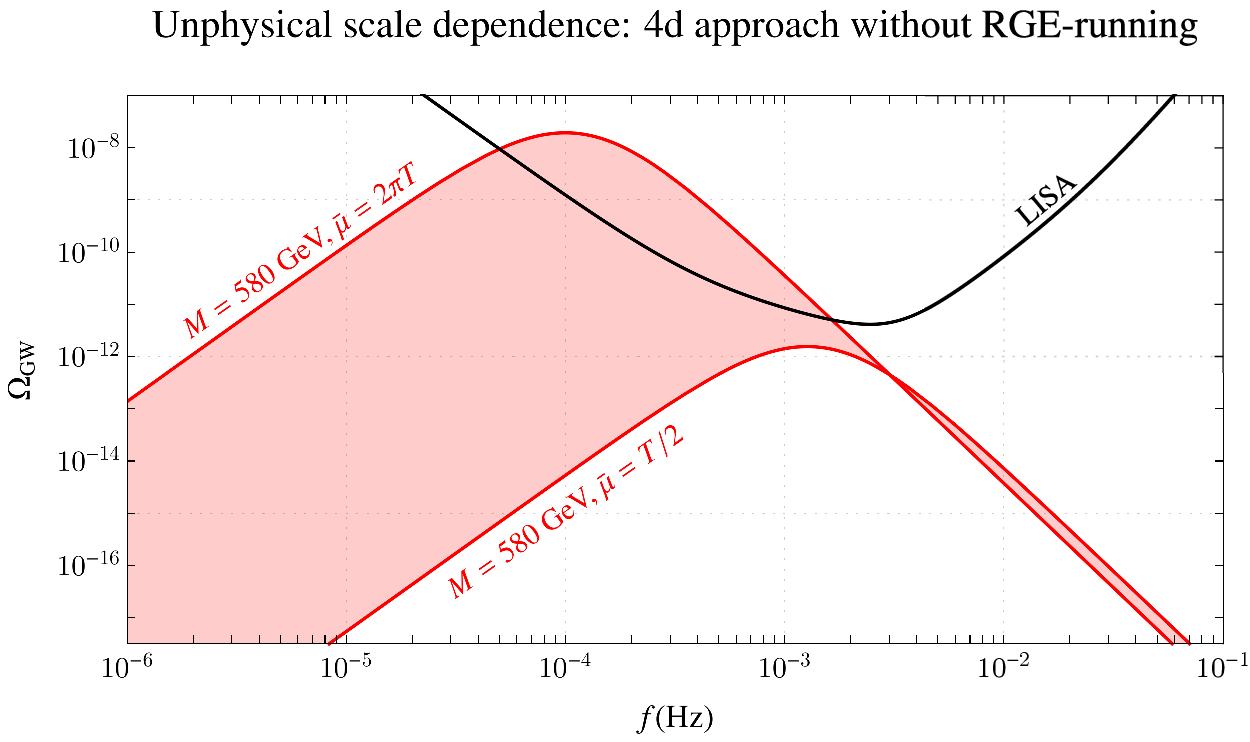}
    \caption{A common method of calculating the thermal parameters of a phase transitions is very sensitive to the choice of renormalisation scale.
    Here we show this dependence in the popular daisy-resummed ``4d~approach'' for a benchmark point of our SMEFT defined by Eq.~\eqref{eq:O6}, without the renormalisation group (RG) running of couplings.
    The LISA signal-to-noise ratios are 6 and 210 for the renormalisation scales $\bmu=T/2$ and $2\pi T$ respectively, for the calculation of which we have used {\em PTPlot}~\cite{Caprini:2019egz} and assumed a three year mission profile.
    }
    \label{fig:scale_dependence_no_running}
\end{figure}

By way of example, Fig.~\ref{fig:scale_dependence_no_running} shows results for the generated gravitational wave spectrum calculated in the daisy-resummed approach with parameters matched at the $Z$-pole. 
The calculations are performed for two different renormalisation scales,
$\mu_{4}=T/2$ and
$2\pi T$.%
\footnote{
Here, unlike in later sections, we neglect the renormalisation group running of the \MSbar-parameters, a common shortcut taken in the literature.
}
Because any dependence on the renormalisation scale is unphysical, the width of the band in Fig.~\ref{fig:scale_dependence_no_running} estimates a corresponding theoretical uncertainty.
The theoretical uncertainty is multiple orders of magnitude such that the sensitivity of LISA to this parameter point is completely ambiguous.

Later, we will show that when one includes the running of such parameters in the daisy-resummed approach, the scale dependence of the gravitational wave peak amplitude reduces by about an order of magnitude.
This compares well both analytically and numerically to the same calculation performed using dimensional reduction at one-loop level.
However, there is a systematic difference due to the breakdown of the gradient expansion in the daisy-resummed approach in the calculation of the nucleation rate.

In dimensional reduction, the inclusion of next-to-leading order terms is comparatively amenable, and somewhat standard.
These terms are essential for fractional uncertainties for many thermodynamic observables to be perturbatively small~\cite{Kajantie:1995dw}.
Further, these terms are precisely what is needed to cancel the leading renormalisation scale dependence.
However, the proof really is in the pudding: by explicit calculation, we find that, with the inclusion of these next-to-leading order terms, the theoretical uncertainties in the dimensionally-reduced approach are numerically {\em much} smaller than in the daisy-resummed approach. 

The outline of this paper is as follows.
Section~\ref{sec:thermodynamics} outlines and compares the daisy-resummed and dimensionally-reduced approaches at a theoretical level.
Daisy resummation is introduced in Sec.~\ref{sec:overview_4d}, and the recipe used to calculate thermodynamic quantities is given in Sec.~\ref{sec:recipe_4d}.
Following this, Sec.~\ref{sec:overview_3d} shows how higher order resummations are incorporated by dimensional reduction, giving an explicit comparison of the effective potentials as series expansions in the couplings.
In Sec.~\ref{sec:recipe_3d} we give the recipe used to calculate thermodynamic quantities in the dimensionally-reduced approach, including an overview of the $\hbar$-expansion.

In Sec.~\ref{sec:theoretical_uncertainties}, we show explicit numerical comparisons of different theoretical uncertainties.
In particular, we study
the importance of scale dependence in
Sec.~\ref{sec:scale_dependence},
gauge dependence in
Sec.~\ref{sec:gauge_dependence},
the high-temperature approximation in
Sec.~\ref{sec:high_temp},
higher loop orders in
Sec.~\ref{sec:higher_loop} and
corrections to bubble nucleation in
Sec.~\ref{sec:nucleation_corrections}.
In Sec.~\ref{sec:nonperturbativity} we also gather existing nonperturbative estimates of the effect of the breakdown of perturbation theory.

Finally, Section~\ref{sec:conclusions} summarises our findings and discusses their consequences for the predicted gravitational wave signal. We have endeavoured to make the main of the document intellectually self-contained, though some topics and many detailed results have been transferred to the appendices.
Appendix~\ref{appendix:smeft:4d} presents the purely 4d parts of our calculations for the SMEFT.
Appendix~\ref{appendix:smeft:3d} provides a hands-on introduction to dimensional reduction, followed by the details of our calculation within the 3d~approach for the SMEFT.
Appendix~\ref{appendix:prefactor} discusses various approximations to the nucleation prefactor in-depth.

\section{Thermodynamics of the phase transition}
\label{sec:thermodynamics}

In studying the thermodynamics of the phase transition, we calculate four thermodynamic parameters that play an important role in the SGWB:
the critical temperature $\Tc$,
the percolation temperature, $\Tp$,
the inverse duration of the phase transition, $\beta/\Hp$, and
the strength of the phase transition, $\alpha$.
In the following, they are defined independently of the methods to calculate them.

The critical temperature for a first-order phase transition defines the temperature at which the free energy of both phases are equal. For homogeneous phases, the free energy density is equal to the effective potential.

To define the percolation temperature, requires first discussing the rate per-unit-volume at which bubbles of the broken phase are nucleated. For first-order phase transitions on cosmological timescales, the bubble nucleation rate takes an exponential, or semiclassical, form,
\beq
\label{eq:rate_exponential}
    \Gamma = A \mbox{e}^{-S_\bub}
    \;.
\eeq
Here $A$ is the nucleation prefactor and $S_\bub$ is the Euclidean action of the critical bubble, or bounce. For a thermal transition this is the Boltzmann suppression
$S_\bub = E_\bub/T$~\cite{Affleck:1980ac,Linde:1981zj},
which we assume throughout. To calculate the nucleation rate from first principles, one can start from the semiclassical result~\cite{Langer:1967ax,Langer:1969bc},
\begin{equation}
\label{eq:nucleation_rate_Z}
    \Gamma = \frac{\freq}{\pi \mc{V}}\frac{\mbox{Im}\,Z[\phi_\bub]}{Z[0]}
    \;.
\end{equation}
Here $Z[\phi_\bub]$ is the contribution to the partition function from the region around the critical bubble, $\phi_\bub$, suitably analytically continued~\cite{Langer:1967ax}.
And $Z[0]$ is the contribution from the region around the high-temperature phase, or false vacuum.
The factor $\mc{V}$ is the volume of space and $\freq$ is a real-time frequency which gives the inverse decay time of the critical bubble.

The percolation temperature, $\Tp$, is taken to be the temperature for which a fraction $1-1/\mbox{e}\approx 0.63$ of the universe has transitioned to the broken phase, following the conventions of Refs.~\cite{Enqvist:1991xw,Caprini:2019egz}. In terms of the action of the critical bubble, this condition can be written as
\begin{equation}
\label{eq:Tp_definition}
    S_\bub \approx 131
    + \ln\left( \frac{A}{\Tp^4} \right)
    - 4 \ln\left( \frac{\Tp}{100\,{\rm GeV}} \right)
    - 4 \ln\left( \frac{\beta/\Hp}{100} \right)+3 \ln (v_w)
    \;,
\end{equation}
where $\beta$ is the inverse time scale of the transition 
\begin{equation}
\label{eq:beta_definition}
    \frac{\beta}{\Hp} = \Tp \left. \frac{{\rm d} S_\bub}{{\rm d}T} \right|_{\Tp}
    \;.
\end{equation}
For strongly supercooled transitions, the second derivative of the tunnelling action can modify the relation between the tunneling action and the inverse time scale of the transition~\cite{Jinno:2017ixd,Hindmarsh:2019phv}.
However, we wish to focus on the theoretical uncertainties arising from using finite temperature quantum field theory perturbatively in estimating the gravitational wave spectrum. The uncertainties arising from other steps in calculating the gravitational wave observables, including an accurate calculation of the mean bubble separation and a precise treatment of the hydrodynamics we leave to future work.

Note that we do not calculate the wall velocity, $v_w$, as it requires a real-time calculation which is beyond the scope of this article. We discard the last term in Eq.~\eqref{eq:Tp_definition}, noting that corrections due to the wall velocity will have small numerical impact on our results as long as $v_w=\mc{O}(1)$ which becomes more likely for stronger phase transitions and even for moderately weak phase transitions, especially if there are no BSM light particles to generate additional friction.

Finally, the appropriate measure of the strength of the transition, $\alpha$, consistent with the conventions of Refs.~\cite{Espinosa:2010hh,Hindmarsh:2017gnf}, is determined by the difference in the trace anomaly, $\Theta$, between the two phases,
\begin{align}
\label{eq:alpha_definition}
    \alpha = \left.  \frac{\Delta\Theta }{\rho_{\rmi{rad}}} \right|_{\Tp}
    \;, \qquad
    \Delta\Theta = \Delta V - \frac{1}{4}\frac{{\rm d}\Delta V}{{\rm d}\ln T}
    \;,
\end{align}
evaluated at the percolation temperature.
Here
$V$ denotes the effective potential of the theory and
$\Delta$ denotes the difference between the broken and symmetric phases.
The numerator, $\rho_{\rmi{rad}}$ is the radiation density of the high-temperature phase, equal to 3/4 of the enthalpy of that phase,
\begin{equation}
    \rho_{\rmi{rad}} = \frac{\pi^2}{30}g_{*}T^4
    \;.
\end{equation}
We take the effective number of relativistic degrees of freedom
$g_{*} = 106.75$, equal to its high-$T$ Standard Model value throughout (without e.g.\ right-handed neutrinos).
A recent work~\cite{Giese:2020rtr} reappraised the correct definition of $\alpha$, and proposed a generalisation of Eq.~\eqref{eq:alpha_definition}. In this article we adopt Eq.~\eqref{eq:alpha_definition} throughout, justified because we focus on quantum field theoretical uncertainties which will be present regardless of the precise definition of $\alpha$.

Concrete calculations will be carried out in the simplest SMEFT truncation.
Its Lagrangian includes the gauge, fermion and Yukawa parts of the Standard Model, and extends the Higgs sector by the single additional operator~\eqref{eq:O6},
\begin{align}
\label{eq:L:SMEFT}
\mc{L}_{\rmii{SMEFT}} &=
      \mc{L}_{\rmii{gauge}}
    + \mc{L}_{\rmii{fermion}}
    + \mc{L}_{\rmii{Yukawa}}
    + \mc{L}_{\rmii{Higgs}}
    \;,\\
\mc{L}_{\rmii{Higgs}} &=
    \frac{1}{2} \left(D_\mu \phi\right)^\dagger\left(D_\mu \phi\right)
    - V_{\rm 0}(\phi)
    \;, \\
\label{eq:Vtree}
V_{\rmii{tree}}(\phi) &=
    \mh^2\, \phi^\dagger \phi
    + \lambda\, (\phi^\dagger \phi )^2
    +\frac{1}{M^2} (\phi^\dagger \phi )^3
    \;.
\end{align}
These parts,
the covariant derivatives $D_{\mu}$,
the gauge fields with corresponding field strength tensors,
the associated gauge couplings, and
ghosts
follow the conventions of Ref.~\cite{Brauner:2016fla}.
In the following, we will also use $c_6\equiv1/M^2$ for the coefficient of the higher dimensional operator, as it is more convenient to work with $c_6$ when carrying out Feynman diagrammatic calculations, but $M$, being related to the energy scale of new physics, aids intuition.
\begin{table}[t]
    \centering
    \begin{tabular}{c|c|c}
    \MSbar-Parameter & Observables & Central Value
    \\ \hline 
    $\mh^2$ & $G_f$ 
    & $G_f=1.1664\times 10^{-5} \ {\rm GeV}^{-2}$
    \\
    $\lambda$ & $M_h$ & $M_h=125.10$~GeV
    \\ 
    $\g$ & $M_W$ 
    & $M_W=80.379$~GeV 
    \\ 
    $\gp$ & $M_Z$ 
    & $M_Z=91.188$~GeV
    \\
    $\gs$ & Various & $\alphas(M_Z) = 0.1179$ 
    \\
    $\gY$ & $M_t$ 
    & $M_t=172.9$~GeV
    \end{tabular}
    \caption{
    Experimental values for observables from Ref.~\cite{Tanabashi:2018oca}.
    Observables and \MSbar-parameters are matched at one-loop order, at the input scale $\bmu=M_Z$
    with details of the matching relations collected in Appendix~\ref{appendix:zero_temp_matching}.
    Multiple observables are involved in calculating a global average of the strong coupling constant $\gs$.
    }
    \label{tab:parameters}
\end{table}
For experimentally measured physical parameters we will use the central values presented in Table~\ref{tab:parameters} throughout the paper, taken from Ref.~\cite{Tanabashi:2018oca}.
Our perturbative calculations use the \MSbar-scheme for renormalisation, with the 4-dimensional renormalisation scale denoted by $\bmu$. We match experimental results to \MSbar-parameters at 1-loop order, matching pole masses using the full 1-loop self-energies. This includes momentum-dependent terms additional to those from evaluating the second derivative of the 1-loop effective potential at the minimum.

\subsection{The 4d~approach: Daisy-resummation}
\label{sec:overview_4d}

The effective potential (the free energy density) encodes the equilibrium properties of a phase transition,
such as its
character (or order),
critical temperature and
latent heat.
While it is possible to compute the effective potential in perturbation theory, the perturbative expansion at high temperatures suffers from problems at low energies.
This is Linde's infamous Infrared Problem~\cite{Linde:1980ts}.
Namely, at high temperatures infrared bosonic modes become highly occupied,
enhancing the effective loop expansion parameter for modes with energy $E\ll T$,
\beq
\label{eq:g:eff}
    g^2 \to g^{2} \nB(E,T) = 
    \frac{g^2}{\mbox{e}^{E/T}-1} \approx
    \frac{g^2 T}{E} \geq
    \frac{g^2 T}{m}
    \;,
\eeq
where $\nB$ is Bose-Einstein distribution and $m$ is mass of the bosonic mode.
At sufficiently high temperatures comparable to $m/g^2$, the infrared bosonic modes become strongly coupled.
Furthermore, infrared divergences appear at finite loop order: at four-loop order for the effective potential~\cite{Linde:1980ts}. 
This means that although the electroweak theory is weakly coupled at zero temperature, massless bosonic modes are nonperturbative at high temperatures and should be treated with appropriate (lattice) techniques. However, it is still possible -- and economical -- to use perturbation theory as a first approximation in studies of phase transitions.%
\footnote{
    Furthermore, in theories with chiral fermions, perturbation theory is required to integrate these out in order to perform lattice simulations for the nonperturbative bosonic fields.
}
For this reason, this section pedagogically describes a recipe for the purely perturbative analysis of cosmological phase transitions. In particular, it describes how to consistently perform resummations to mitigate the Infrared Problem.
We also comment on how to find a nonperturbative solution, after resummations are performed perturbatively in an infrared safe manner.

\subsubsection{Resummation at leading order}
\label{sec:resummation_LO}

We use dimensional regularisation in
$D = d+1 = 4-2\epsilon$ dimensions
and the \MSbar-scheme with renormalisation scale $\bmu$.
We define the notation
$P \equiv (\omega_n,\vec{p})$ for Euclidean four-momenta where the bosonic Matsubara frequency is
$\omega_n = 2\pi n T$ and 
\begin{align}
\label{eq:Tint}
\Tint{P} &\equiv T \sum_{\omega_n} \int_p
\;,\quad\quad
\int_p \equiv \Big( \frac{\bmu^{2}e^\gammaE}{4\pi} \Big)^\epsilon \int \frac{{\rm d}^{d}p}{(2\pi)^d}
\;,\\
\Tint{P}' &\equiv T \sum_{\omega_n \neq 0} \int_p,  \quad \quad 
n_{\rmii{B/F}}(E_p,T) \equiv \frac{1}{e^{E_p/T}\mp 1}
\;.
\end{align}
This last definition is the Bose(Fermi)-distribution with
$E_p = \sqrt{p^2 + m^2}$.
In addition, we parametrise the perturbative expansion in terms of the weak gauge coupling, $g$, and assume the usual power counting for the other coupling constants~\cite{Kajantie:1995dw}
\begin{align}
\label{eq:g_scalings}
    \gp^2 \sim
    \gY^2 \sim
    \lambda \sim
    & \g^{2}
    \;, \nn \\
    c_6 \sim
    & \g^4/\Lambda^2
    \;,
\end{align}
so that the loop expansion and the expansion in powers of $g^2$ are equivalent at zero temperature.
Due to the nonrenormalisability of the $c_6$ term, that relation contains an explicit energy scale, denoted by $\Lambda$, which should be typical of the low energy SMEFT.
At high temperatures we will assume $\Lambda\sim T$.

As an illustrative starting point, let us consider the one-loop correction to the two-point correlator at high temperature.
This contributes to the 1-loop thermal mass of the Higgs field zero-mode.
For a scalar field with \MSbar-mass parameter $m^2$, this correction is of the form (dropping overall symmetry and coupling constant factors)
\begin{align}
\label{eq:hard-loop}
    \TopoST(\Lsci,\Asci) &= 
    I^{4b}_1(m) \equiv \Tint{P} \frac{1}{P^2+m^2}
    \nn\\ &= \underbrace{
    \vphantom{\int_{p}}
    \Big( \frac{\bmu^{2}e^\gammaE}{4\pi} \Big)^\epsilon \int
    \frac{{\rm d}^{D}p}{(2\pi)^D} \frac{1}{p^2+m^2} }_{\equiv I^4_1(m)}
    + \underbrace{
    \int_{p}
    \frac{\nB(E_p,T)}{E_p} }_{I^{\T}_1(m)}
    \;.
\end{align}
Where
$I^{4}_1(m)$ is the UV divergent zero-temperature piece and
$I^{\T}_1(m)$ is the UV finite but IR sensitive finite-$T$ piece.
It is this last temperature-dependent term that leads to problems in the infrared.
In order to see this, it is more useful to write this integral in the form
\begin{align}
\label{eq:I:split:s:h}
I^{4b}_1(m) =
    \underbrace{
    \vphantom{\Tint{P}'}
    \int_{p} \frac{T}{p^2+m^2} }_{\equiv I_{\rmii{soft}}(m)}
    + \underbrace{
    \Tint{P}' \frac{1}{P^2+m^2} }_{I_{\rmii{hard}}(m)}
    \;,
\end{align}
separating the soft zero-mode from the hard non-zero Matsubara modes.
In fact, at high temperature $m/T \sim g$ (note that this choice merely parametrises the high-$T$ limit),
the mass of the zero-mode scales as $\sim gT$ while all non-zero modes exceed this, with masses of $\sim\pi T$ -- this signals a scale hierarchy.

Therefore, in the high-$T$ limit, non-zero mode excitations of the thermal plasma effectively screen the zero mode.
The zero mode acquires an effective thermal mass
$m^2_{\T} = m^2 + \# g^2 T^2$,
where the numerical coefficients, denoted generically by $\#$, depend on the group structure and representation of the fields in question.
Physically this thermal mass arises as a screening mass due to the heat bath. Since $m^2$ can be negative, thermal corrections can trigger a phase transition around where they cancel the zero-temperature contribution, i.e.\ when the effective mass of the zero-mode becomes ultrasoft (we will expand upon this point later).

Similarly, the gauge field zero-mode is also screened by non-zero modes.
Since it is massless in the symmetric phase (or gauge eigenstate basis), its thermal mass is solely dictated by the hard modes and reads
$\mD = \# g T$.
This thermally induced mass is called the gauge field Debye mass, in analogy to Debye screening of the electric plasma.
In Appendix~\ref{appendix:dr_for_beginners} we show the calculation of these thermal masses in detail.

The infrared problem manifests itself when considering higher loop contributions, the so-called daisy diagrams
\begin{align}
\label{eq:daisy}
\TopnSTT(\Lsci,\Asci,\Asaii,N,)
\propto g^{2N} \Biggl[ \int_p
    \frac{T}{(p^2+m_{\T}^2)^N} \Biggr]
    \Biggl[ \Tint{Q}' \frac{1}{Q^2} \Biggr]^N \propto
    m_{\T}^3 T \Bigl(\frac{g T}{m_{\T}}\Bigr)^{2N}
    \;,
\end{align}
where we have omitted an overall combinatorial factor and replaced $\lambda$ by $g^2$ according to its assumed scaling.
In these diagrams
the hard mode contributions (double dashed lines) screen soft zero-modes (single dashed lines).
When the inner loop of Eq.~\eqref{eq:daisy} is a zero-mode, i.e.\ has a soft momenta $P=(0,\vec{p})$ and all $N$ outer loops or petals have hard momenta $Q$ with non-vanishing Matsubara frequencies, this contribution is of order $\mathcal{O}(g^3)$ for any $N$.
Furthermore, it is IR-divergent for $N\geq 2$ in the limit of vanishing mass.
For scalar fields and zero-components of gauge fields this IR-problem can be partially cured.
The recipe is called daisy resummation.
One must first calculate the thermal corrections from the non-zero modes to find the corrected mass of the zero-mode.
Then, in computing the contributions of the zero-mode, its mass is upgraded to the thermally corrected mass~$m_{\T}$.
The following subsection describes this prescription in more detail.

Now, we turn to the effective potential, which at one-loop is of the following form, in terms of the background field $\phi$
\begin{align}
    V_{\rmii{eff}}(\phi,T,\bmu) &=
        V_{\rmii{tree}} +
        V_{\rmii{1-loop}}
    \;,
\end{align} 
where the one-loop piece is composed of the master sum-integral%
\footnote{
For a brief review in a similar context, see Appendix~C of Ref.~\cite{Delaunay:2007wb}.
}
\begin{align}
\label{eq:J}
    V_{\rmii{1-loop}} \simeq J_{\rmii{1-loop}} \equiv
    \frac{1}{2} \Tint{P} \ln \big(P^2 + m^2 \big)
    \;,
\end{align} 
where $m^2$ is a $\phi$-dependent mass eigenvalue, and for the full effective potential all mass eigenvalues are summed over with proper coefficients for scalar, gauge and fermion fields. Additionally, in renormalised perturbation theory, there is a term with counterterms that we have omitted for simplicity.
For the complete effective potential, see Appendix~\ref{appendix:V:1loop}.  
Customarily, the sum-integral \eqref{eq:J} is split into
a zero-temperature (Coleman-Weinberg) piece and
a temperature-dependent piece (thermal function)
\begin{align}
\label{eq:J:split}
J_{\rmii{1-loop}} =
    \underbrace{
    \vphantom{\int_{p}}
    \frac{1}{2} \Big( \frac{\bmu^{2}e^\gammaE}{4\pi} \Big)^\epsilon \int \frac{{\rm d}^{D}p}{(2\pi)^D} \ln(p^2+m^2)}_{\equiv J_{\rmii{CW}}(m)}
    %
    %
    %
    - \underbrace{
     \int_{p} T\ln \Big( 1 \pm n_{\rmii{B/F}}(E_p,T)  \Big)}_{
    J_{\T,b/f}\big(\frac{m^2}{T^2}\big)}
    \;,
\end{align}
with minus (plus) for bosons (fermions) in the thermal functions
$J_{\T,b/f}$
given in Eqs.~\eqref{eq:J:b} and \eqref{eq:J:f}, evaluated in $d=3-2\epsilon$.
Alternatively, it is useful to separate
the soft (zero) mode and
the hard (non-zero Matsubara) mode contributions in the master sum-integral
\begin{align}
J_{\rmii{1-loop}} =
    \underbrace{
    \vphantom{\Tint{P/\{P \} }'}
    \frac{T}{2} \int_{p} \ln(p^2+m^2)}_{\equiv T J_{\rmii{soft}}(m)} +
    \underbrace{\frac{1}{2} \Tint{P/\{P \} }' \ln (P^2 + m^2 )}_{\equiv J_{\rmii{hard}}(m)}
    \;.
\end{align}
Next, daisy-resummation can be performed by replacing the masses of the zero-modes by thermal screening masses%
\footnote{
    Technically, this resummation can be achieved by adding and subtracting thermal masses for soft modes in the Lagrangian, such that
    terms with plus sign contribute to the mass and
    terms with minus sign are treated as counterterm-like interactions~\cite{Karsch:1997gj,Andersen:2008bz}.
    This reorganises the perturbative expansion while the original Lagrangian stays untouched.
} 
\begin{align}
\label{eq:Jsoft}
T J_{\rmii{soft}}(m) =
- \frac{T}{12\pi} (m^2)^\frac{3}{2} \to
T J^{\rmii{resummed}}_{\rmii{soft}}(m) =
- \frac{T}{12\pi} (m^2 + \Pi_{\T})^\frac{3}{2}
\;,  
\end{align}
where $\Pi_{\T}$ is the one-loop thermal contribution to the screening mass. 
By writing 
\begin{align}
    J_{\rmii{daisy}}(m) \equiv
    J^{\rmii{resummed}}_{\rmii{soft}}(m) -
    J_{\rmii{soft}}(m) = - \frac{T}{12\pi} \Big( (m^2 + \Pi_{\T})^\frac{3}{2} - (m^2)^\frac{3}{2} \Big)
    \;,
\end{align}
we end up with the Arnold-Espinosa type~\cite{Arnold:1992rz} -- or ring-improved -- resummed effective potential 
\begin{align}
V_{\rmii{eff}}^{\rmii{A-E res.}}(\phi,T,\bmu) &=
    V_{\rmii{tree}} +
    V_{\rmii{CW}} +
    V_{\T} +
    V_{\rmii{daisy}}
    \;.
\end{align}
where
$V_{\T}\simeq J_{\T,b/f}$ and
$V_{\rmii{daisy}}\simeq J_{\rmii{daisy}}$.
Note, in order to reach this familiar form where the zero-temperature pieces are separated from thermal pieces, we had to subtract the original soft contribution from the resummed one, in order to avoid double counting. Instead of this 
form of resummation -- that is encoded in the daisy term -- we could simply write the resummed effective potential as   
\begin{align}
V_{\rmii{eff}}^{\rmii{resummed}}(\phi,T,\bmu) &=
    V_{\rmii{tree}} +
    V^{\rmii{resummed}}_{\rmii{soft}} +
    V_{\rmii{hard}}
    \;,
\end{align}
with $V_{\rmii{hard}}\simeq J_{\rmii{hard}}$.
This form is equal to the Arnold-Espinosa form in the case where only mass parameters have been resummed. 
However, in the special case of the SMEFT, the new six-leg vertex introduces qualitatively new features to resummation.
In fact, the leading so-called {\em flower} contributions of the dimension-six coupling $c_{6}$ in SMEFT (cf.~Eq.~\eqref{eq:hard-loop} for notation)
\begin{align}
\label{eq:flower}
    \ToptSTTT(\Lsci,\Ascii,\Ascii) &\simeq -36 \; c_6 \; [I^{4b}_1(\mh)]^2
    \;,\\[2mm]
    \TopoVA(fex(\Lsci,\Lsci,\Lsci,\Lcsi),\Ascii) &\simeq -24 \; c_6 \; I^{4b}_1(\mh)
    \;,
\end{align}
appear
at 2-loop order for the mass parameter $\mh^2$ and
at 1-loop order for the scalar self-coupling $\lambda$.
Their effect incorporates thermal screening by resumming not only the mass parameter but also the self-coupling.
Appendix \ref{appendix:V:1loop} fully derives the one-loop effective potential in the SMEFT with leading order daisy resummations. 

Finally, let us comment on gauge invariance.
As Secs.~\ref{sec:recipe_3d} and \ref{sec:gauge_dependence}
explain in detail, in perturbation theory a gauge-invariant treatment requires an $\hbar$-expansion, in which the effective potential is expanded around its tree-level minimum. However, the Arnold-Epinosa type resummed effective potential -- with an inclusion of thermal corrections in the soft parts -- reorganises the perturbative expansion and departs from the strict $\hbar$-expansion, since the thermal correction $\Pi_{\T}$ is of order $\mc{O}(\hbar)$~\cite{Patel:2011th}.
In Ref.~\cite{Patel:2011th}, a prescription to cure this problem to ensure gauge-invariance has been proposed, and we will comment on this proposition in Sec.~\ref{sec:higher_order_resummation}.
We therefore do not implement the $\hbar$-expansion in our 4-dimensional approach for computing thermodynamic parameters. As such, our 4d analysis retains an unphysical gauge-dependence which leads to a theoretical uncertainty we calculate in Sec.~\ref{sec:gauge_dependence}.  
For a recent introductory review of daisy resummation, see Ref.~\cite{Senaha:2020mop}. 

\subsubsection{Daisy-resummed recipe for thermodynamics}
\label{sec:recipe_4d}

Here we outline the calculation of the thermodynamic parameters, which are calculated from the effective potential in 4-dimensional perturbation theory to one loop, including both its scale and gauge dependence. In way of summary, a brief recipe of the approach follows:
\bi
    \item
    Fix the zero-temperature \MSbar-parameters by matching to physical observables at the input scale, here $m_Z$, and then run them to a scale characterising the phase transition, e.g. $\bmu\sim T$.
    Optimise $\bmu$ according to the principle of minimal sensitivity~\cite{Stevenson:1981vj}.
    \item
    Calculate the effective potential of the 4d theory by summing the tree-level potential, the zero-temperature Coleman-Weinberg piece and the finite-temperature piece, with daisy resummation.
    \item
    Numerically find the minima of the real part of the effective potential to determine the phase structure and pattern of phase transitions.
    \item
    Solve the bounce equation with the potential given by the real part of the effective potential.
    From this, solve Eq.~\eqref{eq:beta_definition} to find
    the percolation temperature $\Tp$ and
    the inverse duration of the transition, or $\beta/\Hp$.
\ei

The first step in the daisy-resummed recipe consists of zero-temperature physics, and hence is the same as for the 3d~approach. For the SMEFT, the details are given explicitly in Appendix~\ref{appendix:zero_temp_matching}.

The phases, distinguished by different Higgs vacuum expectation values, are found by numerically minimising the real part of the effective potential,
\beq
\label{eq:ReV}
    V(\phi,T,\bmu)=\mbox{Re}(V_{\rmii{eff}}^{\rmii{A-E res.}}(\phi,T,\bmu))
    \;,
\eeq
with respect to the background field $\phi$.
The imaginary part of the effective potential can be related to the growth rate of long-wavelength modes about a constant background field~\cite{Weinberg:1987vp}. We treat the presence of this nonzero imaginary part as a source of systematic uncertainty in our daisy-resummed calculation. Following standard practice in the literature~\cite{Delaunay:2007wb}, we content ourselves with checking that the imaginary part is much smaller than the real part of the effective potential at its minima.

The nonzero imaginary part of the effective potential, which has to be removed by hand in Eq.~\eqref{eq:ReV}, gives a hint that something is not right. The interpretation of this imaginary part as a decay rate~\cite{Weinberg:1987vp} does little to allay this suggestion, as this decay rate is not exponentially suppressed and hence is generically a much faster process than bubble nucleation.
Further, this decay rate may be nonzero at the broken minimum solved numerically using Eq.~\eqref{eq:ReV}, suggesting that the broken phase itself will decay into another phase with nonhomogeneous Higgs vacuum expectation value (vev).
Both this problem and the problem of gauge dependence are circumvented in the $\hbar$-expansion, which leads to a real effective potential, but this method unfortunately is incompatible with daisy resummation.

To calculate the rate of bubble nucleation, and in particular, the effective tunneling action, we assume O(3) symmetry and solve the bounce equation
\begin{align}
\label{eq:bounce_4d}
    \frac{{\rm d}^{2}\phi}{{\rm d}\rho^{2}} +
    \frac{2}{\rho}\frac{{\rm d}\phi}{{\rm d}\rho} &=
    \frac{{\rm d}V(\phi,\bmu,T) }{{\rm d}\phi}
    \;,
\end{align}
with boundary conditions,
\begin{align}
\label{eq:bounce_boundary}
    \phi(\rho \to \infty)&=0
    \;, \\
    \left.\frac{{\rm d}\phi}{{\rm d}\rho}\right|_{\rho=0}&=0
    \;.
\end{align}
This approach essentially follows Ref.~\cite{Linde:1981zj}. Equation \eqref{eq:bounce_4d} is typically solved using the shooting method, here we employ
{\verb AnyBubble } and
{\verb BubbleProfiler }~\cite{Masoumi:2016wot,Athron:2019nbd}.%
\footnote{
    Very small differences resulting from these different methods are at the percent level and as they are not quantum field theoretic uncertainties.
    We do not present these differences here.
    Where inconsistent, we take the geometric mean of results.
} 
Evaluating the Euclidean action on this solution then yields $S_{\bub}(T,\bmu)$, from which the thermal parameters can be found using Eqs.~\eqref{eq:Tp_definition}--\eqref{eq:alpha_definition}.
In the
4d~approach, we take the prefactor to be $\ln(A/T^4)\sim -14$,
following Ref.~\cite{Caprini:2019egz}.%
\footnote{
    Note that this expression for the nucleation prefactor is a rough guess based on the results of Refs.~\cite{Csernai:1992tj,Carrington:1993ng}. The expression, however, does not reproduce the temperature dependence of the prefactor derived in Refs.~\cite{Csernai:1992tj,Carrington:1993ng}, nor is applicable beyond the parameter point for the SM with light Higgs studied in Ref.~\cite{Carrington:1993ng}. Further, Appendix~\ref{appendix:prefactor} shows that Refs.~\cite{Csernai:1992tj,Carrington:1993ng} contain a significant error in their result for the (statistical part of the) prefactor. Regardless, as we argue in Sec.~\ref{sec:nucleation_corrections}, even the definition of the prefactor is problematic in the daisy-resummed approach, so we adopt this estimate nevertheless.
}

\subsection{The 3d~approach: Dimensional-reduction}
\label{sec:overview_3d}

Dimensional reduction is a general framework for studying the thermodynamic properties of quantum field theories at high temperatures.
It applies widely, and has been particularly fruitful in application to non-Abelian gauge theories.
While its use in hot QCD is standard and by now approaches impressive orders in perturbation theory~\cite{Ghisoiu:2015uza,Laine:2018lgj} (cf.~Refs.~\cite{Laine:2016hma,Ghiglieri:2020dpq} for reviews), 
its success within electroweak theories and studies of EWPT is far less exploited --
even though it proved essential in understanding the phase transition of
the Standard Model~\cite{Kajantie:1995dw,Kajantie:1996mn,DOnofrio:2015gop}
and various bulk thermodynamic properties therein~\cite{Gynther:2005dj,Gynther:2005av}.
Despite featuring in early studies of
supersymmetric extensions of
the SM~\cite{Losada:1996ju,Losada:1998at,Laine:1998qk,Laine:2012jy} and of
the two-Higgs doublet model (2HDM)~\cite{Andersen:1998br}, 
only more recently has the use of dimensional reduction in cosmology been reinvigorated,
in studies of the SM with extended scalar sectors, such as
the real singlet extension (xSM)~\cite{Brauner:2016fla,Gould:2019qek},
the real triplet extension ($\Sigma$SM)~\cite{Niemi:2018asa,Niemi:2020hto}, and
the 2HDM~\cite{Andersen:2017ika,Gorda:2018hvi,Kainulainen:2019kyp}.

High-temperature dimensional reduction (DR) is based on a hierarchical separation of energy scales.
In accordance with the effective expansion parameter \eqref{eq:g:eff},
the underlying scales
\begin{equation}
\label{eq:scale_hierarchy}
    g^{2}T/\pi \ll
    g^{ }T \ll
    \pi T
    \;,
\end{equation}
render the theory
perturbative at
the hard scale ($p\sim\pi T$),
barely perturbative at
the soft scale ($p\sim g^{ }T$), and
non-perturbative at
the ultrasoft scale ($p\sim g^{2}T$).
Here $p=|p|$ denotes a momentum scale of particles in the heat bath.
Note, that related literature~\cite{Kajantie:1995dw} interchangeably refers to
the hard scale as superheavy,
the soft scale as heavy, and
the ultrasoft scale as light.

This hierarchy classifies degrees of freedom when constructing an effective field theory (EFT) for its ultrasoft sector; see Table~\ref{tab:dr:smeft}.
\begin{table}
\centering
\resizebox{\textwidth}{!}{%
\renewcommand{\arraystretch}{1.75}
\begin{tabular}{cccccc}
  \multicolumn{6}{l}{{\sl Start: {\bf $(d+1)$-dimensional SMEFT}}} \\
  \hline
  \textbf{Scale} &
  \textbf{Validity} &
  \textbf{Dimension} &
  \textbf{Lagrangian} &
  \textbf{Fields} &
  \textbf{Parameters} \\
  \hline
  {\sl Hard} & $\pi T$ & $d+1$ &
  $\mathcal{L}_{\rmii{SMEFT}}$~\eqref{eq:L:SMEFT} &
  $A_{\mu},B_{\mu},C_{\mu},\phi,\psi^{ }_{i}$ &
  $\mh^{2},\lambda,c^{ }_{6},
  \g,\gp,\gs^{ },\gY^{ }$ \\
  &&\multicolumn{4}{l}{$\Big\downarrow$ {\sl Integrate out $n\neq 0$ modes and fermions}} \\
  {\sl Soft} & $g T$ & $d$ &
  $\mathcal{L}_{3d}$~\eqref{eq:3d:soft:action} &
  $A_{r},B_{r},C_{r},$ &
  $\mu_{h,3}^{2},\lambda^{ }_{3},c^{ }_{6,3},
  \g^{ }_{3},\mD^{ },$\\
  &&&&
  $A^{ }_{0},B^{ }_{0},C^{ }_{0},\phi$ &
  $\gp_{3},\mD',g^{ }_{\rmi{s},3},\mD''$
  \\
  &&\multicolumn{4}{l}{$\Big\downarrow$ {\sl Integrate out temporal adjoint scalars $A_{0},B_{0},C_{0}$}} \\
  {\sl Ultrasoft} & $g^{2}T/\pi$ & $d$ &
  $\bar{\mathcal{L}}_{3d}$~\eqref{eq:3d:ultrasoft:action} &
  $A_{r},B_{r},C_{r},\phi$ &
  $\bar{\mu}_{h,3}^{2},
  \bar{\lambda}^{ }_{3},\bar{c}^{ }_{6,3},
  \bar{\g}^{ }_{3},\bar{g}_{3}',\bar{g}_{\rmi{s},3}$
  \\\hline
  \multicolumn{6}{l}{{\sl End: {\bf $d$-dimensional Pure Gauge}}} \\
\end{tabular}
}
\caption[Dimensional reduction of SMEFT]{
  Dimensional reduction of $(d+1)$-dimensional SMEFT into
  effective $d$-dimensional theories based on the scale hierarchy at high temperature.
  The effective couplings are functions of the couplings of their
  parent theories and temperature and are determined by a matching procedure.
  The first step integrates out
  all hard non-zero modes.
  The second step integrates out
  the temporal adjoint scalars $A^{ }_{0},B^{ }_{0},C^{ }_{0}$
  with soft Debye masses $\mD^{ },\mD',\mD''$.
  At the ultrasoft scale, only ultrasoft spatial gauge fields $A_{r},B_{r},C_{r}$
  (with corresponding field-strength tensors $G_{rs},F_{rs},H_{rs}$)
  remain along with a light Higgs that undergoes the phase transition.
  }
\label{tab:dr:smeft}
\end{table}
In the Matsubara formalism of thermal field theory
the hard scale screens the purely spatial (static) zero-modes which live at
the soft scale.
At sufficiently high temperature the infinite tower of non-zero modes is integrated out in a conventional EFT sense.
This includes all bosonic non-zero modes and all fermionic modes.
Their effect and temperature dependence is encoded solely in the parameters of the resulting EFT of lower dimension.
Due to the heat bath breaking Lorentz invariance for temporal gauge fields,
the 3-dimensional EFT contains temporal remnants of gauge fields that are adjoint Lorentz scalars ($A_0,B_0,C_{0}$).
They get screened at the scale of their respective Debye masses
$\mD^{ },\mD',\mD''\sim\mc{O}(gT)$.
Furthermore, since the spatial gauge bosons are only Debye screened at the next natural order $\mc{O}(g^{2}T)$,
an additional scale separation emerges between the soft scale of adjoint temporal scalars and the ultrasoft scale.
The effective theory of the ultrasoft scale is then non-perturbative, since
$g^{2}\nB\sim\mc{O}(1)$.
Note that, massive bosonic scalar fields may assume all three scales depending on their zero-temperature mass.

The separation of scales defines the high-temperature regime and
generically holds for phase transitions involving scalars in weakly coupled theories.
With decreasing temperature the zero Matsubara modes of the scalars signal the absolute instability of the high-temperature phase, below some temperature $T_0$.
A scalar thermal mass, of the general form $m_{\T}^2 = m^2 + \# g^2 T^2$, goes through zero at the temperature
\begin{align}
\label{eq:scalar_thermal_mass}
    m_{\T}^2 = 0 \;
    \Rightarrow \; T_0 \sim \frac{\sqrt{-m^2}}{g}\;.
\end{align}
This is generically in the high-temperature regime, at least regarding the scalar field undergoing the transition, because the temperature is larger than the vacuum mass parameter by a factor of $1/g$.
Note also that $T_0 < \Tp < \Tc$,
as the thermal mass is necessarily positive at both $\Tp$ and $\Tc$,
so implying that $\Tp$ and $\Tc$ are generically in the high-temperature regime.
This further suggests that bubble nucleation will almost always take place via a purely spatial, O(3) and not O(4) symmetric, instanton~\cite{Linde:1980tt,Linde:1981zj,Garriga:1994ut}.
In contrast to the scalar undergoing the phase transition, for the temporal gauge bosons the lack of a (negative) vacuum mass implies that they are always of the soft scale, and hence are integrated out in constructing the EFT of the ultrasoft scale.

The Higgs zero Matsubara mode is treated as ultrasoft throughout our analysis of the SMEFT.
At temperatures relevant for the dynamics of the phase transition, between the percolation and critical temperatures, the thermal mass of the scalar zero mode is positive.
But following Eq.~\eqref{eq:scalar_thermal_mass} it is at most of order $gT$, and hence is either of the soft or ultrasoft scale.
An ultrasoft Higgs mass is certainly correct in the vicinity of $T_0$ where the vacuum and thermal mass contributions exactly cancel, but should also hold near $\Tp$ and $\Tc$ due to a remaining partial cancellation of vacuum and thermal mass contributions. Regardless, we do not expect any significant discrepancies between treating the Higgs as soft versus ultrasoft due to the small numerical effects of the temporal gauge fields.

The philosophy of dimensional reduction is to treat
perturbative modes perturbatively and
nonperturbative modes nonperturbatively.
Fermions and bosonic non-zero Matsubara modes are perturbative, and are treated perturbatively when integrated out in the construction of the EFTs.
The bosons of the soft scale are also perturbative, and are treated similarly.
Since only the ultrasoft scale is nonperturbative this scale is then normally treated with non-perturbative lattice studies.
Existing lattice studies utilise the super-renormalisability of the EFT to perform an exact mapping between bare lattice parameters and \MSbar-parameters~\cite{Laine:1995np,Laine:1997dy}. However, in the EFT we consider for the SMEFT, the presence of the marginal, sextic Higgs field operator $\mc{O}_{6}$ means that the EFT is merely renormalisable and not super-renormalisable, aggravating the matching of lattice parameters to known physics.
Nevertheless, recent lattice computations in scalar-extended BSM models~\cite{Kainulainen:2019kyp,Niemi:2020hto} have indicated that, for relatively strong transitions in weakly coupled theories, two-loop perturbation theory within the ultrasoft EFT describes the phase transition with reasonable accuracy; see also Sec.~\ref{sec:nonperturbativity}.
There are a few reasons for this perhaps surprisingly good agreement between lattice and perturbation theory.
On the one hand, by constructing this effective theory for the ultrasoft modes, dimensional reduction makes it easier to hone in on these important modes and to treat them to higher loop order than is otherwise possible.
On the other hand, at least in the case of a strong transition, the transition depends most strongly on the scalar sector, which is, in a concrete sense, less nonperturbative than the spatial gauge bosons, for which there are true IR divergences in the symmetric phase at finite loop order.
Further, the IR divergences of the spatial gauge bosons only arise at higher loop order, for example, at four-loop order for the free-energy.
So when the first few terms of the loop expansion converge well, one can expect the nonperturbative effects to be relatively small.

In practice dimensional reduction is performed along the modern EFT recipe.
One first identifies the most general Lagrangian that respects the symmetries of the full theory, and then matches static Green's functions to determine the parameters of the EFT in terms of temperature and parameters of the parent theory.
For a fuller explanation of dimensional reduction, we refer to our Appendix~\ref{appendix:smeft:3d}, which accounts step-by-step of how to construct such effective theories in phenomenologically relevant models.
Therein, Appendix~\ref{appendix:dr_for_beginners} present a breakdown of the calculation of the ${\rm SU}(2)$ Debye mass, which we hope suitably introduces the nitty-gritty of dimensional reduction.
Appendix~\ref{appendix:matching} presents our explicit results for the dimensional reduction of the SMEFT at full NLO.

\subsubsection{Resummations at higher orders and gauge invariance}
\label{sec:higher_order_resummation}

In dimensional reduction, higher order resummations are systematically incorporated order-by-order in powers of the couplings.
This is achieved by careful power counting, necessary because thermal screening breaks the alignment between the loop and coupling expansions.%
\footnote{
We replace all couplings by appropriate powers of the gauge coupling $\g$ according to Eq.~\eqref{eq:g_scalings}.
}
By contrast, in the 4d~approach, resummation is carried out in a more {\em ad hoc} way, by identifying and resumming infrared sensitive parts at the level of Feynman diagrams.
As has long been recognised~\cite{Arnold:1992rz}, at higher orders it is necessary to resum new classes of diagrams beyond just the daisy diagrams.

One-loop daisy resummation, as presented above, generates the effective potential accurately up to $\mc{O}(g^3)$.
However -- as argued in Sec.~2.2 in Ref.~\cite{Kajantie:1995dw} -- one must go beyond this and achieve $\mc{O}(g^4)$ accuracy in order to obtain perturbatively small fractional uncertainties for many infrared observables.
Further, the RG running of the leading order effective potential starts at $\mc{O}(g^4)$, so one must reach this order to control the RG scale dependence.
This requires two-loop contributions, both to the effective potential and to the resummed thermal masses.
It also requires additional resummations at one-loop order: both to the couplings and to the field itself, the latter due to the momentum dependence of thermal screening at $\mc{O}(g^4)$.
Dimensional reduction provides a systematic means to keep track of these disparate resummations, and is extendable to still higher orders.

The effective potential provides a convenient means to show how the differences between the 3d and 4d~approaches manifest in concrete calculations. Schematically there is a relation of the form
\begin{align} \label{eq:Veff_comparison}
    T \; V^{\rmii{3d}}_{\rmii{eff}} \simeq
    V^{\rmii{4d}}_{\rmii{eff}}
    \;,
\end{align}
which holds up to $\mc{O}(g^3)$.
Note that at leading order in powers of $g^2$, the 3d and 4d fields are related as
$\phi_{\rmii{3d}} = \phi_{\rmii{4d}}/\sqrt{T}$.
At higher orders, momentum dependent thermal screening modifies this relation, as captured in the 3d matching relations.
Here, for simplicity, we compare the 3d effective potential at the soft scale, leaving discussion of the effects of integrating out the soft scale to later in this section.

To understand in more detail where the two approaches differ, we break down Eq.~\eqref{eq:Veff_comparison}, giving
\begin{align}
\label{eq:veffs}
    T\;\bigl(
      V^{\rmii{3d}}_{\rmii{tree}}
    + V^{\rmii{3d}}_{\rmii{loops}}\bigr) \simeq
      V^{\rmii{4d}}_{\rmii{tree}}
    + V^{\rmii{4d}}_{\rmii{hard}}
    + V^{\rmii{4d}}_{\rmii{soft, resummed}}
    \;.
\end{align}
From the construction of the dimensionally-reduced EFT, one can deduce the following approximate equality for the hard contributions
\begin{align} \label{eq:Vhard_comparison}
    T \; V^{\rmii{3d}}_{\rmii{tree}} &\simeq
      V^{\rmii{4d}}_{\rmii{tree}}
    + V^{\rmii{4d}}_{\rmii{hard}}
    \;.
\end{align}
This follows since the effective potential is the generator of one-particle irreducible (1PI) correlation functions and 3d parameters are defined by matching the 1PI correlation functions to the 4d theory.

Utilising $\phi \sim T$ for the dimensionful background field we arrive at the following schematic power counting,
\begin{align}
    T \; V^{\rmii{3d}}_{\rmii{tree}}
    &\approx T^4 ( \# g^2 + \# g^4 + \dots )
    \;.
\end{align}
As we indicate, this equation is free from nonanalytic dependence on $g^2$ because it involves only hard modes.
Our 4d~approach correctly captures only the leading order term in this expansion, that first discussed in Ref.~\cite{Dolan:1973qd}.
Both this leading term and the $\mc{O}(g^4)$ term are captured in the NLO matching relations of the 3d~approach.
Appendix~\ref{appendix:matching} presents the NLO matching relations for the SMEFT.

Daisy resummation is engineered to correctly describe the leading effects of the soft scale. This results in the following approximate equality for the remaining soft parts
\begin{align}
    T\; V^{\rmii{3d}}_{\rmii{loops}} &\simeq
    V^{\rmii{4d}}_{\rmii{soft, resummed}}
    \;.
\end{align}
In the soft sector, the presence of infrared modes leads to a nonanalytic dependence on $g^2$,
\begin{align}
    T\; V^{\rmii{3d}}_{\rmii{loops}} &\approx T^4 (\# g^3 + \# g^4 + \dots )
    \;.
\end{align}
As we have explicitly verified in the SMEFT, the 4d~approach correctly reproduces the $\mc{O}(g^3)$ term.
In the 3d~approach, by including two-loop corrections to the effective potential, we capture also the $\mc{O}(g^4)$ term.
Appendix~\ref{appendix:Veff3d} yields an expression for the 3d two-loop effective potential in the SMEFT.
To compute the full $\mc{O}(g^5)$ term requires a three-loop computation~\cite{Zhai:1995ac,Ghiglieri:2020dpq}, whereas the $\mc{O}(g^6)$ term is nonperturbative~\cite{Linde:1980ts}.


The comparisons made in this section utilise the 3d effective potential of the soft sector.
However, to simplify the thermodynamic calculations in our 3d~approach we integrate out the soft temporal bosons and instead utilise the 3d effective potential of the ultrasoft sector;
see the end of Appendix~\ref{sec:matching-relations}.
This additional step incorporates both the $\mc{O}(g^3)$ and $\mc{O}(g^4)$ effects of thermal fluctuations of the soft sector into the parameters of the ultrasoft theory.
A difference does arise, though, regarding the dependence on the Higgs vev of the masses of the temporal gauge bosons.
For the SMEFT with the 3d EFT truncated at $(\hsq)^3_{\rmii{3d}}$, this difference arises at $O(g^3(\hsq)^4_{\rmii{3d}}/T)$ for the 3d potential.
Although formally of $\mc{O}(g^3)$ for a transition with $\phi\sim T$, this discrepancy is accompanied by a sizable numerical suppression, $\mc{O}(10^{-6})$, due to combinatorial and loop factors.
Thus it is expected to have only a very small numerical effect, though it could become more significant for very strong transitions.

Beyond aiding higher order computations, an additional benefit of the 3d~approach, is that one can achieve exact order-by-order gauge invariance by applying the $\hbar$-expansion {\em inside} the 3d effective theory, cf.\ Sec.~\ref{sec:recipe_3d}.
In this expansion, the value of the effective potential is computed as an expansion around the minimum of $V^{\rmii{3d}}_{\rmii{tree}}$.
This possibility depends upon the gauge invariance of the 3d matching relations.
In Appendix~\ref{appendix:matching} we show this explicitly for the dimensional reduction of the SMEFT:
choosing a general covariant gauge, the $\xi_{i}$-dependencies cancel duly in the matching relations up to $\mc{O}(g^4)$
(and higher order terms can be discarded).

Reference~\cite{Patel:2011th} -- in order to maintain gauge invariance -- proposed an alternative resummation approach.
In this the soft part of the LHS of Eq.~\eqref{eq:veffs} is evaluated at the (temperature dependent) minimum of
$V^{\rmii{3d}}_{\rmii{tree}}$%
\footnote{
    Note that Ref.~\cite{Patel:2011th} essentially performs a leading order dimensional reduction for resummation, i.e.\ just
    one-loop for mass parameters and
    tree-level for couplings.
},
but the remaining tree-level and hard parts are evaluated in an expansion around the (temperature independent) minimum of $V^{\rmii{4d}}_{\rmii{tree}}$.
This approach differs from the approaches presented here already at leading $\mc{O}(g^2)$ order, since the minima of the 4d and 3d tree-level potentials differ at leading $\mc{O}(g^0)$ order.
Had the authors used the 3d minimum also in
$V^{\rmii{4d}}_{\rmii{tree}} + V^{\rmii{4d}}_{\rmii{hard}}$
in addition to the soft, resummed part, their potential would have matched those presented here up to $\mc{O}(g^3)$.

\subsubsection{Dimensionally-reduced recipe for thermodynamics}
\label{sec:recipe_3d}

Here we outline the calculation of the thermodynamic parameters in the dimensionally-reduced approach. In way of summary, a brief recipe of the approach follows:
\bi
    \item
        Fix the zero-temperature \MSbar-parameters by matching to physical observables at the input scale, here $m_Z$, and then run them to a thermal, matching scale $\bmu\sim\pi T$.
        Optimise $\bmu$ according to the principle of minimal sensitivity~\cite{Stevenson:1981vj}.
    \item
        Carry out dimensional reduction at this thermal scale, by matching the static, infrared correlators of the 4d theory to an effective 3d theory.
        Further match to a reduced effective theory at the infrared energy scale $g^2 T/\pi$.
        This amounts to integrating out all modes with energies $\sim\pi T$ and $\sim gT$.
    \item
        Calculate the effective potential of the effective 3d theory, and find its minima.
        If possible, maintain a strict $\hbar$-expansion.
    \item
        By taking derivatives of the 3d effective potential with respect to the parameters of the theory, calculate the gauge-invariant condensates, such as $\langle\hsq\rangle$.
        From this one can find the critical temperature $\Tc$, as well as the strength of the transition.
    \item
        Calculate the bubble nucleation rate in the 3d effective theory.
        If possible, maintain a strict $\hbar$-expansion.
        From this, solve Eq.~\eqref{eq:beta_definition} to find
    the percolation temperature $\Tp$ and
    the inverse duration of the transition, or $\beta/\Hp$.
\ei

The first step in the dimensionally-resummed recipe is the same as for the daisy-resummed approach. For the SMEFT, the details are given explicitly in Appendix~\ref{appendix:zero_temp_matching}.

In the second step of our recipe, the hard and soft modes are integrated out and the 3d effective theory for the ultrasoft modes is constructed. An explanation of this procedure at a synoptic level has been given above, at the beginning of Sec.~\ref{sec:overview_3d}. An example application is worked through in Appendices~\ref{appendix:dr_for_beginners} and \ref{appendix:matching}. In application to the SMEFT, we utilise the high-temperature approximation, partially for simplicity and partially because we expect this approximation to be valid, following the argument given around Eq.~\eqref{eq:scalar_thermal_mass}.
Note, however, that this approximation is not an inherent limitation of dimensional reduction~\cite{Laine:2017hdk,Brauner:2016fla,Laine:2019uua}.

Once we have arrived at the ultrasoft 3d effective theory, the advantages of dimensional reduction manifest themselves.
The 3d effective theory is simpler than the full 4d theory. Not only has the theory a reduced field content: all fermions, plus any bosons with masses of order or greater that $\sim g T$ have been integrated out. More importantly, all sum-integrals have been evaluated and one is effectively studying the zero-temperature vacua of a $2+1$~dimensional theory. The temperature merely enters the parameters of the effective theory. As a consequence, perturbative calculations within 3d can be performed as a vanilla loop expansion, so 3d loop orders are not mixed. This allows one to perform strict $\hbar$-expansions, thereby
maintaining order-by-order gauge invariance (see Sec.~\ref{sec:gauge_dependence}), as well as
avoiding double-counting in computing the bubble nucleation rate (see Sec.~\ref{sec:nucleation_corrections}).
It also makes it more feasible to go to higher loop orders than would otherwise be practical.

The third step of the dimensionally-reduced recipe calculates the effective potential in the 3d effective theory.
For our calculations in the SMEFT, we carry this out to 2-loop accuracy. The calculation utilises previous work of Refs.~\cite{Farakos:1994kx,Laine:1994bf,Kripfganz:1995jx}, and is given explicitly in Appendix~\ref{appendix:Veff3d}.

For the SMEFT, due to the presence of the $c_{6,3}(\hsq)^3$ term in the Lagrangian of the 3d EFT, there is a first-order phase transition at tree-level.%
\footnote{
    Note that this is contrary to the full 4d theory, for which the tree-level potential is temperature-independent.
}
From the perspective of the 3d effective theory, the transition takes place as the effective parameters change with temperature.
At least for transitions that are not too weakly first-order,%
\footnote{
    Such that the 3d loop-expansion parameter is perturbative (see Appendix~\ref{appendix:3d_perturbation_theory}), and loop corrections are small compared with the distance to the second-order point in the phase diagram.
}
higher loop orders will not change the order of the transition, so a strict $\hbar$-expansion should converge well.
On the other hand, without the $c_{6,3}$ term
some one-loop contributions must compete with tree-level contributions to give a first-order phase transition, signalling a breakdown of the $\hbar$-expansion~\cite{Laine:1994bf,Kripfganz:1995jx}.
In this case, spurious imaginary parts arise and ``little constructive information'' can be gained from the $\hbar$-expansion~\cite{Laine:1994bf} (see also Ref.~\cite{Kripfganz:1995jx}).
The difference is the presence or absence of a tree-level barrier.
In the case of a strong first-order transition, one can instead recover a consistent expansion parameter in terms of ratios of couplings (see for example Refs.~\cite{Arnold:1992rz,Weinberg:1992ds,Buchmuller:1993bq,Karjalainen:1996rk,Kajantie:1997tt,Moore:2000jw,Garny:2012cg}) such as $\lambda_3/g_3^2$ for the SM.
However, for very weak
first-order,
second-order or
crossover transitions, even this option is no longer possible
and one must resort to lattice simulations~\cite{Kajantie:1995kf,Kajantie:1996qd,Karjalainen:1996rk,Kajantie:1997tt,Moore:2000jw,Kainulainen:2019kyp,Niemi:2020hto}.

To perform the $\hbar$-expansion in the 3d effective theory, one expands all quantities in powers of $\hbar$, the loop-counting parameter of the 3d EFT.
In particular, the 3d effective potential and scalar vev are expanded as
\beq
    V_3(v_3) = \sum_{n=0}^N \hbar^n V_{3(n)}(v_3)
    \;, \qquad
    v_3=\sum_{n=0}^N \hbar^n v_{3(n)}
    \;.
\eeq
Equations, such as $V_3'(v_3)=0$ for finding the loop-corrected vev, are then solved order-by-order in $\hbar$. Note that, following the notation in vacuum, we denote by $\hbar$ the loop counting parameter, though Planck's constant scales out of the 3d effective theory.

In calculating thermodynamic quantities in the $\hbar$-expansion, one first carries out all the necessary computations at tree-level.
Once completed, including higher-order contributions is a simple algebraic exercise in matching powers of $\hbar$ and solving corresponding linear equations.
The tree-level vev solves,
\beq
    V'_{3(0)}\left(v_{3(0)}\right) = 0
    \;, \qquad
    V''_{3(0)}\left(v_{3(0)}\right) > 0
    \;,
\eeq
where multiple solutions signify a coexistence of phases.
Expanding the effective potential around the tree-level vev, one finds
\begin{align}
\label{eq:V_hbar}
V_3(v_3) = V_{3(0)} &
    + \hbar\Bigl(V_{3(1)} + V_{3(0)}' v_{3(1)}\Bigr)
    \nn\\ &
    + \hbar^2\Bigl(V_{3(2)} + V_{3(0)}v_{3(2)} + V_{3(1)}'v_{3(1)} +  \frac{1}{2}V_{3(0)}''v_{3(1)}^2\Bigr)
    + \mc{O}(\hbar^3)
    \;,
\end{align}
where all the potential terms on the right hand side are evaluated on $v_{3(0)}$. Solving for the broken minimum order-by-order in $\hbar$, the solution to $\mc{O}(\hbar^2)$ is,
\beq
\label{eq:v_hbar}
v_{3(1)} =
    -\frac{V'_{3(1)}}{V''_{3(0)}}
    \;, \qquad
v_{3(2)} =
    - \frac{V_{3(2)}'}{V''_{3(0)}}
    + \frac{V_{3(1)}''V_{3(1)}'}{V''_{3(0)}{}^2}
    - \frac{V_{3(1)}'{}^2V'''_{3(0)}}{2V''_{3(0)}{}^3}
    \;.
\eeq
The two-loop expression for the broken minimum, Eq.~\eqref{eq:v_hbar}, contains infrared (IR) divergences in
$V_{3(1)}''(v_{3(0)})$ and
$V_{3(2)}'(v_{3(0)})$ due to the vanishing Goldstone mass in the tree-level broken minimum.
These Goldstone IR divergences are a feature of the Landau gauge, and do not occur for positive gauge fixing parameters.
However, if we regularise the loop integrals by taking the Goldstone mass to zero from above, the IR divergences in Eq.~\eqref{eq:v_hbar} precisely cancel. This is equivalent to taking the gauge parameters to zero from above in taking the Landau gauge limit. For a more detailed discussion of this point see Refs.~\cite{Laine:1994zq,Laine:1994bf}, and, for an alternative approach see Ref.~\cite{Espinosa:2016uaw}.

At the critical temperature two phases coexist with
equal free energy density,
which for homogeneous phases equals the effective potential.
Thus, to solve for the critical temperature, we have the additional equation to solve,
\beq
\label{eq:criticality}
\Delta V_3 \equiv V_3(v_3)-V_3(0) = 0
\;.
\eeq
Here, in light of our intention to apply this formalism to the SMEFT, we have assumed that one of the two phases lies at the origin.
Following Eq.~\eqref{eq:alpha_definition}, we use $\Delta$ generally to refer to the broken phase value minus the symmetric phase value of some quantity.

As before, Eq.~\eqref{eq:criticality}, is solved order-by-order in $\hbar$.
As the 3d EFT does not directly see the parameter $T$, it makes sense to solve this equation instead for the mass of the scalar undergoing the transition, which is what we do for the SMEFT.
For now, let us denote by
$m_{3,c}$ the value of the scalar mass at which Eq.~\eqref{eq:criticality} holds.
The equation determining criticality then takes the form,
\beq
\label{eq:critical_mass}
m_{3}^2 = m_{3,c(0)}^{2}
    + \hbar^{ }\, m_{3,c(1)}^{2}
    + \hbar^{2}\, m_{3,c(2)}^{2}
    + \dots\;,
\eeq
where the terms on the right hand side are functions of the other parameters in the 3d EFT. Hence this equation defines a surface
in this space of the parameters of the 3d EFT. As the temperature changes, a line is traced out in the space of parameters of the EFT, a line which pierces the critical surface at $\Tc$. Written more explicitly, the $\mc{O}(\hbar)$ and
$\mc{O}(\hbar^2)$ corrections to the critical mass take the form,
\beq
\label{eq:mass_hbar}
m^2_{3,c(1)} =
    - \frac{ \Delta V_{3(1)}}{\Delta V_{3(0)}'}
    \;,\qquad
m^2_{3,c(2)} =
    - \frac{\Delta V_{3(2)}}{\Delta V_{3(0)}'}
    + \frac{\Delta V_{3(1)}' \Delta V_{3(1)}}{\Delta V_{3(0)}'{}^2}
    - \frac{\Delta V_{3(1)}{}^2 \Delta V_{3(0)}''}{2\Delta  V_{3(0)}'{}^3}
   \;.
\eeq
Here the $\Delta V_{3(i)}$ are evaluated at the tree-level minima, and at the tree-level critical mass.

One question then arises: do we also expand $\Tc$ in powers of $\hbar$?
This has been discussed in Refs.~\cite{Laine:1994zq,Patel:2011th}. Given that here $\hbar$ is the loop counting parameter of the 3d EFT, we have chosen not to expand $\Tc$ in $\hbar$, instead solving Eq.~\eqref{eq:critical_mass} for $\Tc$ numerically. Both options yield gauge invariant results. Due to the presence of a tree-level barrier between phases in the SMEFT, we expect any difference between these two approaches to be small, as the difference is formally of higher order in the EFT loop expansion.

The measure of the strength of the transition, $\alpha$, defined in Eq.~\eqref{eq:alpha_definition}, is calculated via the trace anomaly.
Once we have determined $\Delta V_3$ to some order in $\hbar$, following Eq.~\eqref{eq:V_hbar} above, the trace anomaly follows from differentiation with respect to temperature. The 3d EFT does not depend explicitly on temperature, so all temperature dependence must arise through the effective couplings.
Thus, we find that,
\begin{align}
\label{eq:trace_anomaly_3d}
\frac{\Delta\Theta}{T} &=
- \frac{3}{4}\Delta V_3
+ \frac{1}{4}\sum_{\{\kappa_i\}}\frac{{\rm d} \kappa_i}{{\rm d} \ln T}\frac{\partial \Delta V_3}{\partial \kappa_i}
\;,
\end{align}
where the sum over $\{\kappa_{i}\}$ runs over the parameters of the EFT, and the factor of $1/T$ on the left hand side follows from the basic, dimensional relation between 3d and 4d physics. Note that as $\Delta V_3$ is calculated order-by-order in $\hbar$, this expression is completely independent of the gauge fixing in the EFT~\cite{Nielsen:1975fs,Fukuda:1975di}, and, if the effective couplings are themselves gauge invariant, then the whole expression is gauge invariant. Furthermore, the $\hbar$-expansion also ensures that $\Delta V_3$ is manifestly real, so there is no need to {\em ad hoc} take the real part, as in the daisy-resummed approach, Eq.~\eqref{eq:ReV}. However imaginary parts can arise in the $\hbar$-expansion in the absence of a tree-level barrier in $V_3$, at least when $\Tc$ is also expanded in powers of $\hbar$~\cite{Laine:1994zq}.

We note that at 2-loop order, a dependence on the 3d renormalisation scale $\bmu_{3}$ arises. This dependence can be diminished by solving the appropriate renormalisation group equation (RGE) (see e.g.\ Ref.~\cite{Farakos:1994kx}), though the choice
$\bmu_{3}=\mc{O}(g_3 v_{(0)})$
is sufficient to avoid large logarithms.
In Sec.~\ref{sec:scale_dependence} we treat this $\bmu_{3}$ dependence as a source of theoretical uncertainty, and vary it over some appropriate range, to estimate its magnitude.
As the dependence on $\bmu_{3}$ only arises at two-loop order, its effect is expected to be small.

The remaining thermodynamic quantities, $\Tp$ and $\beta/\Hp$, can be determined in terms of the bubble nucleation rate. Just as for the other quantities, this can be calculated in an $\hbar$-expansion within the 3d EFT, and there are significant benefits from doing so. The calculation begins with the semiclassical expression for the bubble nucleation rate, Eq.~\eqref{eq:nucleation_rate_Z}, written in terms of the partition function in 4d. The relation between the partition functions of the full theory and the 3d EFT is
\beq
\label{eq:Z3dZ4d}
Z \approx \mbox{e}^{-f_{\mathbb{1}}V}Z_{\rm 3}
\;,
\eeq
where $f_{\mathbb{1}}$ is the coefficient of the unity operator in the dimensional reduction, and (the logarithm of) the equation holds up to some order in powers of the coupling constant (see e.g.\ Refs.~\cite{Braaten:1995cm,Braaten:1995jr}).
A semiclassical evaluation of the partition function expands around a background configuration. For a background configuration that varies, at most, on the long length scales of the ultrasoft theory, Eq.~\eqref{eq:Z3dZ4d} follows directly from the matching of correlation functions carried out in DR. As the coefficient of the unit operator, $f_{\mathbb{1}}$, is independent of the field configuration, Eq.~\eqref{eq:nucleation_rate_Z} takes the same form in the 3d EFT as in the full 4d theory,
\beq
\label{eq:nucleation_3d}
\Gamma \approx \frac{\freq}{\pi V}\frac{\mbox{Im}\,Z_{3}[\phi_{\bub}]}{Z_{3}[0]}
\;,
\eeq
where $Z_{\rm 3}[\phi_{\bub}]$ contributes to the partition function of the EFT from the region around the critical bubble, or bounce, and $Z_{3}[0]$ contributes from the region around the symmetric phase, or false vacuum.
The only factor which cannot be computed purely within the 3d EFT is $\freq$.
In principle it requires a real-time calculation, though it has been estimated in the literature (see Appendix~\ref{appendix:dynamical_prefactor}).
Since the bubble nucleation rate is an intrinsically real-time quantity, it is quite surprising that all but $\freq$ can be calculated within a timeless EFT.

Starting from Eq.~\eqref{eq:nucleation_3d}, the calculation of the bubble nucleation rate is now an unambiguous application of the original, vacuum bounce formalism~\cite{Coleman:1977py,Andreassen:2016cff}, except in three Euclidean dimensions. The temperature only enters the couplings of the EFT, so that one only needs to calculate the vacuum tunnelling rate as a function of the couplings.
Thus, one can derive from first principles the following tree-level bounce equation for the critical bubble, $\phi_{\bub}$, of the 3d EFT,
\beq
\label{eq:bounce_3d}
 \frac{{\rm d}^2 \phi}{{\rm d} \rho^2}
+ \frac{2}{\rho}\frac{{\rm d} \phi}{{\rm d} \rho}
= \frac{{\rm d}V_{3(0)}(\phi) }{{\rm d}\phi}
\;,
\eeq
with the usual boundary conditions (see Eq.~\eqref{eq:bounce_boundary}).
As discussed in greater depth in Sec.~\ref{sec:nucleation_corrections}, the apparently innocuous difference between Eq.~\eqref{eq:bounce_3d} and Eq.~\eqref{eq:bounce_4d} makes all the difference, as only Eq.~\eqref{eq:bounce_3d} avoids the common pitfalls of thermal bubble nucleation calculations:
double-counting,
stray imaginary parts and
an uncontrolled derivative expansion.
In the following we solve this bounce equation by implementing the recently proposed ``Fresh Look'' method
of Refs.~\cite{Espinosa:2018hue,Espinosa:2018szu},%
\footnote{
    We thank J.R.~Espinosa and T.~Konstandin for their help with this.
}
which we have crosschecked against the numerical package
{\verb CosmoTransitions }~\cite{Wainwright:2011kj}.

Solving Eq.~\eqref{eq:bounce_3d} gives precisely the $\mc{O}(\hbar^0)$ contribution to the nucleation rate. This follows because Eq.~\eqref{eq:bounce_3d} contains the tree-level potential in the 3d EFT, and should be contrasted with the 4d~approach, in which tree-level and one-loop contributions are mixed in the calculation of the tunnelling action.
In the 3d~approach, the $\hbar$-expansion of the nucleation rate reads
\beq
\label{eq:tunneling_action_3d_hbar}
\ln\left(\Gamma\right) =
    - S_\bub
    + \hbar \ln\left(A\right)
    + \mc{O}(\hbar^2)
    \;.
\eeq
One-loop fluctuations around the critical bubble make up the nucleation prefactor, $A$, and do not enter $S_\bub$.

The calculation of the nucleation prefactor, the $\mc{O}(\hbar)$ contribution to the nucleation rate, is, in general, a significantly more difficult task than calculating the tunnelling action, $S_\bub$. Nevertheless, in the 3d~approach, this difficult task is significantly easier than in the 4d~approach.
This is mostly because loop orders are not mixed in the calculation, and hence a vanilla semiclassical analysis applies out-of-the-box~\cite{Coleman:1977py,Callan:1977pt,Baacke:2003uw,Dunne:2005rt}.
The difficult task simplifies further because the temperature has already been eliminated from the calculation, and because the field content is reduced. As a consequence, we are able to reasonably estimate the nucleation prefactor, which we carry out in Appendix~\ref{appendix:prefactor}.

Once the rate of bubble nucleation has been calculated, one can solve Eq.~\eqref{eq:Tp_definition} for the percolation temperature, and evaluate Eq.~\eqref{eq:beta_definition} at this temperature to find $\beta/H$.
Finally, one also evaluates Eqs.~\eqref{eq:trace_anomaly_3d} and \eqref{eq:alpha_definition} at the percolation temperature to find $\alpha$.

\section{Sources of theoretical uncertainty}
\label{sec:theoretical_uncertainties}

\begin{table}[ht]
\centering
\begin{tabular}{ r|c|c || c } \rule{0pt}{2ex}
    $\Delta \Omega_{\rm GW}/\Omega_{\rm GW}$ &
    4d~approach &
    3d~approach &
    Demonstrated in
    \\ \hline \rule{0pt}{3ex}
    \hyperref[sec:scale_dependence]{RG scale dependence} &
    $\mathcal{O}(10^{2}-10^{3})$ &
    $\mathcal{O}(10^{0}-10^{1})$ &
    Fig.~\ref{fig:scale_dependence} 
    \\[1ex]
    \hyperref[sec:gauge_dependence]{Gauge dependence} &
    $\mathcal{O}(10^{1})$ &
    $\mathcal{O}(10^{-3})$ &
    Fig.~\ref{fig:gauge}
    \\[1ex]
    \hyperref[sec:high_temp]{High-$T$ approximation} &
    $\mathcal{O}(10^{-1}-10^{0})$ &
    $\mathcal{O}(10^{0}-10^{2})$ &
    Figs.~\ref{fig:high_T}--\ref{fig:c8_dependence}
    \\[1ex]
    \hyperref[sec:higher_loop]{Higher loop orders} &
    unknown &
    $\mathcal{O}(10^{0}-10^{1})$ & Figs.~\ref{fig:matching}--\ref{fig:eft_convergence}
    \\[1ex]
    \hyperref[sec:nucleation_corrections]{Nucleation corrections} &
    unknown &
    $\mathcal{O}(10^{-1}-10^{0})$ &
    Fig.~\ref{fig:prefactor}
    \\[1ex]
    \hyperref[sec:nonperturbativity]{Nonperturbative corrections} &
    unknown &
    unknown &
    Sec.~\ref{sec:nonperturbativity}
\end{tabular}
\caption{Sources of theoretical uncertainty and relative importance quantified by the parameter
$\Delta \Omega_{\rm GW}/\Omega_{\rm GW}$ defined in Eq.~\eqref{eq:deltaOmega} over the range $M=\{580-700\}$~GeV in the SMEFT. 
Although we do not have reliable estimates for the uncertainties of the 4d~approach due to higher loop orders and nucleation corrections, they are expected to be much larger than the corresponding uncertainties of the 3d~approach (see the relevant subsections).
}
\label{tab:uncertainties}
\end{table}

To determine the relative merits of the two approaches outlined in the previous section, in this section we critically examine a range of sources of theoretical uncertainty.
Concrete numerical comparisons are made for the four thermodynamic parameters
which are most important for determining the SGWB spectrum: $\Tc$, $\Tp$, $\alpha$ and $\beta/\Hp$.%
\footnote{
    In studies of the dynamics of gravitational wave spectrum, it is customary to use the subscript $\ast$ to refer to the time of peak gravitational wave production. We will assume throughout that this can be replaced by the percolation time, and hence
    $T_\ast \approx \Tp$ and
    $\beta/\Hs \approx \beta/\Hp$. 
}
Further, in order to compare the various sources of uncertainty, we combine these thermodynamic parameters into a single parameter,
\begin{equation}
\label{eq:deltaOmega}
    \Delta\Omega/\Omega_{\rm min} = \frac{\Omega_{\rm max}-\Omega_{\rm min}}{\Omega_{\rm min}}
    \;,
\end{equation}
where $\Omega_{\rm max,min}$ are the maximum and minimum peak of the SGWB spectrum due to sound waves (sw)~\cite{Hindmarsh:2017gnf,Weir:2017wfa,Guo:2020grp}.
These are predicted by varying the thermodynamic parameters across the theoretical uncertainty band and utilising the following,
\begin{align}
\label{eq:Omega_sw}
h^2 \Omega_{\rm sw} (f) &=
    8.5 \times 10^{-6} \left( \frac{100}{g_\ast} \right)^{1/3}
    \kappa_f^2\,
    \frac{\alpha^2}{(1+\alpha)^2}
    \left( \frac{\Hs}{\beta} \right) 
    S_{\rm sw}(f)
    \left(1-\frac{1}{\sqrt{1+2\Hs t_{\rm sw}}} \right)
    \;, \\
\kappa_f &= \frac{\alpha}{\alpha+0.083 \sqrt{\alpha}+0.73}
    \;,
\end{align}
where $S_{\rm sw}(f)$ encodes the frequency dependence;
at the peak, $S_{\rm sw}(f)=1$.
We estimate the outstanding factor in Eq.~\eqref{eq:Omega_sw}
, the timescale on which acoustic waves are active, from~\cite{Ellis:2019oqb,Guo:2020grp}%
\begin{equation}
\Hs t_{\rm sw} = \frac{2(8 \pi )^{1/3}  \sqrt{1+\alpha }}{\sqrt{3\alpha \, \kappa_{f}}\, \beta/\Hs }
\;.
\end{equation}
Since we are interested in the variation of the theoretical predictions and not in the magnitude of the spectrum, the latter is of minor importance here.

For those theoretical uncertainties which can be assessed numerically, Eq.~\eqref{eq:deltaOmega} gives a relatively good measure of the corresponding uncertainty in predictions for upcoming GW experiments, such as LISA. We apply this measure in estimating the following theoretical uncertainties:
renormalisation scale dependence in Sec.~\ref{sec:scale_dependence},
gauge dependence in Sec.~\ref{sec:gauge_dependence},
high-$T$ approximation with truncation of the 3d EFT in Sec.~\ref{sec:high_temp},
higher loop orders in Sec.~\ref{sec:higher_loop},
and for some corrections to the bubble nucleation rate in Sec.~\ref{sec:nucleation_corrections}.

The construction of the dimensionally-reduced theory naturally entangles three different sources of errors:
(i) higher-dimensional operators,
(ii) loop expansions, and
(iii) high-temperature expansion.
Concretely, the information from integrating out the hard scale fermionic and bosonic modes is distributed over all three of the above. 

Not all the uncertainties that we consider can be reliably quantified.
In particular, internal inconsistencies in the 4d~approach cannot be estimated by simply varying a parameter.
These stem from the relatively {\em ad hoc} implementation of thermal resummation in the 4d~approach.
The most notable of these internal inconsistencies arises in the bubble nucleation calculation. 
Section~\ref{sec:nucleation_corrections} discusses these, in particular:
(i) double-counted degrees of freedom, and
(ii) an uncontrolled derivative expansion.
Therefore, at the level of principle, the 3d~approach should always be preferred over the 4d~approach, as it does not suffer from the same internal inconsistencies of the bubble nucleation calculation.
A similar point could be made regarding gauge independence, which in principle should be maintained order-by-order in perturbation theory, as is only possible in the 3d~approach.

Finally, we should note that any purely perturbative calculation suffers from an irreducible uncertainty due to the nonperturbativity of the IR modes of magnetic gauge bosons in the symmetric phase. We discuss this nonperturbativity in Sec.~\ref{sec:nonperturbativity}, and collect some estimates of its expected magnitude present in the literature. 

Table~\ref{tab:uncertainties} summarises all the assessed sources of uncertainty and refers to the relevant sections and figures. 

\subsection{Renormalisation scale dependence}
\label{sec:scale_dependence}

Perturbative approximations to physical results generally depend on the renormalisation scale, signalling a source of theoretical uncertainty in the approximation. 
Thus, a strong dependence on renormalisation scale, as in  Fig.~\ref{fig:scale_dependence_no_running}, reflects the inadequacy of the approximation%
\footnote{
    Not to mention its incorrectness.
    As Fig.~\ref{fig:scale_dependence_no_running} neglects the running of \MSbar-couplings, the scale dependence only arises explicitly from the Coleman-Weinberg potential.
}.
Dependence on the renormalisation scale can be ameliorated by performing renormalisation group improvement, but it also can be exploited to probe the magnitude of higher order perturbative corrections.
Without large hierarchies of scale in the problem, all logarithms 
can be made small by an appropriate choice of $\bmu$, of the same order as the energy scale of the process under consideration.
In that case, a variation of $\bmu$ by an $\mc{O}(1)$ factor will induce a correction which is formally of higher order in the perturbative expansion. 

\begin{figure}[t]
    \includegraphics[width=\textwidth]{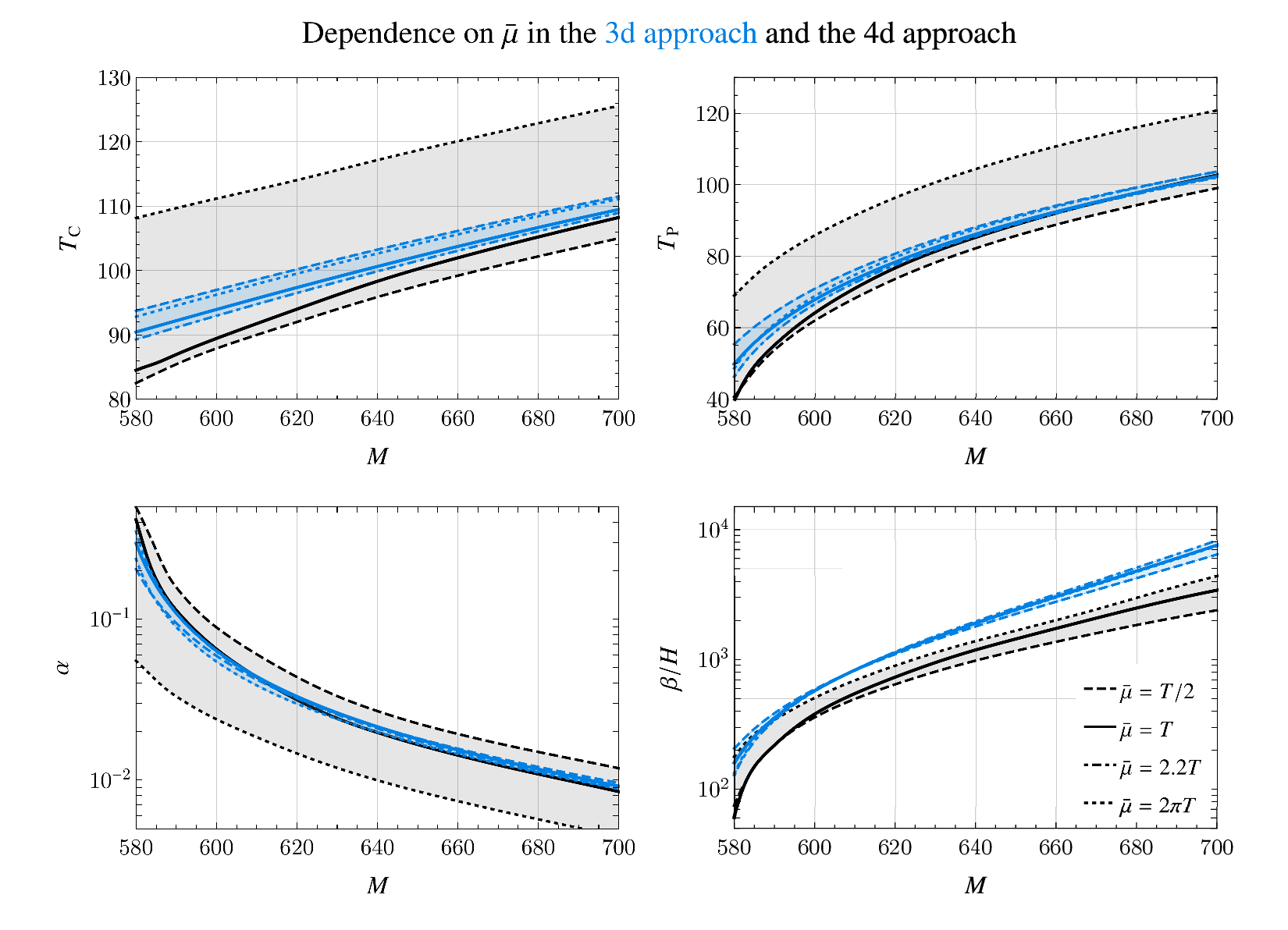}
    \caption{A comparison of the dependence on the renormalisation scale, $\bmu$, in
    the daisy-resummed (black) and
    dimensionally-reduced (blue) approaches.
    The thermodynamic quantities are calculated for different choices of $\bmu$, with uncertainty bands indicating the envelope spanned by these choices.
    The optimal $\bmu = 2.2 T$ is established in Eq.~\eqref{eq:bmu4:dr}.
    }
    \label{fig:scale_dependence}
\end{figure}

When studying a thermal first-order phase transition, it seems natural to choose a 4d renormalisation scale $\bmu$ which depends on temperature.
A question arises though, as to determine the precise numerical coefficient,
$\bmu = \# T$.
In the Matsubara formalism the temperature enters via
the thermal frequencies $\omega_n = n \pi T$, with $n$ an odd integer for fermions and an even integer for bosons.
As a consequence, one might consider any of the following energy scales as possible choices for $\bmu$,
\begin{align*}
    T&
    && \text{for the $n=0$ modes, no factor of $\pi$ is present},\\
    \pi {\rm e}^{-\gammaE} T &\approx 1.76 T
    &&\text{the weighted sum of fermionic } \omega_n,\\
    \pi T &\approx 3.14 T
    &&\text{the lowest fermionic } \omega_n, \\
    2\pi T &\approx 6.28 T
    &&\text{the lowest nonzero bosonic } \omega_n,  \\
    4\pi {\rm e}^{-\gammaE} T &\approx 7.05 T
    &&\text{the weighted sum of nonzero bosonic } \omega_n.
\end{align*}
The ``weighted sums'' here are those that arise within logarithms at one-loop order in a range of quantities, such as the free-energy.
In a theory such as the SMEFT with a large and varied particle content, any of these choices for the renormalisation scale can be equally motivated. 
To estimate the magnitude of higher order corrections, we vary the renormalisation scale over an order of magnitude in the range
$\bmu = (0.5 \dots 2 \pi)T$.
Figure~\ref{fig:scale_dependence}
shows the result of this calculation for both the 4d and 3d~approaches.

\begin{figure}[t]
    \centering
    \includegraphics[width=.7\textwidth]{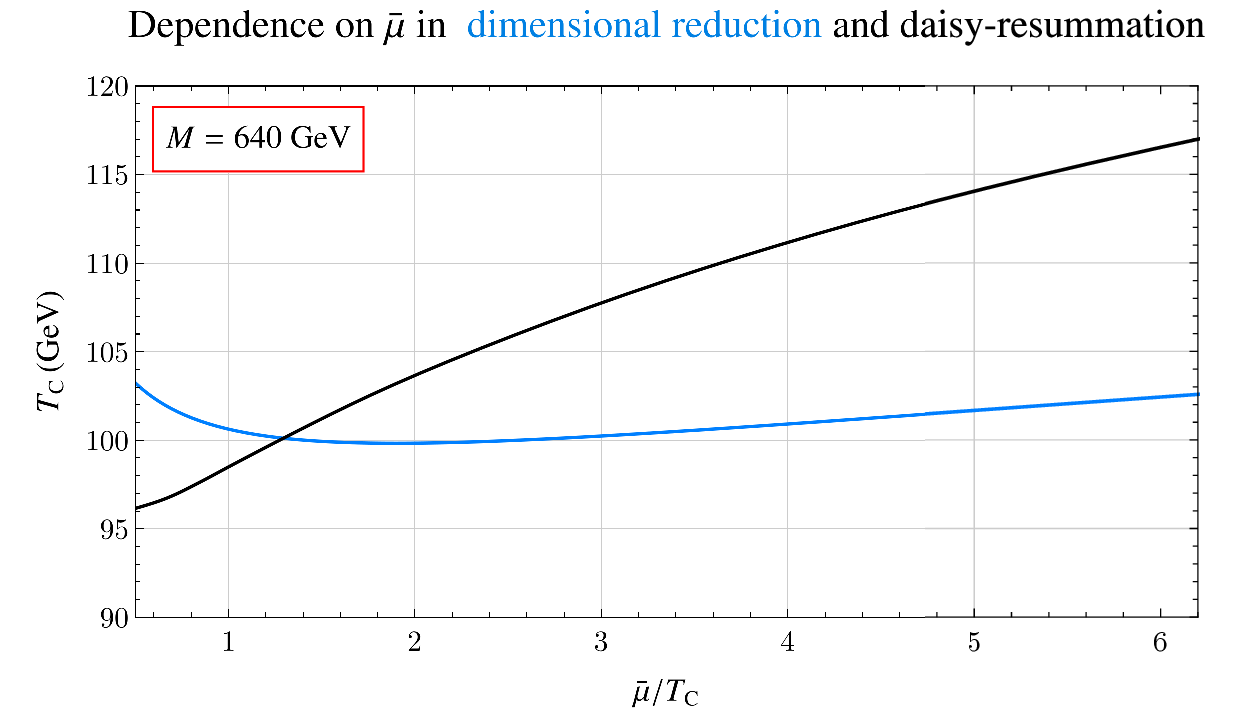}
    \caption{The remormalisation scale dependence of the critical temperature, $\Tc(\bmu)$, in
    the 4d (black) and
    3d (blue) approaches at the benchmark point $M=640$~GeV.
    As discussed in the text, in the 3d~approach the other thermodynamic parameters show a similar $\bmu$ dependence to $\Tc$.
    }
    \label{fig:scale_dependence_bm}
\end{figure}
An optimal choice of scale, $\bmu_{\rmii{opt}}$ can be found according to
the ``principle of minimal sensitivity''~\cite{Stevenson:1981vj}.
This principle demands that at $\bmu=\bmu_{\rmii{opt}}$ some approximation to a physical quantity, $\mc{O}$, is independent of the renormalisation scale,
\beq
\label{eq:rg_improvement}
\frac{{\rm d} \mc{O}}{{\rm d}\bmu}\bigg|_{\bmu_{\rmii{opt}}} = 0
\;.
\eeq
In the 4d~approach, choosing $\mc{O}=\Tc$, this equation finds no solution for $\bmu_{\rmii{opt}}$, $\Tc(\bmu)$ being a monotonic function over the range of $\bmu$ considered.
In contrast, in the 3d~approach, solutions to Eq.~\eqref{eq:rg_improvement} exist for the SMEFT for the whole range of $M$ that we consider. In general, at finite order in a perturbative expansion, different physical quantities may satisfy Eq.~\eqref{eq:rg_improvement} at different scales.
In the 3d~approach we find that for
$\mc{O}=\Tc$, Eq.~\eqref{eq:rg_improvement} is satisfied around
$\bmu_{\rmii{opt}} \approx (2.0-2.1)T$, whereas for
$\mc{O}=\Tp$ Eq.~\eqref{eq:rg_improvement} is satisfied around
$\bmu_{\rmii{opt}} \approx (2.2-2.3)T$.
In both cases the dependence on $M$ is mild.
For $\mc{O}=\alpha$ and $\mc{O}=\beta/H$, the solution to Eq.~\eqref{eq:rg_improvement} depends more strongly on $M$, but is still centred around
\begin{equation}
\label{eq:bmu4:dr}
    \bmu_{\rmii{opt}} \approx 2.2 T
    \;,
\end{equation}
which we choose as the default renormalisation scale in the 3d~approach.
Figure~\ref{fig:scale_dependence_bm} shows the scale dependence of $\Tc$ at the benchmark point $M=640$~GeV.

As seen in Fig.~\ref{fig:scale_dependence}, the scale dependence in the 4d~approach (shown as the grey bands) is very significant:
about 20-30\% for $\Tc$,
20-75\% for $\Tp$,
200-800\% for $\alpha$
and 40-200\% for $\beta/H$.
For the gravitational wave peak amplitude, the corresponding uncertainty is $\Delta\Omega/\Omega_{\rm min}=O(10^{2}-10^{3})$.
This shows that higher order perturbative corrections to the 4d~approach are large, and the calculation is not well under control at this order.
In notable contrast, the scale dependence of the 3d~approach (shown as the blue bands) is much smaller:
2-5\% for $\Tc$,
2-20\% for $\Tp$,
10-70\% for $\alpha$
and 10-60\% for $\beta/H$.
For the gravitational wave peak amplitude, the corresponding uncertainty is $\Delta\Omega/\Omega_{\rm min}=O(10^{0}-10^{1})$.
Thus, at the order that we work to, the 3d~approach appears much better under control.
The uncertainty in the 4d~approach becomes more dramatic when ignoring the scale dependence of the couplings.
This is a corner often cut in the literature, and was discussed in this context in Ref.~\cite{Cai:2017tmh}.
An example of the resulting scale uncertainty is shown in Fig.~\ref{fig:scale_dependence_no_running} in the introduction.

The 3d effective theory depends on its own additional renormalisation scale, $\bmu_{3}$.
Just as with the 4d renormalisation scale, dependence on $\bmu_{3}$ is unphysical.
As the effective theory is applicable to energy scales of order $g^2T\approx g_3^2$ and below, we choose simply
$\bmu_{3,\rmii{opt}} = g_3^2$ as our default scale.
Taking this default scale as the geometric midpoint, we vary the 3d renormalisation scale over the range
$g^2_3/\sqrt{10}$ to
$g^2_3\sqrt{10}$.
The dependence on $\bmu_{3}$ is in all cases much weaker than that on $\bmu$ shown in Fig.~\ref{fig:scale_dependence}, so we do not plot the $\bmu_{3}$-dependence explicitly.
The uncertainty related to the $\bmu_3$ dependence amounts to $\Delta\Omega/\Omega_{\rm min}=O(10^{-2}-10^{-1})$, and increases monotonically with $M$.
Note that, due to its numerical insignificance, we do not solve the renormalisation group equations for $\bmu_{3}$, and neither do we use the more optimal choice $\bmu_{3,\rmii{opt}} = g_3 v_{3(0)}$.
For more discussion of this, see Ref.~\cite{Farakos:1994kx}. 

\subsection{Gauge dependence}
\label{sec:gauge_dependence}

At the leading $\mc{O}(g^2)$ order the effective potential is gauge invariant~\cite{Dolan:1973qd}, motivating some researchers to truncate their calculations at this order, and thereby compromising accuracy.
Gauge dependence enters the effective potential at $\mc{O}(g^3)$, through the one-loop contributions of (soft) Goldstone bosons.
We demonstrate this for the SMEFT in Appendix~\ref{appendix:4Dmasses},
adopting the class of general covariant (or Fermi) gauges,
with gauge parameters $\xi_i$.
As a consequence, if one numerically evaluates the one-loop effective potential at its minima, there is a residual gauge dependence.

In perturbation theory a gauge-invariant treatment requires a suitable power expansion, such as the
$\hbar$-expansion.%
\footnote{
    We postpone discussion of the gauge dependence of the bubble nucleation rate to Sec.~\ref{sec:nucleation_corrections}.
}
Therein the Nielsen identities ensure gauge invariance order-by-order~\cite{Nielsen:1975fs,Fukuda:1975di,Laine:1994zq,Kripfganz:1995jx,Patel:2011th}.
By contrast, without performing such an expansion, perturbative results in general bear a residual gauge dependence.
Daisy resummation conflicts with a strict $\hbar$-expansion.
This is because the tree-level minima are not perturbatively close to the one-loop minima.
The tree-level minima are temperature independent, and differ at leading order from the temperature-dependent minima which are relevant for the phase transition.
Thus, it is not clear how to arrange for the cancellation of gauge dependence in the 4d~approach while keeping consistent to perturbation theory.

\begin{figure}[t]
    \centering
    \includegraphics[width=\textwidth]{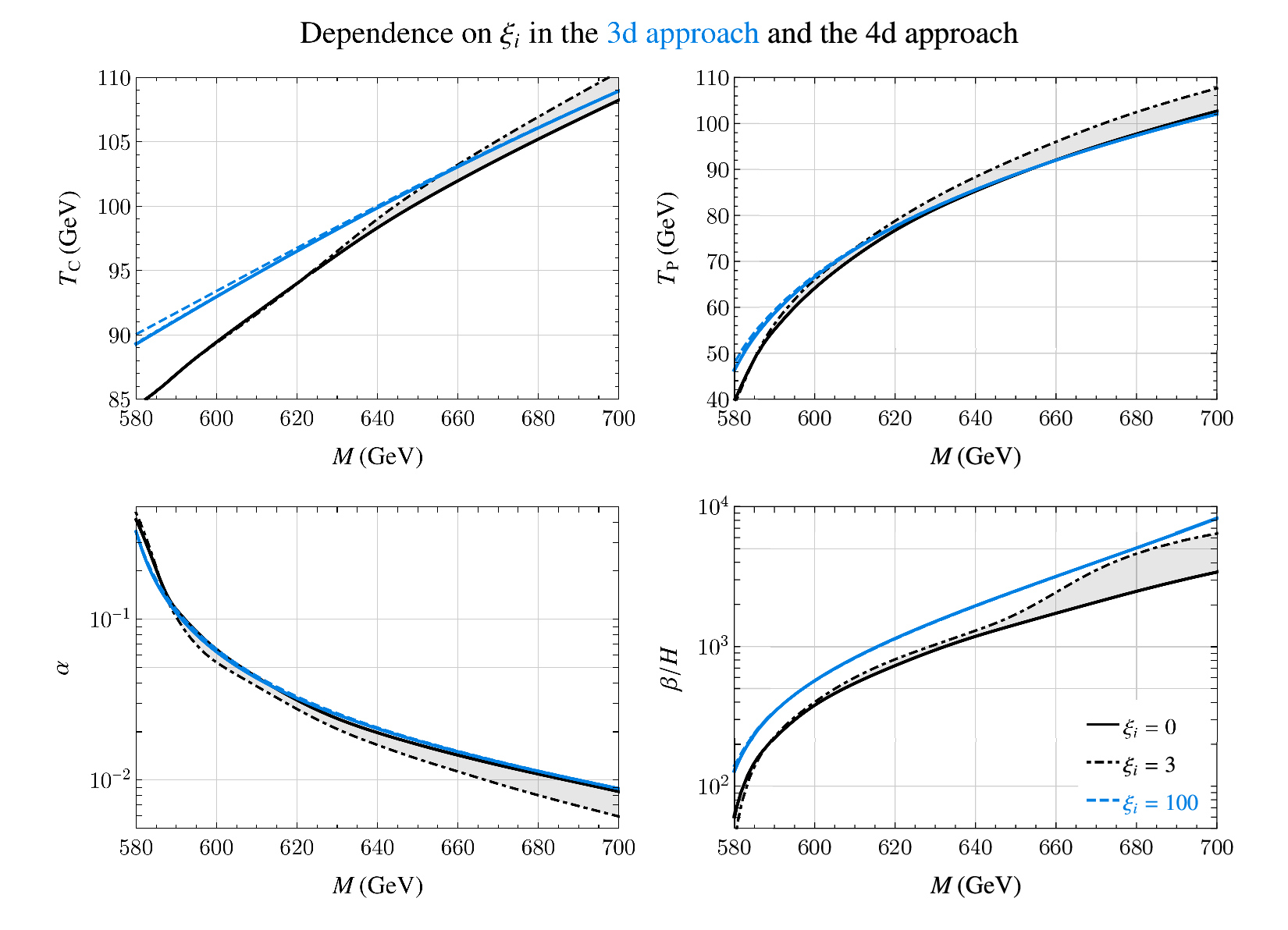}
    \caption{Gauge dependence of thermal parameters
    for the 4d~approach at $\bmu = T$ (black) and
    the 3d~approach at $\bmu = 2.2T$ (blue).
    In both cases the continuous lines denote $\xi_1=\xi_2=0$ and the dot-dashed lines denote $\xi_1 = \xi_2 = 3$.
    At $\xi_1=\xi_2=10$ the 4d~approach breaks down, whereas
    in the 3d~approach the artificial gauge dependence is largely indiscernible, even to very large values of $\xi_i$.
    A dashed blue line demonstrates this at $\xi_1=\xi_2=100$ for the 3d~approach.
    Note that the residual gauge dependence in the 3d~approach is an artifact of an incomplete operator basis in this EFT, and as such is not morally equivalent to the inherent gauge dependence in the 4d~approach. 
    }
    \label{fig:gauge}
\end{figure}

The 3d~approach exhibits two sets of gauge-fixing parameters:
one within the 4d theory itself and
another within the dimensionally-reduced 3d EFT.
As shown in Appendix~\ref{sec:matching-relations}, gauge dependence from the hard scale cancels in the matching relations at the order that we work, $\mc{O}(g^4)$, thus implying that one can consistently truncate the relations at this order.
However, we choose to include a subset of the $\mc{O}(g^6)$ corrections, in particular the $\mc{O}(g^6)$ corrections to $c_{6,3}$, as these are expected to be relatively numerically important given that $c_{6,3}$ is $\mc{O}(g^4)$ at LO (in fact we find this amounts to an $\mc{O}(10\%)$ correction to $c_{6,3}$).
These $\mc{O}(g^6)$ corrections show an explicit dependence on the 4d gauge parameters, though this gauge dependence is numerically very small -- even for very large gauge parameters, as is shown in Fig.~\ref{fig:gauge}.
To remove this gauge dependence would require matching to a complete operator basis in the 3d EFT, and perhaps even a complete matching at $\mc{O}(g^6)$.
However, we choose not to, given the difficulty of such a calculation, as well as its numerical insignificance (cf.\ end of Sec.~\ref{sec:higher_order_resummation}).
Gauge dependence from the soft scale is expected to cancel in the same way as that for the hard scale, though we have not demonstrated this explicitly.
Regarding the gauge parameters of the ultrasoft-scale EFT, since the thermal nature of the 4d theory manifests itself only in the matching parameters, computations within the 3d EFT are carried out in its vacuum, simplifying matters greatly.
This allows a strict $\hbar$-expansion, as long as the tree-level potential of the 3d EFT qualitatively agrees with the full effective potential.
For the specific 3d EFT we study this is indeed the case, since there is a (temperature-dependent) tree-level barrier between the phases.
Therefore, Nielsen identities are recycled in the $\hbar$-expansion, as outlined in Sec.~\ref{sec:recipe_3d}, ensuring independence of the gauge fixing parameters $\xi_{i,3}$ of the EFT, order-by-order.

In a gauge-invariant analysis, the gauge parameters can take any value, as dependence on them cancels exactly.
However, in a gauge-dependent analysis this is no longer true.
Sufficiently large values of $|\xi_i|$ scale as inverse powers of the coupling constants and violate perturbativity~\cite{Laine:1994bf,Kripfganz:1995jx,Garny:2012cg}.
Since generic gauge-dependent loop corrections take the form
$\sim g^{2}\xi_{i}$ at zero temperature, this reduces to
$\sim g^{ }\xi_{i}$ at high temperature (cf.\ Eq.~\eqref{eq:scale_hierarchy}),
where $g$ is some dimensionless coupling.%
\footnote{
    Extending this argument to the ultrasoft modes of the magnetic SU(2) gauge bosons might be seen to impose a stronger constraint, $|\xi_i|\ll 1$, though imposing this does not make sense as the nonperturbativity of these modes means that perturbativity is already violated in all gauges.
}
Thus, perturbativity at high temperature constrains the gauge parameters by,
\begin{equation}
    |\xi_i|\ll\frac{1}{g}
    \;,
\end{equation}
dropping numerical factors.
Hence, by varying the gauge parameter in a range around $\mc{O}(1)$, we can estimate the magnitude of the uncertainty in our results due to gauge dependence.
A calculation of when perturbativity breaks down in general covariant gauges in the ${\rm SU}(2)$-Higgs theory indeed demonstrated that it does so for
$\xi_i\sim\mc{O}(1)$~\cite{Laine:1994bf}.
The focus on small gauge parameters is also supported by various observations in specific models, in which gauge-invariant analyses appear to agree well with the Landau gauge~\cite{Laine:1994zq,Karjalainen:1996rk,Wainwright:2011qy}.

Ideally, we would like to compare the relative magnitudes of the uncertainty due to
gauge dependence and
renormalisation scale dependence, considered in Sec.~\ref{sec:scale_dependence}.
Both arise in perturbative calculations multiplied by coupling constants in essentially the same way, though the latter in logarithms.
So for a relatively fair comparison, we vary the gauge parameters over the range $\xi_{i}=\{0,3\}$, which corresponds approximately with
$\ln(\bmu_{\rmi{max}}/\bmu_{\rmi{min}})=\ln(4\pi)\approx 2.53$.

Finally, for values of $\xi_i$ as large as 10,
the imaginary part of the 4d effective potential exceeds the real part at temperatures close to the critical temperature.
Consequently, the 4d~approach breaks down completely there due to the argument of the square root in the Goldstone modes (Eqs.~\eqref{eq:m:gs1}--\eqref{eq:m:gs2}) which grows with $\xi_{i}$.
Our calculation of the critical temperature may therefore underestimate the theoretical error, accounting for the behaviour in the top left panel of Fig.~\ref{fig:gauge}.
Specifically, for large values of $\xi_{i}$ an extra minimum forms in the real part of the potential, and the interplay of the two minima accounts for the behaviour of $\Tc$. Fortunately, the imaginary part of the effective potential decreases with temperature and tends to be relatively small near the percolation temperature.

In summary, in the 4d~approach the gauge dependence which arises at $\mc{O}(g^3)$ remains uncancelled in the thermal parameters.
By contrast, in the 3d~approach the gauge dependence at this order cancels due to the use of the $\hbar$-expansion at $\mc{O}(\hbar)$.
In addition, in the 3d~approach all gauge dependence at $\mc{O}(g^4)$ cancels:
the gauge dependence of the hard and soft scales cancels in the matching relations,
while the gauge dependence of the ultrasoft scale cancels due to the use of the $\hbar$-expansion at $\mc{O}(\hbar^2)$.
The calculation in the 3d~approach may be truncated at $\mc{O}(g^4)$ yielding exact gauge invariance.
Thus we may conclude that while in the 4d~approach gauge dependence represents a limiting theoretical uncertainty in the gravitational wave spectrum of the phase transition, it is conceptually absent in the 3d~approach. 

\subsection{High temperature approximation}
\label{sec:high_temp}

Both daisy resummation -- \`a la Arnold-Espinosa -- and dimensional reduction at least implicitly rely on the high-temperature approximation, as they are predicated upon a hierarchy of energy scales.
The hard thermal scale $\sim\pi T$ is assumed much larger than the masses of bosonic zero modes. 
This assumption generically holds for thermally driven phase transitions near the critical temperature.
This is ensured by the structure of the loop expansion near the critical temperature; see Eq.~\eqref{eq:scalar_thermal_mass}.
However, for transitions dominated by vacuum (rather than thermal) physics, in which there is a tree-level barrier between phases at $T=0$, a lot of supercooling
can occur between
the critical temperature $\Tc$ and
the percolation temperature $\Tp$.
In this case the high-temperature approximation can break down at $\Tp$.

In the 4d~approach, the high-temperature approximation enters explicitly in the thermal (Debye) masses, Sec.~\ref{appendix:4Dmasses}, but also implicitly through the singling out of the bosonic $n=0$ Matsubara modes for resummation.
Hence our daisy resummation relies on the high-temperature approximation even though we numerically evaluate the full $m/T$ dependence of the thermal functions in the effective potential, Eqs.~\eqref{eq:VT} and \eqref{eq:arnold-espinosa}.
Alternative approaches which do not rely on the high-temperature approximation have been developed in e.g.\ Refs.~\cite{Parwani:1991gq,Karsch:1997gj,Laine:2017hdk}.

\begin{figure}[t]
	\centering
    \includegraphics[width=\textwidth]{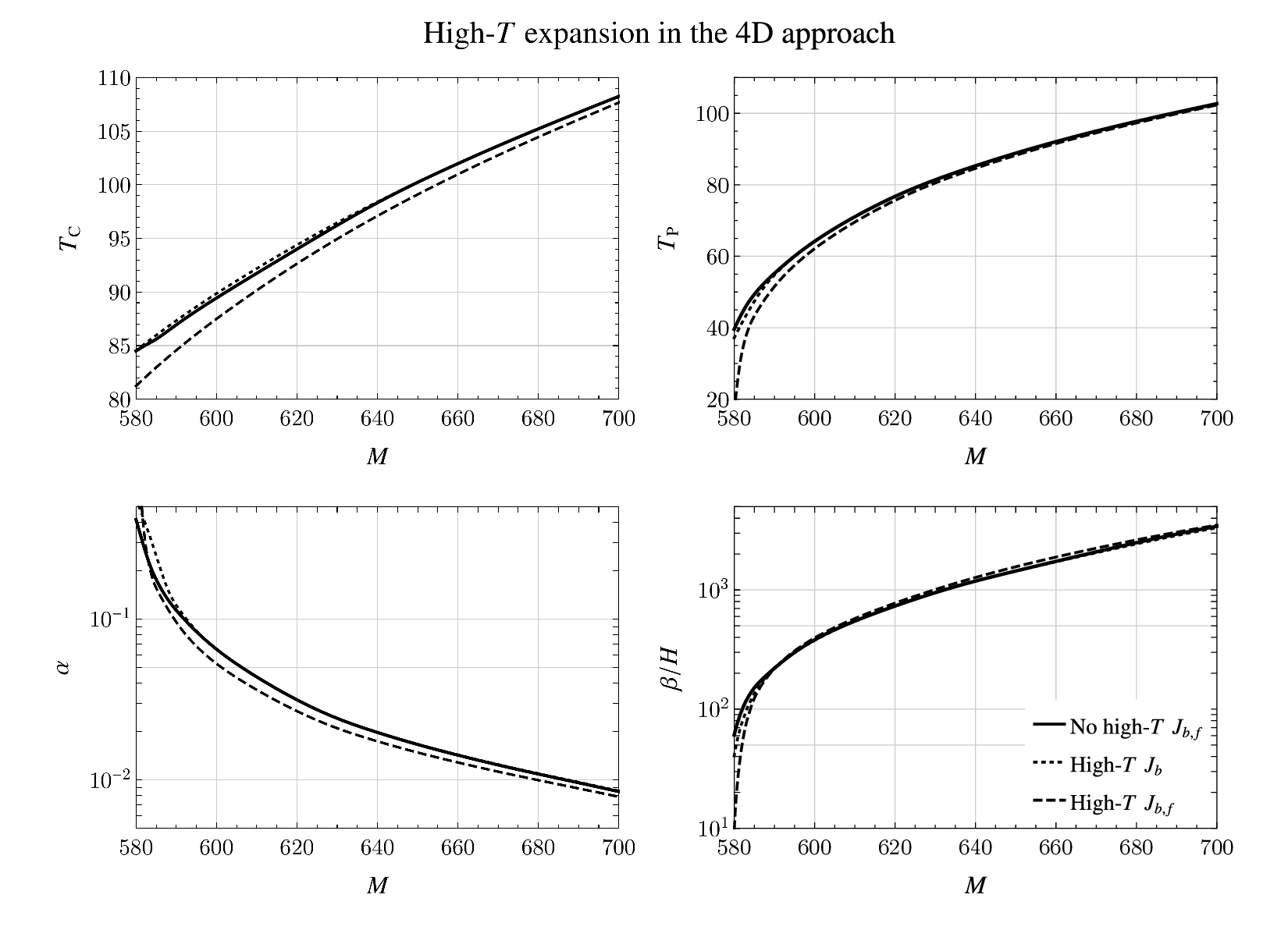}
    \caption{
    $\alpha$, $\beta/\Hp$, each with two lines:
    (i) full thermal functions at $\bmu=T$ (your existing results) and
    (ii) high-$T$ approximation (up to and including logarithms) for all thermal functions (both fermions and bosons), again at $\bmu=T$.
    Thermodynamic parameters in the 4d~approach, with and without high-$T$ expansion of thermal functions, with a choice $\bmu=T$ for the RG-scale.}
    \label{fig:high_T}
\end{figure}

While it is not straightforward to quantify the accuracy of the high-temperature approximation in the 4d~approach, we can estimate its effect by the size of the error introduced by approximating the bosonic thermal functions by the first few terms of their expansion in $m/T$ (up to and including the logarithm);
see Eq.~\eqref{eq:J:b}.
The results of this calculation -- with a choice $\bmu=T$ for the RG-scale -- are plotted in Fig.~\ref{fig:high_T} as the dotted line.
The discrepancy with the full line in that figure indicates our estimate for the size of the uncertainty in the 4d~approach which stems from the high-temperature approximation.
In this estimate, we do not also expand the fermionic thermal functions, as only the bosonic degrees of freedom are resummed, so the high-temperature approximation does not enter the fermionic sector.
We find a difference in the overall gravitational wave spectrum of order $\Delta\Omega/\Omega=O(10^{-1}-10^{0})$ with the larger uncertainties at smaller $M$.

Within the 3d~approach of dimensional reduction, the high-temperature approximation is closely related to the truncation of the 3d effective theory.
The coefficients of higher dimensional operators in the 3d effective theory are related to the coefficients in the high-temperature expansion of the hard mode parts of thermal functions. 
Thus, large contributions from the addition of higher dimensional operators to the 3d effective theory signals the breakdown of the high-temperature approximation. 

\begin{figure}[t]
	\centering
    \includegraphics[width=\textwidth]{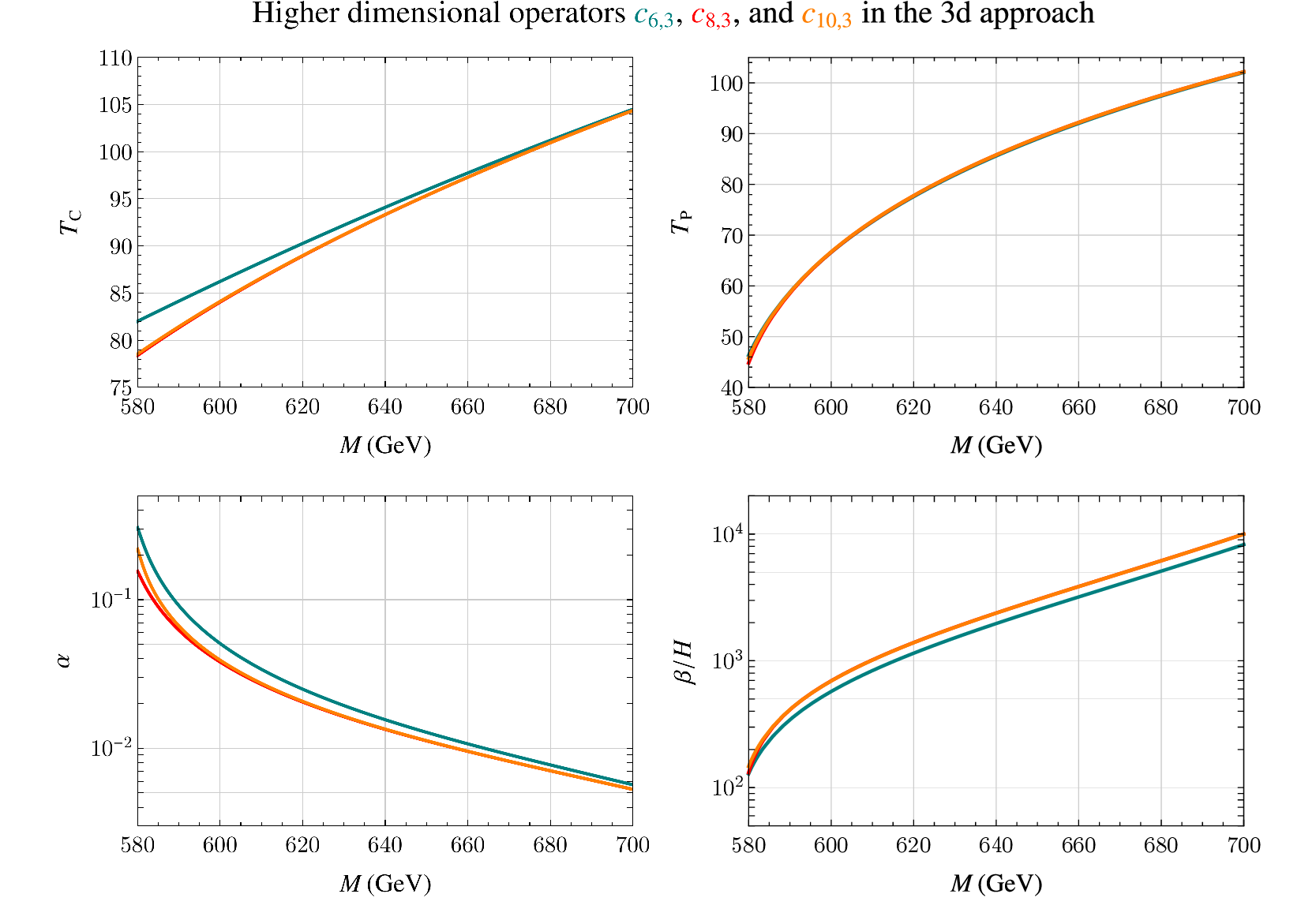}
    \caption{Estimate of the uncertainty in our 3d~approach due to truncating the effective theory at $(\hsq)^3$.
    The cyan lines are the LO results in this truncation,
    the red lines include also the $(\hsq)^4$ operator and
    the orange lines include both the $(\hsq)^4$ and $(\hsq)^5$ operators. 
    }
    \label{fig:c8_dependence}
\end{figure}

To estimate the leading corrections to our truncation of the high-temperature expansion, we match the scalar dimension-8 and -10 operators,
\begin{equation}
\label{eq:higher_dim_operators}
    \mc{O}_{8}^{ } = c^{ }_{8,3}(\hsq)^4
    \;,\quad
    \mc{O}_{10}^{ } = c^{ }_{10,3}(\hsq)^5
    \;.
\end{equation}
These operators enter the Higgs potential directly, and hence we can estimate the order of magnitude of their effects by analysing the tree-level potential in 3d with these operators.
By a direct extension of the tree-level analysis in
Sec.~\ref{sec:recipe_3d} and
Appendix~\ref{appendix:3d_thermodynamics}, we calculate
$\Tc$, $\Tp$, $\alpha$, and $\beta/\Hp$ including the operators in Eq.~\eqref{eq:higher_dim_operators}.
The difference between this and the tree-level result without operators
$\mc{O}_{8}^{ }$ and 
$\mc{O}_{10}^{ }$ estimates our uncertainty.
Fig.~\ref{fig:c8_dependence} shows this uncertainty estimate for the four thermodynamic parameters.
The effect of $c_{8,3}$ is relatively small, and the additional effect of $c_{10,3}$ even smaller, suggesting good convergence of the expansion. Note that the effect of these higher dimensional operators is seen to be larger for stronger transitions.
Since the discontinuities of the scalar condensates are larger for stronger transitions, in general $(\phi^\dagger \phi)^n$-operators give larger effects. 

Higher dimensional operators in the 3d EFT do not just arise from the high-$T$ expansion, but also arise necessarily at higher loop orders.
Powers of $m/T$ from the high-$T$ expansion compete with powers of the coupling constants which arise from the loop expansion~\cite{Appelquist:1974tg}.
Thus, in a rigorous power-counting scheme, as one increases the accuracy of the calculation, and includes more loop orders, one will also need to match to correspondingly higher dimensional operators~\cite{Landsman:1989be}.
This is also necessary in order to render soft and ultrasoft observables finite~\cite{Laine:2018lgj}.

While the high-$T$ expansion of thermal integrals is utilised in our matching relations for the 3d~approach and simplifies the matching significantly, it can be avoided at the cost of tougher master integrals; see Refs.~\cite{Laine:2000kv,Laine:2019uua}.
Rather than attempting this, we utilise the 4d~approach to estimate the error introduced in the 3d~approach from the use of the expansion of thermal integrals in the matching relations.
To do this we calculate the change in thermal parameters as calculated in the 4d~approach when all bosonic and fermionic thermal functions are expanded (up to and including the logarithm);
see Eqs.~\eqref{eq:J:b} and \eqref{eq:J:f}.
The results of this are show as the dashed lines in Fig.~\ref{fig:high_T}.
For weaker transitions, at larger $M$, the resulting changes of the thermal parameters are relatively small; for $M\gtrsim 590$~GeV this results in $\Delta\Omega/\Omega=\mc{O}(10^{-1}-10^{0})$.
However, for strong transitions with significant supercooling, this becomes the dominant theoretical uncertainty of the 3d~approach, with $\Delta\Omega/\Omega$ growing to $\mc{O}(10^{0}-10^{2})$ at $M\lesssim 590$~GeV.

\subsection{Higher loop orders}
\label{sec:higher_loop}

At zero temperature, the convergence of the loop expansion is dictated by the smallness of the dimensionless coupling constants. Large zero-temperature couplings will correspond to large theoretical uncertainties.
Being interested in studying finite temperatures, the convergence of the loop expansion is more delicate due to the Infrared Problem.

In the 4d~approach,
our calculations reach one-loop level.
There is unfortunately no tree-level result to compare with our one-loop calculations, as the phase transition takes place due to thermal fluctuations, which appear first at one-loop.
Already Refs.~\cite{Parwani:1991gq,Arnold:1992rz} extended the daisy-resummed approach to two-loops, though realistic Standard Model extensions were only recently tackled for the two-Higgs doublet model~\cite{Laine:2017hdk} and the SM with a lighter Higgs mass~\cite{Patel:2011th}.
In the daisy-resummed approach, the extension to two-loops is complicated due to massive two-loop sum-integrals, as well as the large number of different particles contributing 
As such, we refrain from
two-loop corrections in the 4d~approach.

In the 3d~approach,
next-to-leading order (NLO) calculations become more amenable.
This is because, for NLO matching, the only two-loop corrections required are
the thermal mass corrections, and thermodynamic properties are analysed within the {\em simpler} 3d effective theory.
This approach expands separately:
(i) the matching relations in powers of $g$ and
(ii) the 3d effective theory in powers of $\hbar$.
In order to calculate some physical quantity to a given order in $g$, both expansions should reach that same order.
It is nevertheless instructive to test their convergence separately.
Further, by not expanding the 3d couplings, the results benefit from resummations of some higher order terms, as well as additional cancellations of scale dependence~\cite{Blaizot:2003iq,Laine:2006cp}.

\begin{figure}[t]
	\centering
    \includegraphics[width=\textwidth]{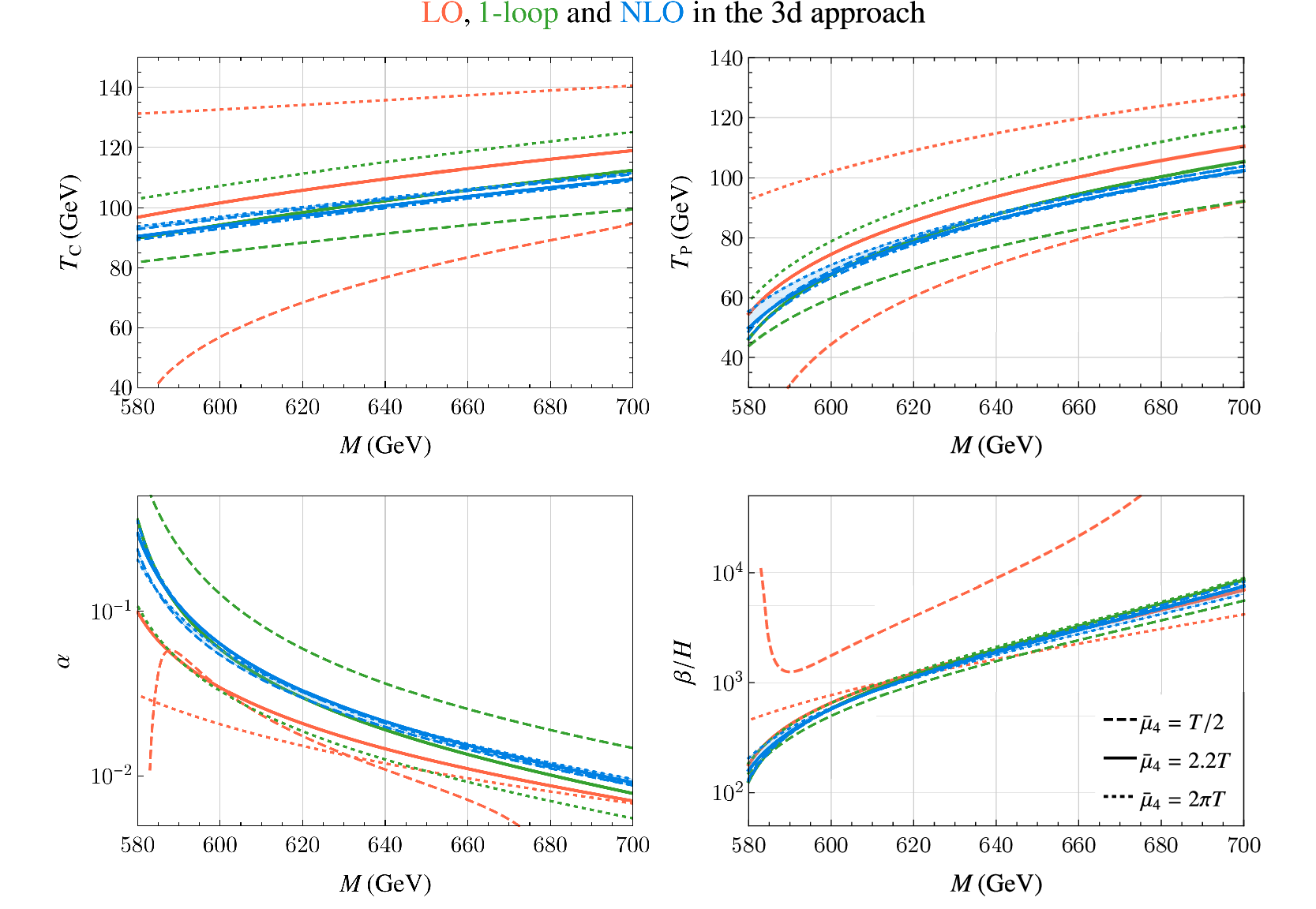}
    \caption{
    Comparison of LO, 1-loop and NLO matching relations in dimensional reduction, showing their $\bmu$ scale dependence. The LO matching relations, though gauge invariant, depend strongly on the scale.
    The 1-loop matching relations significantly reduce the scale dependence with respect to LO, and give an uncertainty band which is relatively constant with $M$.
    The NLO matching relations significantly reduce scale dependence further with respect to 1-loop. It is heartening to see that the NLO results lie entirely within the 1-loop scale uncertainty bands.
    The optimal $\bmu = 2.2 T$ is established in Eq.~\eqref{eq:bmu4:dr}.
    }
    \label{fig:matching}
\end{figure}

For the matching relations in the 3d~approach, we compare three different approximations:
LO matching,
one-loop matching and
NLO matching; see Fig.~\ref{fig:matching}.
The LO and NLO approximations are both consistent truncations of the perturbative series in powers of $g^2$, and as such are gauge invariant%
\footnote{
Excepting the caveat regarding the incomplete basis of operators in our truncation of the SMEFT, discussed in Sec.~\ref{sec:gauge_dependence} and Appendix~\ref{sec:matching-relations}.
}.
The LO approximation consists of
one-loop matching of masses, and
tree-level matching of couplings.
In this we include the LO effects of both the hard~$\mc{O}(g^2)$ and the soft~$\mc{O}(g^3)$ scales, and hence the calculation is accurate to $\mc{O}(g^3)$ but contains nothing at $\mc{O}(g^4)$.
The NLO approximation is accurate to $\mc{O}(g^4)$ and consists of
two-loop matching of masses, and
one-loop matching of couplings, and this is what we utilise elsewhere in the paper.
As argued in Ref.~\cite{Kajantie:1995dw}, the LO approximation is not expected to be quantitatively accurate as fractional corrections to observables from NLO corrections are $\mc{O}(1)$.
One must carry out NLO matching for higher order corrections to observables to be perturbatively suppressed.
In between the LO and NLO approximations lies the full one-loop approximation.
This constitutes an incomplete calculation at any order in $g$, and consequently is not gauge invariant (e.g. Eq.~\eqref{eq:phi:2pt:1l:1d}), just like the 4d~approach.
For example, similarly to the 4d~approach discussed in Sec.~\ref{sec:higher_order_resummation}, a full one-loop matching includes some $\mc{O}(g^4)$ corrections to the effective potential, but misses others.
However it is simpler than the full NLO matching and includes logarithms which cancel the running of couplings (but not masses); see Refs.~\cite{Laine:1998wi,Laine:2000rm,Laine:2012jy,Brauner:2016fla}.

The results for the SMEFT shown in Fig.~\ref{fig:matching} inspire some confidence in the NLO matching relations which we have utilised.
For $\Tc$ and $\Tp$, as one progresses from lower to higher order approximations, from LO to 1-loop and then NLO,
the scale dependence bands shrink from
$\sim 20-50\%$ to
$\sim 10\%$ to
$\sim 2\%$, while the centre of the bands, at the optimal scale $\bmu=2.2T$, change relatively little.
This suggests good convergence as one increases the order of the approximation for the matching relations.
It also shows the importance of higher order corrections for reducing unphysical scale dependence.
For $\alpha$ and $\beta/H$, while the results at the optimal scale show good convergence, the LO approximation appears to break down at small $M$ and $\bmu=T/2$.
Nevertheless, that the NLO scale dependence bands lie entirely within the 1-loop bands still suggests relatively good convergence.
We can naively estimate the size of the unknown NNLO corrections to $\Tc$ as $(1-{\rm LO/NLO})^2\sim (10\%)^2\sim 1\%$, with all quantities evaluated at the optimal scale.
For $\Tp$ this estimate for the NNLO corrections is $\sim 1-4\%$,
for $\beta/H$ it is $\sim 4-15\%$ and
for $\alpha$ it is $\sim 10-50\%$.
These values naively estimate NNLO corrections to the matching relations to result in
$\Delta\Omega/\Omega=\mc{O}(10^{0}-10^1)$, with the larger uncertainties at smaller $M$.
This estimate is supported by its agreement with the magnitude of the scale uncertainty of the NLO result.

In the 3d~approach, we also assess the convergence of the $\hbar$-expansion within the 3d EFT.
For purely equilibrium quantities, such as
the transition strength evaluated at the critical temperature, $\alpha_c$, we can $\hbar$-expand up to $\mc{O}(\hbar^2)$, or NNLO.
This utilises the two-loop effective potential computed in Appendix~\ref{appendix:Veff3d}.
However, for the bubble nucleation rate, the spatial dependence of the bubble profile severely complicates the computation of higher loop orders, so we are only able to compute the LO in $\hbar$ and estimate the NLO in $\hbar$.
Due to this distinction, we defer the discussion of the nucleation rate to Sec.~\ref{sec:nucleation_corrections}.

The presence of the single higher dimensional operator in the truncation of the 3d effective theory leads to an interesting structure of the perturbative series.
Essentially, if $c_{6,3}$ is the smallest coupling in the EFT, it determines the convergence of the entire loop expansion for the coupled 3d gauge-Higgs theory, due to it being the only interaction of dimension~3.
Appendix~\ref{appendix:3d_perturbation_theory} proves this.
As $c_{6,3}$ is naturally the smallest effective coupling (with not too low cut off, $M$), the $\hbar$-expansion within the EFT is expected to converge well.

\begin{figure}[t]
	\centering
    \includegraphics[width=\textwidth]{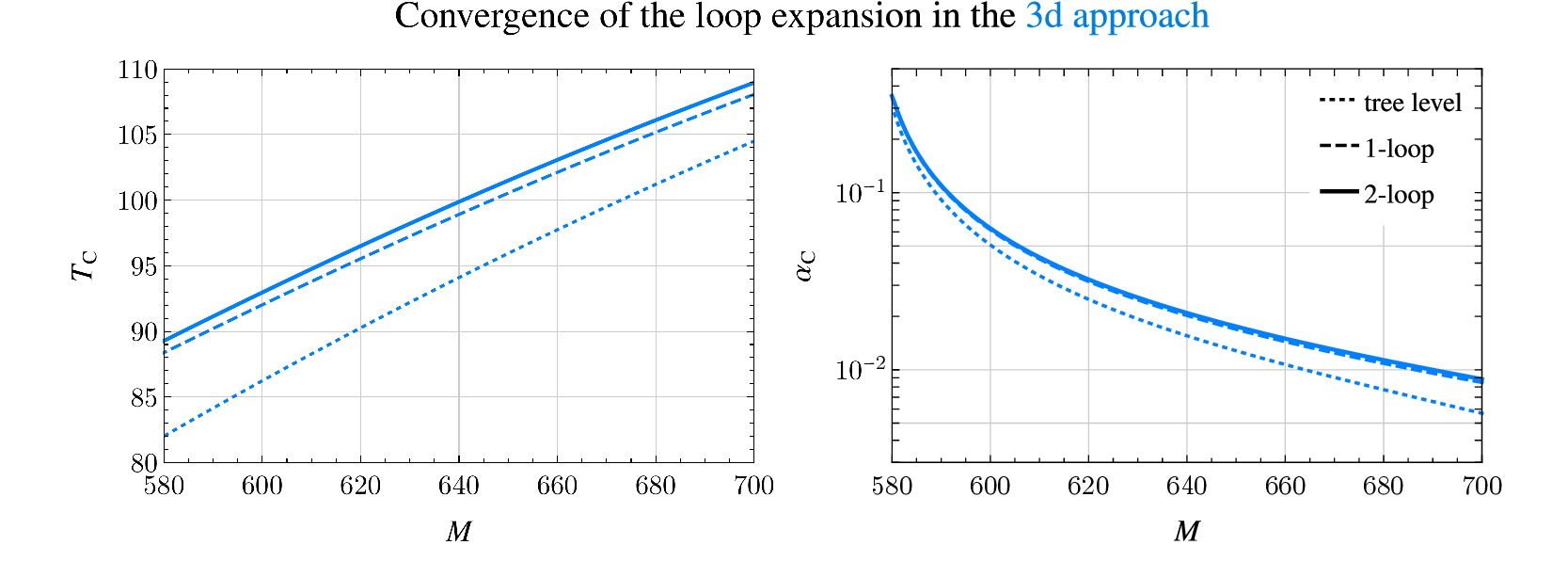}
    \caption{Convergence of the loop expansion in the 3d effective field theory at
    tree-level (dotted),
    one-loop (dashed), and
    two-loop (solid)
    showing the purely equilibrium quantities $\Tc$ and $\alpha_{\rm c} = \alpha(\Tc)$.
    The discussion of the nucleation rate is left to Sec.~\ref{sec:nucleation_corrections}.}
    \label{fig:eft_convergence}
\end{figure}

Computing the purely equilibrium thermodynamic quantities in the SMEFT, we find that across the entire range $M\in [575,750]$~GeV, the tree-level contributions in the 3d effective theory dominate, and the higher loop contributions converge well.
Numerically
one-loop (NLO) contributions are $\sim 5-25\%$ of the tree-level and
two-loop (NNLO) contributions are $\sim 0.5-2.5\%$.
Demonstrating this explicitly, we take a closer look at the benchmark point $M=600$~GeV, for which
\begin{align}
\label{eq:Tc_loops}
\Tc &= 86.2\left( 
    1^{\rmii{(tree-level)}}
    + 0.067^{\rmii{(1-loop)}}
    + 0.011^{\rmii{(2-loop)}}
    + \dots \right)~{\rm GeV}
\;, \\
\label{eq:alphac_loops}
\alpha(\Tc) &= 0.00809 \left(
    1^{\rmii{(tree-level)}}
    + 0.228^{\rmii{(1-loop)}}
    + 0.025^{\rmii{(2-loop)}}
    + \dots \right)
\;.
\end{align}
Here the matching relations are fixed at NLO.
The convergence of the loop expansion within the 3d EFT is fairly consistent across the considered range of $M$ (cf.\ Fig.~\ref{fig:eft_convergence}).

From Eqs.~\eqref{eq:Tc_loops} and \eqref{eq:alphac_loops} we would naively expect three loop (N$^3$LO) and higher loop corrections in this expansion to give fractional corrections of order
$(1-{\rm LO/NLO})^3\sim (10\%)^3\sim 0.1\%$.
However, it should be remembered that, in the symmetric phase the ultrasoft spatial gauge bosons are nonperturbative.
We discuss this issue in Sec.~\ref{sec:nonperturbativity}.

Finally, we comment on what we can learn regarding the size of unknown higher order corrections from studies in other models.
In QCD the $\mc{O}(g^5)$ corrections have been calculated in Refs.~\cite{Braaten:1995jr,Zhai:1995ac}, and were found to be rather large.
At values of the coupling similar to the electroweak gauge coupling at the electroweak scale, it was found that while in general the perturbative expansion appeared convergent, the $\mc{O}(g^5)$ corrections were anomalously large, and comparable in magnitude to the $\mc{O}(g^4)$ corrections.
If this carries over to our case, then we would have to revise our estimates for the theoretical uncertainty upwards by around an order of magnitude.
However, while the convergence of perturbation theory is slow in the high temperature symmetric phase, the screening of IR contributions in the broken phase typically leads to better convergence.
Only a full $\mc{O}(g^5)$ calculation can settle this question.

\subsection{Nucleation corrections}
\label{sec:nucleation_corrections}

The nucleation rate is by far the most technically challenging quantity for which to calculate higher loop corrections. In fact, even calculating the leading order self-consistently is nontrivial, something which is often unappreciated. It is the intersection of the following two points which lead to this technical difficulty:
\begin{itemize}
    \item[(i)]
    The phase transition occurs due to thermal fluctuations.
    These appear first at one loop, hence the critical bubble (or bounce) cannot be solved for at tree-level.%
    \footnote{
        In other words, the transition is {\em radiatively induced}.
    }
    \item[(ii)]
    The fields should be loop-expanded around the critical bubble, i.e.\ around an inhomogeneous classical background field, $\phi=\phi(x)$.
\end{itemize}
These two points lead to a catch-22: the critical bubble only exists in the background of one-loop corrections and yet the one-loop corrections should be made around the background of the critical bubble. This, coupled with the inhomogeneity of the critical bubble, constitute the main technical challenges in consistently calculating the nucleation rate.

An intuitive solution to (i) above is to use the perturbatively computed effective potential, rather than the tree-level potential, to compute the critical bubble~\cite{Linde:1980tt,Linde:1981zj}.
Starting from the tree-level, vacuum bounce equation of Coleman~\cite{Coleman:1977py},
the suggested prescription to modify it for the thermal case
is the following:
\beq
\label{eq:bounce_naive}
    \nabla^{2}\phi -
    \frac{{\rm d}V_{\rm tree}}{{\rm d}\phi} = 0
    \quad\to\quad
    \nabla^{2}\phi -
    \frac{{\rm d}\mbox{Re}(V_{\rm eff})}{{\rm d}\phi} = 0
    \;. 
\eeq
This prescription gives exactly Eq.~\eqref{eq:bounce_4d} once assuming O(3) symmetry and forms the basis of our nucleation calculations in the 4d~approach, as discussed in Sec.~\ref{sec:overview_4d}.
It is also the most common approach taken in the literature.

While plausible, and clearly a step in the right direction, the naive replacement of Eq.~\eqref{eq:bounce_naive} is not derived from first principles.
It suffers from inconsistencies since, in general, the bounce action thus calculated is not the correct result at leading order in any expansion, as has have been discussed in Refs.~\cite{langer1974metastable,Weinberg:1992ds,Buchmuller:1992rs,Gleiser:1993hf,Alford:1993br,Buchmuller:1993bq,Bodeker:1993kj,Berges:1996ib,Surig:1997ne,Strumia:1998nf,Garbrecht:2015yza}.%
\footnote{
    See also Ref.~\cite{Andreassen:2017rzq} for the resolution of an analogous issue at zero temperature.
}
The presence of the extraneous nonzero imaginary part of the effective potential, which must be discarded by hand, perhaps offers a clue that something has gone wrong.
The nucleation rate calculated this way is also gauge dependent.

In Eq.~\eqref{eq:bounce_naive}, by using the effective potential in the bounce equation as an attempt to solve (i), one evades the catch-22 by inconsistently integrating over the fields twice.
Integrating first in generating the effective potential and then
second in solving the bounce equation and evaluating the nucleation prefactor.
In computing the effective potential all nonzero momentum modes of the fields are integrated out, including those of the scalar. The remaining degree of freedom, the constant mode of the scalar field, cannot describe a localised, inhomogeneous critical bubble (or bounce), hence the practical necessity for {\em ad hoc} promoting the constant mode back up to a full spatially-dependent field, to be integrated over a second time. Furthermore, in computing the effective potential, the spatial dependence of the scalar field, the basis of (ii), is still wrongly ignored.
One might think that a derivative expansion could justify this, and that the effective action evaluated on a background bounce solution might be approximated by the naive bounce action,
\begin{align}
\label{eq:bad_derivative_expansion}
\Gamma[\phi(x)] \stackrel{?}{=} \int {\rm d}^3 x &\bigg[
    \mbox{Re}\left(V_{\rm eff}\left(\phi\right)\right)
    + \frac{1}{2}\left(\nabla\phi\right)^2
    \nn \\ &
    \hphantom{\bigg[\mbox{Re}\left(V_{\rm eff}\left(\phi\right)\right)}
    + \sum_{i,n}C_{i,n}m_\phi^2\left(\frac{m_\phi}{m_i}\right)^{2n}\phi \left(\frac{\nabla}{m_\phi}\right)^{2n}\phi
    + \dots \bigg]
    \;,
\end{align}
where we have indicated the size of the simplest derivative corrections, denoting by \dots\ all other possible terms. Here
$m_\phi$ is the mass of the $\phi$ particle,
$i$ runs over the modes with masses $m_i$ which have been integrated out in deriving $V_{\rm eff}$,
$n$ runs over $2,3,4,\dots$ and
the $C_{i,n}$ are $\mc{O}(1)$ constants.
The solution to Eq.~\eqref{eq:bounce_naive}, will exhibit a virial-type theorem implying that the bubble wall has a width of order $\sim 1/m_\phi$, so that $\nabla/m_\phi \sim 1$ when evaluated on the bubble wall.
By deriving the effective potential the $\phi$ field itself has been integrated out, so that $m_\phi\in\{m_i\}$. Therefore, even ignoring the inconsistencies of double counting, the derivative expansion in this case is, at best, an expansion in powers of $\sim m_\phi/m_\phi = 1$, and even worse if there are any bosons lighter than $\phi$.

A consistent resolution to these issues was given by Langer~\cite{Langer:1969bc,langer1974metastable} in the context of classical statistical mechanics, which has been formulated in quantum field theoretic terms in Refs.~\cite{Weinberg:1992ds,Alford:1993br,Buchmuller:1993bq,Bodeker:1993kj,Berges:1996ib,Strumia:1998nf}.%
\footnote{
    For two alternative approaches, see Refs.~\cite{Moore:2000jw,Moore:2001vf} and \cite{Surig:1997ne,Garbrecht:2015yza}.
}
The resolution depends upon the existence of a certain hierarchy of scales, with UV and IR modes well separated in energy scales, i.e.\
$\Lambda_{\rmii{IR}} \ll \Lambda_{\rmii{UV}}$.
In essence, one first integrates over UV modes, resulting in a temperature-dependent, course-grained effective action, $\Gamma_{\rmii{IR}}$, for the remaining IR modes.
One can then derive a bounce equation for the IR modes, which stationarises this course-grained effective action.
\footnote{
A course-grained procedure for exact renormalization using the functional renormalization group equation only yields a non-convex potential if unstable (long-wavelength) states are disregarded, through the introduction of an IR cutoff scale $k_{\rm IR}$~\cite{Berges:1996ib,Strumia:1998nf,CroonHallMuruyama}.
}
This resolves (i) without double counting. Further, as only the UV modes have been integrated out, the course-grained effective action is only nonlocal on length scales $1/\Lambda_{\rmii{UV}} \ll 1/\Lambda_{\rmii{IR}}$. Hence a derivative expansion of $\Gamma_{\rmii{IR}}$ is applicable, and amounts to an expansion in powers of $\Lambda_{\rmii{IR}}/\Lambda_{\rmii{UV}}$. This consistently resolves (ii).
The action from which one determines the bounce equation is 
\begin{align}
\label{eq:good_derivative_expansion}
\Gamma_{\rmii{IR}}[\phi_{\rmii{IR}}(x)] = 
\int {\rm d}^3 x &\bigg[
    V_{\rmii{IR}}\left(\phi_{\rmii{IR}}\right)
    + \frac{1}{2}Z_{\rmii{IR}}\left(\nabla\phi_{\rmii{IR}}\right)^2
    \nn \\ &
    \hphantom{\bigg[V_{\rmii{IR}}\left(\phi_{\rmii{IR}}\right)}
    + \sum_{n}C_{n}\Lambda_{\rmii{IR}}^2\left(\frac{\Lambda_{\rmii{IR}}}{\Lambda_{\rmii{UV}}}\right)^{2n}\phi_{\rmii{IR}} \left(\frac{\nabla}{\Lambda_{\rmii{IR}}}\right)^{2n}\phi_{\rmii{IR}}
    + \dots \bigg]
    \;,
\end{align}
where $\phi_{\rmii{IR}}$ are the IR modes and
$V_{\rmii{IR}}$ and $Z_{\rmii{IR}}$ are the potential and field renormalisation for $\phi_{\rmii{IR}}$, after having integrated out only the UV modes.
Again we have indicated the size of the simplest derivative corrections, with $C_n$ being $\mc{O}(1)$ Wilson coefficients. In this case the bubble wall width is of order $1/\Lambda_{\rmii{IR}}$ and hence these higher order derivative corrections are suppressed by powers of $\Lambda_{\rmii{IR}}/\Lambda_{\rmii{UV}}$.
These terms, as well as those omitted as \dots\ in Eq.~\eqref{eq:good_derivative_expansion}, can thus be neglected when there is a sufficiently large separation of scales.

Dimensional reduction is built around just such a separation of scales, summarised in Table~\ref{tab:dr:smeft}.
It therefore provides a natural framework for which to perform a self-consistent calculation of the bubble nucleation rate.
After integrating out the modes on length scales
$\Lambda_{\rmii{UV}}\sim 1/\pi T$ and then $\sim 1/gT$, we are left with an effective theory on length scales
$\Lambda_{\rmii{IR}}\sim \pi/g^2 T$, the ultrasoft theory. 
Note that the explicit $Z_{\rmii{IR}}$ factors are removed by a redefinition of the infrared fields according to Eq.~\eqref{eq:dr_field_matching}.
Thus, by identifying $\phi_{\rmii{IR}}$ with $\phi_{\rmii{3d}}$ of the 3d EFT and $\Gamma_{\rmii{IR}}$ with the {\em tree-level} action of the 3d EFT,
we can self-consistently calculate the tunnelling action using Eq.~\eqref{eq:good_derivative_expansion}, obtaining the correct result at LO in the 3d $\hbar$-expansion (see Eq.~\eqref{eq:tunneling_action_3d_hbar}), and to NLO in powers of $\Lambda_{\rmii{IR}}/\Lambda_{\rmii{UV}}\sim g/\pi$. Terms with more powers of $\phi_{\rmii{IR}}$ or more derivatives are suppressed by powers of the coupling, as long as the hierarchy of scales in Table~\ref{tab:dr:smeft} holds.

The resulting tunnelling action is independent of the gauge fixing within the 3d EFT, being the leading order in a consistent $\hbar$-expansion. This extends also to the nucleation prefactor, which is gauge invariant when evaluated on a solution to the tree-level equations of motion~\cite{Nielsen:1975fs,Fukuda:1975di}.%
\footnote{
At zero temperature, the gauge invariance of vacuum tunnelling rates has been discussed in Refs.~\cite{Metaxas:1995ab,Metaxas:2000cw,Plascencia:2015pga}.
}
Therefore, if the matching relations are independent of the gauge fixing within the 4d theory (as, for example, we have shown them to be up to $\mc{O}(g^4)$ in the SMEFT), the calculation is then gauge invariant from end to end. Conversely the small gauge dependence introduced into the matching relations in the SMEFT due to the incomplete basis of operators will carry through to the tunnelling calculation.
 
The picture formed is that the hard, UV modes cause the 3d effective parameters to run with temperature, driving the light, IR modes through the transition. Note that it would be incorrect to equate $V_{\rmii{IR}}$ with the effective potential in the 3d effective theory, after having integrated out the ultrasoft fields, as that would amount to double-counting and an uncontrolled derivative expansion akin to Eq.~\eqref{eq:bad_derivative_expansion}.
The correct identification for $V_{\rmii{IR}}$ is the tree-level potential of the ultrasoft EFT.

Returning now to the naive recipe, Eqs.~\eqref{eq:bounce_naive} and \eqref{eq:bad_derivative_expansion}: despite its inconsistencies, in certain circumstances it
approximates Eq.~\eqref{eq:good_derivative_expansion}.
This is the case when the scalar undergoing nucleation is much lighter than all other particles in the theory and has much smaller self-couplings than couplings to other fields. This occurs, for example, in the Standard Model in the region of parameter space where the Higgs is much lighter than the $W$ bosons. In this case the derivative expansion of the effective action is justified for all diagrams except those containing scalar loops, however scalar loops are subdominant by assumption. Further, it is the scalar loops which lead to the erroneous imaginary part of the effective potential, and this again will be subdominant. Nevertheless, Refs.~\cite{Weinberg:1992ds,Gleiser:1993hf,Buchmuller:1993bq,Bodeker:1993kj,Moore:2000jw} have pointed out that rather than accepting a subdominant inconsistency in the calculation, one should instead omit the scalar loops in the potential used in the bounce calculation.

The NLO correction in $\hbar$ to the bubble nucleation rate is the nucleation prefactor, the term $A$ in Eq.~\eqref{eq:rate_exponential}. Its most important contribution is essentially the entropy of the critical bubble, given by integrations over all field fluctuations in the vicinity of the critical bubble. At this point the inconsistency of Eq.~\eqref{eq:bounce_naive} becomes very clear, as all field fluctuations have already been integrated over in computing $V_{\rm eff}$, mixing the LO and NLO terms in $\hbar$. Hence it is not possible even to define the nucleation prefactor in our 4d~approach, resulting in a fundamental theoretical uncertainty beyond just the breakdown of the derivative expansion implied by Eq.~\eqref{eq:bad_derivative_expansion}. As a consequence, most of our analysis of the nucleation prefactor is carried out in the 3d~approach, though through this we are also able shed some light on theoretical uncertainties in the 4d~approach.

The nucleation prefactor can be split into a product of two terms, called the dynamical and statistical prefactor respectively~\cite{langer1973hydrodynamic},
\beq
\label{eq:prefactor_split}
    A =
    A_{\rmi{dyn}}\,
    A_{\rmi{stat}}\;.
\eeq
Both of these terms are independently technically challenging to compute from first principles, and we refrain therefrom in this paper. The dynamical prefactor requires a real-time computation in the presence of a thermal bath, while the statistical prefactor requires computing functional determinants. To estimate the magnitude of corrections related to the nucleation prefactor we compare various approximations present in the literature. This is discussed at length in Appendix~\ref{appendix:prefactor}.

We consider two different approximations to the dynamical prefactor, $A_{\rmi{dyn}}$. These are
(A) an approximation in which the ultrasoft gauge bosons are assumed to dominate the dynamics and
(B) a hydrodynamic approximation, which also assumes the bubbles are thin walled.
Our default choice in the 3d~approach is (A), the infrared gauge boson dominance approximation.

We also consider three different approximations to the statistical prefactor, $A_{\rmi{stat}}$. For the statistical prefactor, the approximations are
(a) a thick wall approximation, correct up to a multiplicative function of $g_3'^2/g_3^2$ and $\lambda_3/g_3^2$,
(b) a thin wall approximation, which correctly gives the dominant terms in the thin wall limit and
(c) LO in an uncontrolled derivative expansion (comparable to Eq.~\eqref{eq:bad_derivative_expansion}) with {\em ad hoc} treatment of the zero modes.
This last approximation, (c), gives exponentially large errors in general but is expected to become more reliable in the limit that the Higgs is much lighter and more weakly coupled than the gauge bosons, and in the thin wall limit.
Our default choice in the 3d~approach is (a), the thick wall approximation.

\begin{figure}[t]
    \centering
    \includegraphics[width=0.99\textwidth]{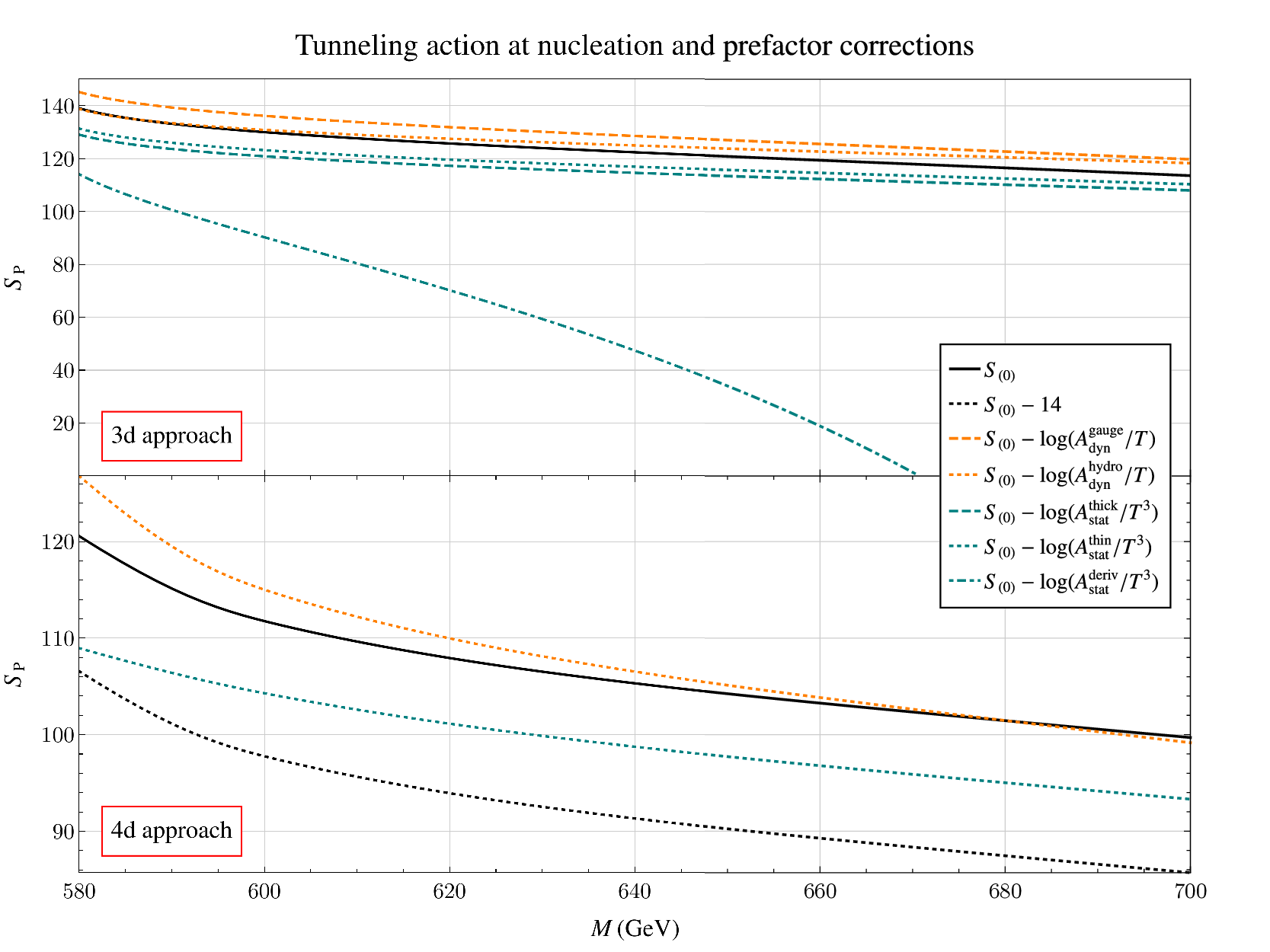}
    \caption{The tunnelling action at nucleation in the 3d (top panel) and 4d (bottom panel) approach, shown in black at LO in the $\hbar$-expansion, $S_{(0)}$.
    Also shown are prefactor corrections which, for each $M$, are all evaluated at the nucleation temperature calculated using our default prefactor approximation.
    The full prefactor factorises into
    dynamical (orange) and statistical (cyan) parts;
    see Eq.~\eqref{eq:prefactor_split}.
    More details of these approximations can be found in Appendix~\ref{appendix:prefactor}.
    Note the large deviations from the LO result for the derivative approximation to the statistical prefactor.
    This aligns with the expected breakdown of this approximation, which underlies the 4d~approach.
    }
    \label{fig:prefactor}
\end{figure}
The tunnelling action at nucleation is plotted in Fig.~\ref{fig:prefactor}, together with prefactor corrections in our different approximations.
These should be small corrections if the $\hbar$-expansion is under control.
Further, agreement between different approximations to the dynamical and statistical parts of the prefactor respectively would suggest that theoretical uncertainties in these quantities are relatively under control.
We remind the reader that the full prefactor is the product of the dynamical (orange) and statistical (cyan) parts.

The most obvious anomaly in Fig.~\ref{fig:prefactor}, is the statistical prefactor in the uncontrolled derivative approximation, (c), which shows $\mc{O}(1)$ multiplicative deviations from the LO tunnelling action. This uncontrolled derivative approximation underlies the 4d~approach, being essentially the same as that carried out in Eq.~\eqref{eq:bad_derivative_expansion}.
The numerical results shown in Fig.~\ref{fig:prefactor} suggest the approximation breaks down badly, especially for larger $M$, where the transitions are weaker. Although the 4d~approach relies on this derivative expansion, by incorporating these corrections directly in Eq.~\eqref{eq:bounce_naive}, rather than in the prefactor, it effectively resums the expansion, so mitigating (but not avoiding) the breakdown of the expansion.
The consequences of this uncontrolled error cannot easily be quantified in
$\Delta\Omega / \Omega_{\rm min}$, so we denote them as {\em unknown} in Table~\ref{tab:uncertainties}.
However, the systematic difference for $\beta/\Hs$ with respect to the 3d~approach may give some indication (see e.g. Fig.~\ref{fig:scale_dependence}).
Just this amounts to a systematic error $\Delta\Omega / \Omega_{\rm min}$ of around $\mc{O}(10^{1})$.

For the 4d~approach, leaving aside the unknown effect due to the application of the uncontrolled derivative expansion, the different approximations for the prefactor $A$ result in a modest change in the effective action compared to the value recommended in the recent LISA review, $\log(A/T^4)\sim-14$~\cite{Caprini:2019egz}. This in turn results in a modest change in the peak gravitational wave amplitude between
$\mc{O} (10^{-1}-10^{0})$.
However, since the true error is dominated by the systematic error, we do not include this uncertainty in the final analysis.

On the other hand, for the 3d~approach, we face no such difficulties and so can give a reasonable estimate for the theoretical uncertainty related to the nucleation prefactor. Fig.~\ref{fig:prefactor} shows that the thin and thick wall approximations to the statistical prefactor roughly agree,
as do the two approximations of the dynamical prefactor.
Better agreement than this is not to be expected, as these are all crude estimates. However, their rough agreement encourages us to suggest that they could be used to estimate the magnitude of $\mc{O}(\hbar)$ corrections in the 3d~approach. Combining the various approximations to the dynamical and statistical prefactors (excluding the uncontrolled derivative expansion) in all possible ways leads to a range of different estimates. These different estimates amount to an uncertainty
$\Delta\Omega / \Omega_{\rm min}$ for the 3d~approach, which varies monotonically from $\mc{O}(10^{-1})$ at $M=700$~GeV to
$\mc{O}(10^{0})$ at $M=580$~GeV.

\subsection{Nonperturbativity}
\label{sec:nonperturbativity}

In the symmetric phase the perturbative mass of the magnetic gauge bosons is zero. As a consequence, the effective coupling of their zero Matsubara modes, $\sim g^2T/m$, appears to be infinite. This leads to IR divergences at finite loop order~\cite{Linde:1980ts}, and the consequent complete breakdown of perturbation theory.%
\footnote{
    This breakdown of perturbation theory is not solved by resummation.
}
Physically, these divergences in perturbation theory are softened by a nonpertubatively generated mass for the gauge bosons, of order $g^2T$~\cite{Kajantie:1995kf}.
Based on power-counting arguments, one would expect the resulting softened IR divergences to contribute to the free energy density at
$\mc{O}(g^6T^4/(2\pi)^4)$~\cite{Shaposhnikov:1993jh,Kajantie:1994xp}, or the typical size of a 4-loop order term. However, this estimate is perhaps misleading, as the effective gauge coupling becomes large for momentum-transfers $Q^2\sim (g^2T/\pi)^2$, so perturbative estimates cannot be relied upon. The nonperturbative nature of the symmetric phase leads to an irreducible uncertainty in our perturbative calculations which in principle only lattice simulations can resolve.

In the broken phase, by contrast, perturbation theory may work well, as the troublesome magnetic gauge bosons acquire a mass via the Higgs mechanism.
For a sufficiently strong transition, this broken phase mass is large, and hence the effective coupling of the magnetic gauge bosons, $\sim g^2T/m$, is perturbative. Of course, both phases are relevant for calculating the thermodynamic properties of the transition. However, if for some thermodynamic quantity the contribution of the broken phase exceeds the contribution of the symmetric phase, then one would expect perturbation theory to be an at least qualitatively reliable guide. This scenario should take place for sufficiently strong transitions, in which the Higgs vev in the broken phase is large, and gives correspondingly large contributions to thermodynamic quantities.

Explicit comparisons between lattice and perturbation theory have indeed shown good {\em qualitative} agreement for strong first-order transitions in the
xSM, 2HDM, $\Sigma$SM~\cite{Gould:2019qek,Kainulainen:2019kyp,Niemi:2020hto},
Abelian Higgs model~\cite{Karjalainen:1996rk}
and also in the SM~\cite{Kajantie:1995kf,Kajantie:1996qd,Moore:2000jw} (with artificially light Higgs).
In particular, the scan of the parameter space of the xSM in Ref.~\cite{Gould:2019qek} showed good qualitative agreement between lattice and perturbative calculations for the phase diagram of the theory.
For the two benchmark points in the 2HDM considered in Ref.~\cite{Kainulainen:2019kyp} it was found that $\Tc$ calculated in a dimensionally-reduced approach%
\footnote{
    Just as we have done in this paper, in Refs.~\cite{Kainulainen:2019kyp,Niemi:2020hto}
    the dimensional reduction was carried out to NLO and
    the perturbative calculations within the 3d EFT were carried out to 2-loop order.
}
differed by 4--7\% from the lattice result, whereas $\alpha_c$
differed by 5--25\%.
By contrast, calculations using a daisy-resummed approach showed discrepancies from the lattice result
for $\Tc$ of 20--45\% and
for $\alpha_c$ of 45--75\%.
For the two benchmark points considered in the $\Sigma$SM in Ref.~\cite{Niemi:2020hto} it was found that in a dimensionally-reduced approach $\alpha_c$ differed from the lattice result by 30--40\%, whereas in a daisy-resummed approach the discrepancy was more than 50\%.
Bubble nucleation was studied on the lattice in the Standard Model with light Higgs in Ref.~\cite{Moore:2000jw}. There it was found that perturbative calculations of the value of
$(m_{3,c}^2-m_3^2)/g_3^4$ at nucleation, essentially $(\Tc/\Tp)^2-1$, differed from the lattice result by around 30\%. Note that some of these discrepancies between lattice and perturbative calculations are significantly larger than one might guess based on perturbative power counting, highlighting the importance of carrying out nonperturbative computations.

\section{Discussion}
\label{sec:conclusions}

This work systematically examines theoretical sources of uncertainty in the finite-temperature calculation of thermal tunneling rates, and the resulting uncertainty in thermal parameters of a first-order phase transition.
In particular, we have compared the sources of uncertainty in two different methods that address the breakdown of perturbation theory due to the long-wavelength modes at high temperature:
resummation of daisy diagrams in 4d, and
dimensional reduction to 3d effective theory. 
The benchmark model we have used is the Standard Model aided with a dimension-six operator, $\mathcal{O}_6 = \left( \phi^\dagger \phi \right)^3/M^2$, which we expect represents qualitatively a large set of models of EWSB, which is of particular interest for the planned LISA experiment.

Section~\ref{sec:thermodynamics} and the Appendices include a thorough and comprehensive review of the 4d and the 3d~approaches, and the calculation of the thermal parameters capturing the dynamics of the phase transition.
We develop the 3d~approach, building upon previous works utilising dimensional reduction, to include also a gauge invariant treatment up to $\mc{O}(g^4)$ of bubble nucleation and the calculation of thermodynamic quantities.
The two approaches are outlined in general, and are fleshed out for the specific example of our benchmark model.

The main sources of uncertainty were categorised in Section~\ref{sec:theoretical_uncertainties}, as follows:
renormalisation scale dependence,
gauge dependence,
the high temperature approximation,
the unknown contributions of higher loop orders,
corrections to the bubble nucleation rate,
and nonperturbativity or the Infrared Problem.

\begin{figure}[t]
\centering
 \includegraphics[width=0.99\textwidth]{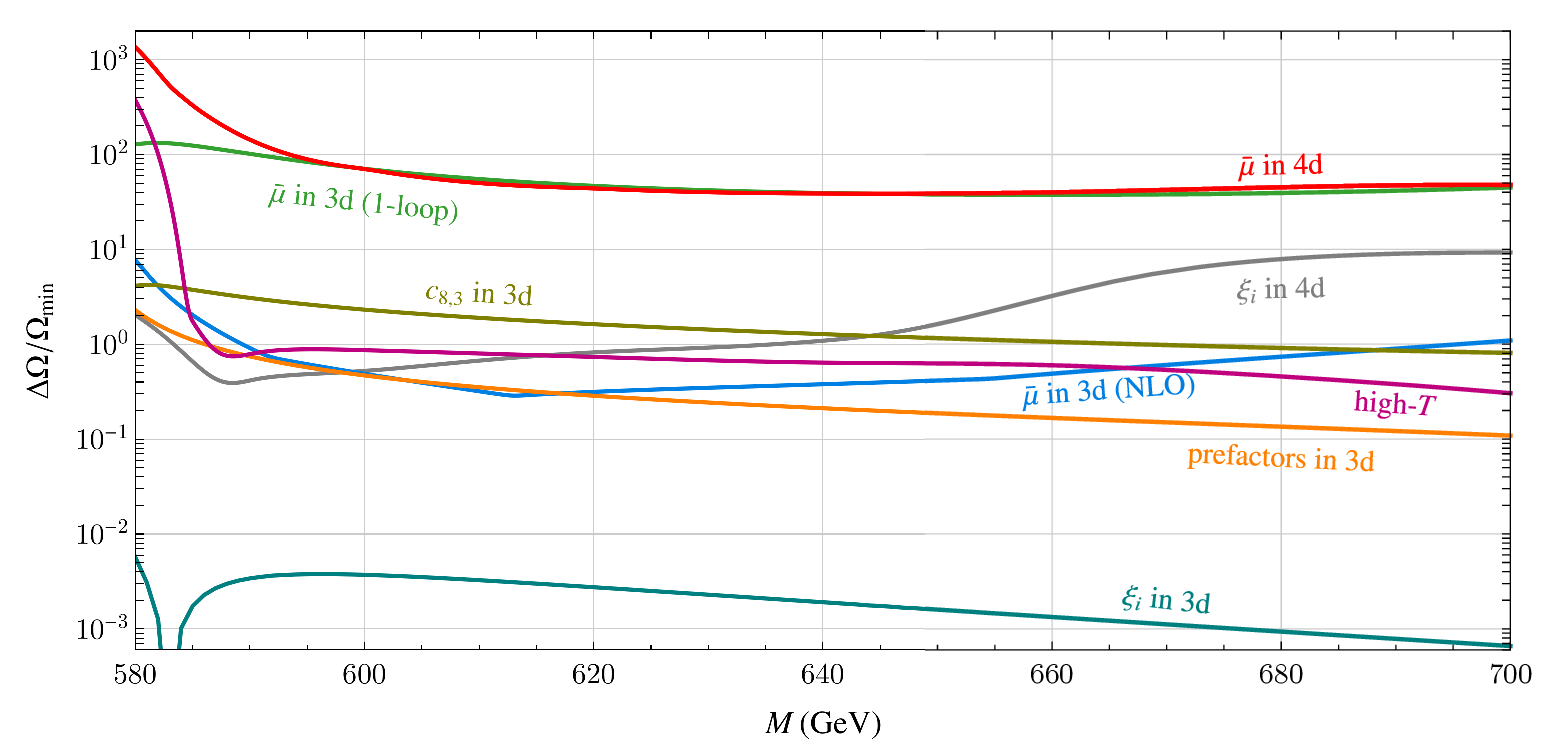}
 \caption{Effects of variation for different sources of theoretical uncertainty on the parameter $\Delta\Omega/\Omega_{\rm min}$, defined in~\eqref{eq:deltaOmega}.
 The renormalisation scale is varied between $\bmu=\{T/2,2\pi T\}$.
 The gauge parameters are varied in the range $\xi=\{0,3\}$ in both the 3d and the 4d~approach.
 The chartreuse line labeled ``$c_{8,3}$ in 3d'' gives the difference between the result including that parameter and the result with only $c_{6,3}$.
 The purple line labeled ``high-$T$'' gives the variation due to the high-temperature expansion of
 the thermal functions $J_{b/f}$.
 Note that theoretical uncertainties due to nonperturbativity and due to the breakdown of the derivative expansion in the 4d~approach are not included in this plot, as we were unable to give reliable and detailed estimates of the magnitude of these uncertainties.
 }
 \label{fig:DeltaOmega}
\end{figure}

We include a qualitative discussion of the origin of these uncertainties, and where possible explicitly calculate their magnitudes over the relevant range in $M$ in our benchmark model.
We summarise the results of these calculations in terms of the parameter $\Delta \Omega/\Omega_{\rm min}$ defined in \eqref{eq:deltaOmega} in Fig.~\ref{fig:DeltaOmega}.
Importantly, the extent of the uncertainty calculated in this paper can not be taken as a direct predictor of the correctness of the result, as is most obviously noticeable from the apparently vanishing uncertainty for certain $M$ in some of our figures.
However, we expect that the variation of the results may be incorporated as an integral part of the diagnosis of the appropriateness of the two methods. 

From the analysis in this paper, we may draw several conclusions.
Firstly, a direct comparison between the 4d and 3d~approaches indicates that the latter generically implies a significantly smaller theoretical uncertainty.%
\footnote{
    As the calculations in both approaches here were done semi-analytically and did not utilise lattice Monte-Carlo simulations, there remains a nonperturbative uncertainty associated with the magnetic gauge bosons in the symmetric phase; see Sec.~\ref{sec:nonperturbativity}.
}
Such a comparison is shown in Fig.~\ref{fig:DeltaOmega}, and yields a difference in $\Delta\Omega/\Omega_{\rm min}$ of several orders of magnitude. 
This may be taken as support for the choice of the 3d~approach over the more commonly used 4d~approach.
Secondly, of those uncertainties which we were able to estimate, the most important driver of theoretical uncertainty in the 4d~approach is the dependence on the choice of renormalisation scale.
This effect dwarfs the uncertainty introduced by other effects, such as gauge dependence -- as uncomfortable as the latter is.
In light of the results presented here, we recommend forthcoming works to include an analysis of the variation of the renormalisation scale.
Thirdly,
the error introduced by applying the high-$T$ approximation to thermal functions grows sharply for the strongest transitions, below about $M=590$~GeV.
This was traced back to the contribution of the top quark.
As these strong transitions are those most relevant to gravitational wave experiments such as LISA, we suggest that at least the full, numerical $m_{t}/T$ dependence should be accounted for in future calculations of the SGWB of first-order phase transitions.

In addition to those theoretical uncertainties for which we were able to give reliable estimates,
we demonstrated that the 4d~approach generically depends upon an uncontrolled derivative expansion.
The breakdown of this expansion is remedied in the 3d~approach, and hence may account for the systematic discrepancy in $\beta/\Hs$ between the two approaches.
We advise that care should be taken in future studies to avoid this particular stumbling block and to ensure that calculations of the tunnelling rate are self-consistent.

Prior to this work, there have been many studies of various different theoretical uncertainties in calculations of the thermodynamics of phase transitions; recent examples include Refs.~\cite{Patel:2011th,Wainwright:2011qy,Wainwright:2012zn,Curtin:2016urg,Chiang:2017nmu,Chiang:2017zbz,Jain:2017sqm,Laine:2017hdk,Cai:2017tmh,Chiang:2018gsn,Prokopec:2018tnq,Gould:2019qek,Carena:2019une,Kainulainen:2019kyp,Senaha:2020mop}.
However, for the first time, in this work we have comprehensively analysed and compared a wide variety of relevant theoretical uncertainties across the full range from weakly to strongly first-order transitions.
In doing so, we have focused on the implications for the resulting gravitational wave spectra.
Previous works have almost exclusively focused on the theoretical uncertainties of the transition temperature and the corresponding vev, which can disguise the severity of the theoretical uncertainty in the gravitational wave spectrum.%
\footnote{
     An exception was Ref.~\cite{Wainwright:2011qy} which studied the gauge dependence in the bubble collision term of the gravitational wave spectrum within the Abelian Higgs model.
     See also Ref.~\cite{Carena:2019une} in which the gravitational wave spectrum of the $Z_2$-symmetric xSM was studied, and within which a $\Delta\Omega/\Omega$ of $\mc{O}(10^4)$ was observed in the comparison of different approximation schemes.
}
For example, as $\alpha$ is inversely proportional to the fourth power of the percolation temperature, and the peak gravitational wave amplitude in turn depends quadratically on $\alpha$, an apparently innocuous uncertainty $\Delta \Tp/\Tp = 0.1$ in the percolation temperature will result in an uncertainty $\Delta \Omega /\Omega  \approx 1$ in the gravitational wave spectrum.
In fact this argument may significantly underestimate the uncertainty, as additionally the trace anomaly grows strongly as the percolation temperature decreases.
To the extent that a comparison is possible, the results presented in this paper are qualitatively consistent with the existing literature. 

The results in this paper will be of particular interest for gravitational wave studies in anticipation of the LISA experiment and other space-based interferometer experiments with sensitivity in the milli-Hertz range. In the benchmark model studied in this paper, the strongest phase transitions -- leading to the gravitational wave spectra most likely to be observable -- are found for smaller $M$. Importantly, this is also the range in which the diagnosis presented in this paper indicates the poorest theoretical control, cf.\ Fig.~\ref{fig:DeltaOmega}.
This implies that the amplitude of the gravitational wave spectrum, but also its peak frequency, can only be predicted to a level of accuracy which depends on the method used. This has serious implications for the prospect of model differentiation and complementary studies with collider probes. The sheer magnitude of the theoretical uncertainty and the comparative success of the 3d~approach at NLO, motivates its use in studying GW phenomenology in specific BSM models. 
Further, we advocate that the theoretical uncertainty should be taken seriously and analyzed in all future phenomenological gravitational wave studies.

\section*{Acknowledgements}
The authors acknowledge discussions with P.~Athron, M.~Bardsley, C.~Balazs, J.R.~Espinosa, A.~Fowlie, T.~Konstandin, J.~Kozaczuk, M.~Laine,  D.~Morrissey, L.~Niemi,  H.H.~Patel, A.~Perko, M.~Ramsey-Musolf, K.~Rummukainen, Y.~Schr\"oder, Chen~Sun, D.J.~Weir and  Y.~Zhang.
In addition, TT is grateful for enlightening discussions with S.~Baum, C.~Cosme, D.~Cutting, A.~Kobakhidze, T.~Tenkanen and J. van de Vis.
GW (2018) and DC (2019) thank the Aspen Center for Physics, which is supported by National Science Foundation grant PHY-1607611, for their hospitality while this work was being completed.
TRIUMF receives federal funding via a contribution agreement with the National Research Council of Canada.
OG was supported by the Research Funds of the University of Helsinki.
This work was partly supported by the Swiss National Science Foundation (SNF) under grant 200020B-188712.
PS has been supported
by the European Research Council, grant no.~725369, and 
by the Academy of Finland, grant no.~1322507.
This work was supported by World Premier International Research Center Initiative (WPI), MEXT, Japan.

\appendix 

\section{SMEFT in four dimensions}
\label{appendix:smeft:4d}

This appendix collects multiple technical details of our computation in the four-dimensional SMEFT, as defined in Eq.~\eqref{eq:L:SMEFT}.
The conventions and notations of the SM parts follow Ref.~\cite{Brauner:2016fla}; see Sec. 2 of that reference. 

\subsection{Renormalisation: counterterms and running}
\label{appendix:zero_temp_matching}

Excluding the example in Fig.~\ref{fig:scale_dependence_no_running}, all calculations in this paper include RGE running of the parameters.
While our main focus is the SMEFT with only the inclusion of the sextic Higgs operator $c_6 (\phi^\dagger \phi)^3$, below we do include similar operators of dimension-8 and -10 with coefficients $c_8$ and $c_{10}$. For example, note that below the counterterm for the dimension-8 operator -- which is non-zero even if the coefficient $c_8$ itself vanishes -- is crucial in order to cancel divergences related to the dimension-6 coefficient $c_6$. This is due to the non-renormalisable nature of the SMEFT, where new higher dimensional operators have to be included to cancel divergences related to operators of lower dimension. For the same reason, the running of higher dimensional operators is non-zero, so even if these operators vanish at some initial scale, with RGE running they are non-zero at other scales. For this reason, one has to ensure that this running does not ruin the numerical analysis in the EFT without these operators. However, in most of our numerical analysis we simply neglect dimension-8 and -10 operators altogether.

For renormalisation, we use dimensional regularisation in the \MSbar-scheme, and for gauge fixing we adopt the class of general covariant (or Fermi) gauges, introducing gauge-fixing parameters
$\xi_1$ for the ${\rm U}(1)$ and
$\xi_2$ for the ${\rm SU}(2)$ sector.
In Landau gauge $\xi_{i} = 0$.
The one-loop counterterms that are affected by $c_{6}, c_{8}$ and $c_{10}$ read
\begin{align}
\delta\mh^2 &= \frac{1}{(4\pi)^2 \epsilon}  \mh^2 \bigg(6 \lambda - \frac{1}{4} \Big(3 \xi_2 \g^2 + \xi_1 {\gp}^2 \Big)  \bigg)
\;, \\
\delta\lambda &= \frac{1}{(4\pi)^2\epsilon} \frac{1}{2} \bigg( \frac{3}{8} \Big( 3\g^4 + {\gp}^4 + 2 \g^2 {\gp}^2 \Big) 
\nn \\
& - \lambda \Big(3 \xi_2 \g^2 + \xi_1 {\gp}^2 \Big)  + 24 \lambda^2 - 6 \gY^{4} + 24 c_6 \mh^2 \bigg)
\;, \\
\delta c_6 &= \frac{1}{(4\pi)^2 \epsilon} \bigg[ c_6 \bigg( 54\lambda - \frac{3}{4} \Big(3\xi_{2}^{ }\g^2 + \xi_{1}^{ }{\gp}^2 \Big)  \bigg) + 20\mh^{2} c_{8}^{ } \bigg]
\;, \\
\delta c_8 &= \frac{1}{(4\pi)^2 \epsilon} \bigg( 63 c^2_6 +30 \mh^2 c_{10} + c_8 \Big( 96 \lambda - 3 \xi_2 \g^2 - \xi_1 {\gp}^2 \Big) \bigg)
\;, \\
\delta c_{10} &= \frac{1}{(4\pi)^2 \epsilon} \bigg( 228 c_6 c_8 - \frac{5}{4} c_{10} \Big( -120 \lambda + 3 \xi_2 \g^2 + \xi_1 {\gp}^2 \Big) \bigg)
\;.
\end{align}
The corresponding $\beta$-functions read
\begin{align}
\label{eq:RGE_higgs_mass}
\bmu\frac{{\rm d}}{{\rm d}\bmu} \mh^2 &=
    \frac{1}{(4\pi)^2} \mh^2 \bigg( -\frac{3}{4} \Big(3\g^{2} + {\gp}^{2} \Big) + 12 \lambda + 6 \gY^{2}  \bigg)
    \;, \\
\bmu\frac{{\rm d}}{{\rm d}\bmu} \lambda &=
    \frac{1}{(4\pi)^2} \bigg( \frac{3}{8} \Big(3\g^{4} + {\gp}^{4} + 2 \g^{2} {\gp}^{2} \Big) + 24 \lambda^2 - 6\gY^4
    \nn \\ &
    + 24 c_6 \mh^2 - 3 \lambda \Big(3\g^{2} + {\gp}^{2} - 4 \gY^{2} \Big) \bigg)
    \;, \\
\bmu\frac{{\rm d}}{{\rm d}\bmu} c^{ }_6 &=
    \frac{1}{(4\pi)^2} \bigg[ c_6 \Big( 108 \lambda - \frac{9}{2}(3\g^{2} + {\gp}^{2}) + 18 \gY^{2} \Big) + 40 c_8 \mh^2 \bigg]
    \;, \\
\bmu\frac{{\rm d}}{{\rm d}\bmu} c^{ }_8 &=
    \frac{1}{(4\pi)^2} \bigg[ c_8 \Big( 198 \lambda - 6 (3\g^{2} + {\gp}^{2}) + 24 \gY^{2} \Big) + 126 c^2_6 + 60 c_{10} \mh^2 \bigg]
    \;, \\
\bmu\frac{{\rm d}}{{\rm d}\bmu} c^{ }_{10} &=
    \frac{1}{(4\pi)^2} \bigg[ c_{10} \Big( 300 \lambda - \frac{15}{2}(3\g^{2} + {\gp}^{2}) + 30 \gY^{2} \Big) + 456 c_6 c_8 \bigg]
    \;. 
\end{align}
These counterterms and $\beta$-functions unaffected by $c_{8}$ and $c_{10}$ are collected for example in
Refs.~\cite{Jenkins:2013zja,Jenkins:2013wua,Alonso:2013hga}.

Note that since the presence of a nonzero-temperature preserves the UV structure of the theory, these same counterterms and $\beta$-functions are used in the renormalisation of unbroken phase correlators for dimensional reduction and for the broken phase vacuum renormalisation calculation of pole mass corrections.

\subsection{Relations between \MSbar-parameters and physical observables}
\label{appendix:MSbar-relations}

We relate the \MSbar-parameters of the Lagrangian
to physical observables, that serve as input parameters
\begin{equation}
    (M_h, M_W, M_Z, M_t, G_f,\alphas) \mapsto (\mh,\lambda,\g,\gp,\gs,\gY) \;. 
\end{equation}
Note that the physically observed masses are the pole masses.
These relations also depend on the new \MSbar-scheme BSM parameters $c_6, c_8$ and $c_{10}$, which we also treat as input parameters.
For the values of the physical observables used in this work, we refer the reader to Table~\ref{tab:parameters}.
We define the shorthand notation
$g^2_0 \equiv 4 \sqrt{2}\,G_f M^2_W$ for the tree-level coupling and
$v^2_0 \equiv 4 M^2_W/g^2_0 \approx (246.22~ {\rm GeV})^2$ for the tree-level minimum.

At tree- and one-loop level only the Higgs mass parameter and self-coupling are affected by $c_6$, and the tree-level relations can be solved from ($V_{\rmi{tree}}$ is defined in Eq.~\eqref{eq:Vtree})
\begin{align}
\frac{\partial^2}{\partial v^2} \big(V_{\rmi{tree}}(v) \big)|_{v=v_0} &= M^2_h
\;,\\
\frac{\partial}{\partial v} \big(V_{\rmi{tree}}(v) \big)|_{v=v_0} &= 0
\;,
\end{align} 
resulting in
\begin{align}
\mh^2 &=
    - \frac{1}{2} M^2_h
    + \frac{3}{4} c_6 v^4_0
    \;, \\
\lambda &=
    \frac{1}{2} \frac{M^2_h}{v^2_0}
    - \frac{3}{2} c_6 v^2_0
    \;.
\end{align}
At tree-level, the relations for gauge and Yukawa couplings are unaffected by $c_6$ and read
\begin{align}
\g^{2} &= g^2_0
\;, \\
{\gp}^{2} &= g^2_0 \Big(\frac{M^2_Z}{M^2_W} - 1 \Big)
\;, \\
\gY^{2} &= \frac{1}{2} g^2_0 \frac{M^2_t}{M^2_W}
\;.
\end{align}

For an accurate numerical analysis of the thermodynamics, the above tree-level relations 
can
be improved by their one-loop corrections (cf.\ Refs.~\cite{Kajantie:1995dw,Kainulainen:2019kyp,Laine:2017hdk}).
These corrections are necessary for the complete $\mc{O}(g^4)$ accuracy of our 3d~approach.
Regarding the masses, this can be achieved with a standard one-loop pole mass renormalisation at zero temperature. 
With the Minkowski metric at zero temperature,
propagators are schematically dressed as
$1/\big(p^2 - m^2 + \Pi(p^2,\bmu) \big)$,
where
$\Pi$ is a self-energy function with external momentum $p$,
$\bmu$ is the \MSbar-scale, and
$m^2$ is the \MSbar-mass eigenvalue (see Eqs.~\eqref{eq:m:phi}--\eqref{eq:m:top}) at the tree-level minimum $v_0$. Diagrammatically $\Pi$ consists of
\begin{equation}
    \TopoST(\Lsai,\Asai) +
    \TopoSB(\Lsai,\Asai,\Asai) +
    \TopoSTT(\Lsai,\Lsai,\Asai)
    \;.
\end{equation}
The self-energy functions capture the momenta dependence of the two-point correlators, and in addition include one-particle reducible tadpole contributions.
Those are generated at one-loop by the non-zero vev in the broken phase computation, since the minimisation condition is imposed only at tree-level. Identifying the physical pole mass at $p^2 = M^2$ leads to a condition
$m^2 = M^2 + \Pi(M^2)$.
We denote
physical masses with capital letters, and
\MSbar-mass eigenvalues -- that are functions of running couplings -- by lower case letters.
Note that self-energies $\Pi$ are functions of \MSbar-masses $m^2$, but one may linearise these pole equations by replacing \MSbar-masses by the corresponding physical pole masses inside the one-loop computation, since the difference is formally of higher order.%
\footnote{
    In the presence of large coupling constants, this linearisation might be insufficient; see discussions in Refs.~\cite{Kainulainen:2019kyp,Laine:2017hdk}.
}
We employ this linearisation, as it suffices to reach the desired $\mc{O}(g^4)$ accuracy.

The one-loop correction $\delta\g^{2}$ to the ${\rm SU}(2)$ gauge coupling can be obtained by computing the one-loop 4-fermion correlator related to muon decay
$\mu^- \to e^- + \bar{\nu}_e + \nu_\mu$
at zero external momenta.
This correlator is proportional to
\begin{align}
\VtxvVD(\Lsai,\Lsai,\Lsai,\Lsai) = 
    - \frac{g_{0}^2}{2 M_{W}^{2}}
    \Bigl( 1 + \frac{\delta \g^{2}}{g_{0}^2}\Bigr)
    \;.
\end{align}
Diagrammatically, there is a tree-level contribution
\begin{equation}
\label{eq:4-fermion-tree}
\VtxvS(\Lsai,\Lsai,\Lsai,\Lsai,\Lglii)
    \propto
    - \frac{g_{0}^2}{2 M_{W}^{2}}
    \Bigl( 1 + \frac{\Pi_{W}(M_{W})}{M_{W}^{2}}\Bigr)
    \;,
\end{equation}
where the $W$-boson self-energy appears via the one-loop improved propagator, and the following classes of one-loop diagrams:
\begin{equation}
\label{eq:4-fermion-loops}
    \VtxvSB(\Lsai,\Lsai,\Lsai,\Lsai,\Lsai) +
    \VtxvSt(\Lsai,\Lsai,\Lsai,\Lsai,\Lsai,\Lsai,\Lsai,\Lsai) +
    \TopoVD(fex(\Lsai,\Lsai,\Lsai,\Lsai),\Asai,\Asai,\Asai,\Asai) +
    \VtxvSSB(\Lsai,\Lsai,\Lsai,\Lsai,\Lsai)
\end{equation}
where blobs denote possible one-loop attachments.
Note that all $c_6$-dependent contributions cancel between the $W$ self-energy piece in the above tree-level diagram of Eq.~\eqref{eq:4-fermion-tree} and the first diagram class of Eq.~\eqref{eq:4-fermion-loops}, such that the final result is equal to the SM. Furthermore, this correction is numerically small.

We adopt the approximation of Ref.~\cite{Kajantie:1995dw}, where the tree-level value of the U(1) gauge coupling is not improved at one-loop, due to its small numerical significance.
Note that one could include one-loop corrections to the U(1) gauge coupling for example by computing the correction to Thomson scattering (cf.~Ref.~\cite{Kainulainen:2019kyp}).

In total, we have the conditions:
\begin{align}
\label{eq:zero_temp_matching}
m_{\phi}^{2} &= M^2_h + \mbox{Re} \; \Pi^{ }_{h}(M^2_h)
\;, \\
m_{W}^{2} &= M^2_W + \mbox{Re} \; \Pi^{ }_{W}(M^2_W)
\;, \\
m_{t}^{2} &= M^2_t \bigg( 1 + 2 \text{Re} \; \Big(
    \Sigma^{ }_{v}(M^2_t) +
    \Sigma^{ }_{s}(M^2_t) \Big) \bigg) 
\;, \\
\g^{2} &= g^2_0 + \delta\g^{2}
\;.
\end{align} 
Here the top quark self energy consists of
a vector part $\Sigma_v$ and
a scalar part $\Sigma_s$;
see Sec.~5 in Ref.~\cite{Kajantie:1995dw}.
Its axial and axial vector parts do not contribute to the pole mass condition.

From these equations we can now solve for
the one-loop improved \MSbar-parameters in terms of
the physical parameters:
\begin{align}
\mh^{2} &=
    - \frac{1}{2} M_{h}^{2} \bigg( 1 + \frac{\mbox{Re} \; \Pi_{h}^{}(M^2_h)}{M^2_h} \bigg)
    + 12 M_{W}^{4} \frac{c_6}{g^4_0} \bigg( 1 - 2 \frac{\delta \g^{2}}{g^2_0} + 2 \frac{\text{Re} \; \Pi_{W}(M^2_W)}{M^2_W}  \bigg)
    \;, \\
\lambda &= \frac{1}{8} g^2_0 \frac{M^2_h}{M^2_W} \bigg( 1
    + \frac{\delta \g^{2}}{g^2_0}
    +  \frac{\mbox{Re} \; \Pi_h(M^2_h)}{M^2_h} 
    -  \frac{\mbox{Re} \; \Pi_W(M^2_W)}{M^2_W} \bigg)
    \nn \\ &
    - 6 M^2_W \frac{c_6}{g^2_0} \bigg(  1 - \frac{\delta \g^{2}}{g^2_0} + \frac{\mbox{Re} \; \Pi_W(M^2_W)}{M^2_W} \bigg)
    \;, \\
\gY^{2} &= \frac{1}{2} g^2_0 \frac{M^2_t}{M^2_W} \bigg( 1 + \frac{\delta \g^{2}}{g^2_0} - \frac{\mbox{Re} \; \Pi_W(M^2_W)}{M^2_W} + 2\mbox{Re}  \Big( \Sigma_v(M^2_t) + \Sigma_s(M^2_t) \Big) \bigg)
\;, \\
\g^{2} &= g^2_0 \Big(1 + \frac{\delta \g^{2}}{g^2_0} \Big)
\;.
\end{align}
The final task is to evaluate the self-energy functions. This calculation is straightforward since there are no new fields compared to the pure SM computation:
there are no new diagrams involved, and
$c_6$ only modifies the mass eigenvalues and vertices. 
By direct computation -- and by adopting the shorthand notations
$h \equiv M_h/M_W$,
$t \equiv M_t/M_W$,
$z \equiv M_Z/M_W$, and
$s \equiv g_3/g_0$ -- we obtain 
\begin{align}
\Pi_h(M^2_h) &=
    \frac{3}{8} \frac{g^2_0 M^2_h}{(4\pi)^2} \bigg(
    - \frac{4}{3}
    - 8 \frac{1}{h^2}
    - 2h^2
    + 16 \frac{t^4}{h^2}
    - \frac{2}{3} z^2
    - 4\frac{z^4}{h^2}
    \nn \\ &
    + 3h^2 F(M_h,M_h,M_h)
    + 4t^2\Big( 1 - 4 \frac{t^2}{h^2} \Big) F(M_h,M_t,M_t)
    \nn \\ &
    + \frac{2}{3}\frac{h^4 - 4h^2 + 12}{h^2} F(M_h,M_W,M_W)
    \nn \\ &
    + \Big( \frac{1}{3} \frac{1}{h^2} - \frac{4}{3}z^2 + 4\frac{z^4}{h^2} \Big) F(M_h,M_Z,M_Z)
    \nn \\ &
    - 2h^2 \ln(h)
    - 8t^2 \ln(t)
    + \Big( -\frac{2}{3}h^2 + 4 z^2 \Big) \ln(z)
    \nn \\ &
    + \Big( -4 +2h^2 + 4 t^2 - 2 z^2 \Big) \ln\Big( \frac{\bmu^2}{M^2_W} \Big)
    \nn \\ &
    + 64 c_6 \frac{M^2_W}{g^4_0 h^4} \bigg[ -2 + 12 t^4 - z^4 +  3 h^4 F(M_h,M_h,M_h)
    - 6 h^4 \ln(h)
    \nn \\ &
    - 24 t^4 \ln(t)
    + 6z^4 \ln(z)
    + \Big( -2 + h^4 + 4 t^4 - z^4 \Big) \ln\Big( \frac{\bmu^2}{M^2_W} \Big) \bigg]
    \nn \\ &
    + 3072 c_6^{2} \frac{M^4_W}{g^4_0 h^2} \bigg[ -1 + F(M_h,M_h,M_h) \bigg]\bigg)
    \;,
\end{align}
\begin{align}
\Pi_W(M^2_W) &=
    \frac{3}{8} \frac{g^2_0 M^2_W}{(4\pi)^2} \bigg(
    - \frac{212}{9}
    - \frac{8}{3}\frac{1}{h^2}
    - \frac{22}{9} h^2
    + \frac{4}{27} (40\Nf - 17)
    - \frac{4}{3}t^2
    + 16 \frac{t^4}{h^2}
    + \frac{14}{9}z^2
    - \frac{4}{3} \frac{z^4}{h^2}
    \nn \\ &
    + \frac{4h^2 (h^2-2)}{h^2-1} \ln(h)
    - 8\Big( \frac{2}{3} - t^2 + 4\frac{t^4}{h^2} \Big) \ln(t)
    + 4\Big(2 \frac{z^4}{h^2} - \frac{z^4 - 4 z^2 - 8}{z^2-1} \Big) \ln(z)
    \nn \\ &
    + \frac{2}{9}\Big( 12 - 4 h^2 + h^4 \Big) F(M_W,M_h,M_W)
    - \frac{4}{3}(t^2+2)(t^2-1) F(M_W,M_t,0)
    \nn \\ &
    - \frac{32}{3}\frac{z^2-1}{z^2} F(M_W,M_W,0)
    + \frac{2}{9}\frac{(z^4 + 20z^2 + 12)(z^2-4)}{z^2} F(M_W,M_W,M_Z)
    \nn \\ &
    + 2\bigg[-1 + \frac{2}{h^2} + \Big( -\frac{59}{9} - 6 \frac{1}{h^2} - h^2 + \frac{16}{9}\Nf - 2 t^2 + 8 \frac{t^4}{h^2} \Big) + z^2 - 2\frac{z^4}{h^2} \bigg] \ln\Big( \frac{\bmu^2}{M^2_W} \Big)
    \nn \\ &
    - \frac{8}{9}\pi i (4\Nf - 3)
    - 64 c_6 \frac{M^2_W}{g^4_0} \bigg[1 - \ln\Big( \frac{\bmu^2}{M^2_h} \Big) \bigg]  \bigg)
    \;, 
\end{align}
\begin{align}
\Sigma_v(M^2_t) + \Sigma_s(M^2_t) &=
    \frac{3}{16} \frac{g^2_0}{(4\pi)^2} \bigg( -2 - 4 \frac{1}{h^2} - 2 h^2 - \frac{256}{9}s^2 + 2 t^2 + 16 \frac{t^4}{h^2}
    \nn \\ &
    - \frac{2}{27} \Big( 39 - \frac{64}{z^2} + 25 z^2 + 18\frac{z^4-1}{h^2} \Big)
    \nn \\ &
    + \Big(4 h^2 - \frac{8}{3}t^2 + \frac{4}{3} t^2 \frac{2t^2 + h^2}{t^2-h^2} \Big) \ln(h) - \frac{8}{9} \Big( -9 \frac{z^4}{h^2} + 4 \frac{(4-5z^2+z^4)}{t^2-z^2} \Big) \ln(z)
    \nn \\ &
    + \Big( \frac{128}{3}s^2 -32 \frac{t^4}{h^2} - \frac{4}{3} \frac{t^2(2t^2 + h^2)}{t^2-h^2} - \frac{32}{9} \frac{(z^2-1)(t^2-4) }{z^2-t^2} \Big) \ln(t)
    \nn \\ &
    + \frac{2}{3}\Big(4t^2 -h^2  \Big) F(M_t,M_t,M_h)
    + \frac{2}{3}\frac{(t^2+2)(t^2-1)}{t^2} F(M_t,M_W,0)
    \nn \\ &
    - \frac{2}{27}\Big( \frac{64 - 80z^2 + 7z^4}{z^2}
     + \frac{32-40z^2 + 17z^4}{t^2} \Big) F(M_t,M_t,M_Z)
    \nn \\ &
    + \bigg[ 2\Big(-6 \frac{1}{h^2} - h^2 -\frac{32}{3}s^2 + t^2 + 8 \frac{t^4}{h^2} \Big)
    \nn \\ &
    - \frac{4}{9}\frac{(z^2-1)(9+4h^2+9z^2)}{h^2} \bigg] \ln\Big( \frac{\bmu^2}{M^2_W} \Big) \bigg)
    \nn \\ &
    - 64 c_6 \frac{M^2_W}{g^4_0} \bigg[1 - \ln\Big( \frac{\bmu^2}{M^2_h} \Big) \bigg] \bigg)
    \;,\\[3mm]
\frac{\delta \g^{2}}{g^2_0} &=
    \frac{1}{(4\pi)^2} g^2_0  \bigg(
    - \frac{257}{72}
    - \frac{1}{24}h^2
    + \frac{20}{9} \Nf
    + \frac{1}{4}t^2 
    - 2\ln(t)
    \nn \\ &
    + \frac{1}{12}(12 - 4 h^2 + h^4) F(M_W,M_h,M_W)
    - \frac{(t^2+2)(t^2-1)}{2} F(M_W,M_t,0)
    \nn \\ &
    - \frac{33}{4}F(M_W,M_W,M_W)
    + \Big(\frac{4}{3}\Nf -\frac{43}{6} \Big)\ln\Big( \frac{\bmu^2}{M^2_W} \Big)
    \bigg)\;.
\end{align}
The one-loop integral function $F(k,m_1,m_2)$ and its various limits are given in Eqs.~(187--188) in Ref.~\cite{Kajantie:1995dw}.
Also note that in the above self-energies, U(1) gauge contributions can be turned off by taking
the limit $z\to 1$ and $M_Z \to M_W$.
Apart from the new $c_6$ terms and U(1) gauge contributions, these results agree with Eqs.~(184), (191--193) in Ref.~\cite{Kajantie:1995dw}, apart from a $-\frac{8}{3} t^2 \ln(h)$ term which is missing in the top quark self energy in Eq.~(193) therein.%
\footnote{
    The erroneous result of Ref.~\cite{Kajantie:1995dw} for the top quark self-energy is also pointed out in Ref.~\cite{Laine:2017hdk}, which performed a similar computation in the Inert Doublet Model.
}  
From the above self-energy functions and relations between \MSbar-parameters and physical parameters we observe that $\g$ and $\gY$ are independent of $c_6$ also at one-loop order.
Even though self-energies for both the $W$-boson and the top quark have $c_6$-dependent pieces, these cancel each other, and the 4-fermion correlator of muon decay is explicitly independent of $c_6$.

Note that often~\cite{Bodeker:2004ws,Chala:2018ari} one-loop improved relations for the scalar mass parameter and self coupling are obtained from the conditions
\begin{align}
\frac{\partial^2}{\partial v^2} \big(V_{\rm tree} + V_{\rmii{CW}} \big)|_{v=v_0} &= M^2_h
\;, \\
\frac{\partial}{\partial v} \big(V_{\rm tree} + V_{\rmii{CW}} \big)|_{v=v_0} &= 0
\;,
\end{align} 
where the tree-level potential is accompanied with the one-loop Coleman-Weinberg potential at zero temperature (see Eq.~\eqref{eq:V:CW}), and $\mh^2$ and $\lambda$ are solved numerically from these equations. However, these conditions should be taken only as a heuristic approximation, since the physical pole mass lies at nonzero momentum and hence cannot be obtained from the effective potential (see also Ref.~\cite{Delaunay:2007wb}).
\begin{figure}
  \centering
  \includegraphics[width=\textwidth]{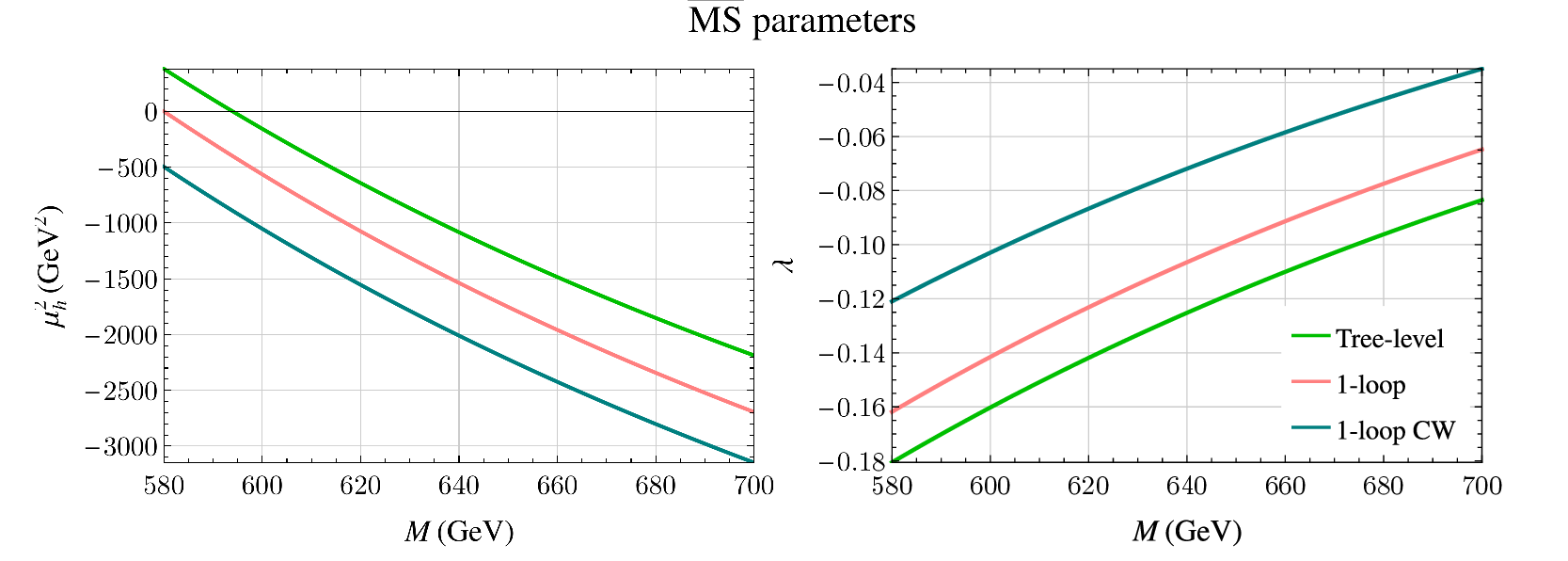}
  \caption{
    Higgs \MSbar-parameters as a function of the cutoff scale $M$ in different approximations.
  }
  \label{fig:MSbar-comparison}
\end{figure}
Figure~\ref{fig:MSbar-comparison} compares the scalar parameters as functions of $c_6=1/M^2$, at tree-level versus one-loop -- computed both from the Coleman-Weinberg potential and from pole conditions. This comparison shows that in fact the estimated 1-loop effect from the Coleman-Weinberg potential overestimates the difference with respect to the tree-level result, compared to the computation from pole masses that consistently include momentum-dependent contributions and tadpole diagrams.

\subsection{Mass eigenvalues and thermal screening}

\label{appendix:4Dmasses}
Calculating the effective potential to one-loop level by using field-dependent mass eigenvalues, requires fixing a gauge.
In this context it is common to use an $R_{\xi}$-style gauge.
However this fixes a different gauge for each field value; regarding the gauge-dependent effective potential, it is not clear {\em a priori} that this is permissible~\cite{Arnold:1992fb,Laine:1994bf}. We therefore follow Ref.~\cite{Andreassen:2013hpa} by using a general covariant (or Fermi) gauge,%
which does not include the vev in the gauge-fixing Lagrangian, to calculate the loop corrections to the effective potential.
Gauge parameters are denoted as $\xi_2$ and $\xi_1$ for the SU(2) and U(1) fields respectively.
In this case the field-dependent mass eigenvalues are given by
\begin{align}
\label{eq:m:phi}
m_\phi^2 &=
    \mh^2
    + 3\lambda \phi^2
    + \frac{15}{4}c_6 \phi^4
    \;,\\
(m_1^\pm )^2 &=
    \frac{1}{2}\left( m_\chi^2
    \pm \sqrt{ m_\chi^2 (m_\chi^2 - \xi_{1}{\gp}^2 \phi^2 - \xi_{2} \g^{2} \phi^2)} \right)
\label{eq:m:gs1}    \;,\\
(m_2^\pm )^2 = (m_3^\pm )^2 &=
    \frac{1}{2} \left( m_\chi^2 
    \pm \sqrt{m_\chi^2  (m_\chi^2 - \xi_{2}\g^{2} \phi^2)} \right)
\label{eq:m:gs2}    \;,\\
m_W^2 &=
    \frac{1}{4}\g^{2} \phi^2
    \;,\\
m_Z^2 &=
    \frac{1}{4} \Big( {\gp}^2+\g^{2} \Big) \phi^2
    \;,\\
\label{eq:m:gamma}
m^2_{\gamma} &= 0
\;, \\
\label{eq:m:top}
m_t^2 &=
    \frac{1}{2} \gY^{2} \phi^2
    \;,
\end{align}
where
$m_\chi^2= \mh^2 + \lambda  \phi^2 +\frac{3}{4} c_{6}^{ } \phi^4$ is the Goldstone mode mass eigenvalue in the Landau gauge. 
Note the complicated gauge dependence in the Goldstone mass eigenvalues
$m^{2\pm}_{1}$,
$m^{2\pm}_{2}$, and
$m^{2\pm}_{3}$.
A very illuminating and thorough derivation for the gauge dependent parts of these mass eigenvalues can be found in Ref.~\cite{Andreassen:2018xx} (see also Ref.~\cite{Espinosa:2016uaw}).

Turning to high temperatures, the most straightforward way to obtain thermally resummed mass eigenvalues, is to calculate thermal corrections from hard modes to the zero modes of the fields in the gauge eigenstate basis of the theory; we demonstrate this explicitly in Appendix ~\ref{appendix:matching}.
Gauge fields $A^a_0$ and $B_0$ obtain Debye masses
\begin{align}
\Pi_{A_0}(T) &= T^2 \g^{2}  \Big(\frac{5}{6} + \frac{1}{3}\Nf \Big) \equiv \mD^{2}\;, \\
\Pi_{B_0}(T) &= T^2 {\gp}^2  \Big(\frac{1}{6} + \frac{5}{9}\Nf \Big) \equiv \mD'^{2}
\;. 
\end{align}
The number of kinematically active families is $\Nf=3$ at electroweak-scale temperatures. For temporal gauge fields, the bilinear part of the Lagrangian reads
\begin{align}
\frac{1}{2}
  \begin{pmatrix}
    A^1_0 & A^2_0 & A^3_0 & B_0
  \end{pmatrix}
  \begin{pmatrix}
    \frac{1}{4} \g^{2} \phi^2 + \mD^{2} & 0 & 0 & 0  \\
    0 & \frac{1}{4} \g^{2} \phi^2 + \mD^{2} & 0 & 0 \\ 
    0 & 0 & \frac{1}{4} \g^{2} \phi^2 + \mD^{2} & \frac{1}{4}\g\gp\phi^2  \\
    0 & 0 & \frac{1}{4} \g\gp \phi^2 & \frac{1}{4} {\gp}^2 \phi^2 + \mD'^{2}
  \end{pmatrix}
  \begin{pmatrix}
    A^1_0  \\
    A^2_0  \\ 
    A^3_0  \\
    B_0 
  \end{pmatrix}
  \;.
\end{align}
A rotation by the Weinberg angle%
\footnote{
$
\begin{pmatrix}
     \cos(\theta) & \sin(\theta)  \\
    -\sin(\theta) & \cos(\theta)
\end{pmatrix}%
\begin{pmatrix}
    Z_0  \\
    A_0 
\end{pmatrix} =
\begin{pmatrix}
    A^3_0  \\
    B_0
\end{pmatrix}\;,$
where
$
  \cos(\theta) = \g/\sqrt{\g^{2}+{\gp}^{2}},
  \; \sin(\theta) = {\gp}/\sqrt{\g^{2}+{\gp}^2}
  \;.
$
} 
yields
the resummed gauge field masses
\begin{align}
m^2_{W,\rmi{res.}} &= m^2_W + \Pi^{ }_W(T)
\;,  \\
m^2_{Z,\rmi{res.}} &= m^2_Z + \Pi^{ }_Z(T)
\;, \\
m^2_{\gamma, \rmi{res.}} &= \Pi_\gamma(T)
\;,
\end{align}
where thermal corrections read
\begin{align}
\Pi_W(T) &= \mD^{2}
\;, \\ 
\Pi_Z(T) &= \Big( \g^{2} \mD^{2} + {\gp}^2 \mD'^{2} \Big)/(\g^{2} + {\gp}^2)
\;, \\
\Pi_\gamma(T) &= \Big( {\gp}^{2} \mD^{2} + \g^{2} \mD'^{2} \Big)/(\g^{2} + {\gp}^2)
\;.
\end{align}
However, due to thermal screening this is not an eigenstate basis, as in fact there remains a mixing term 
\begin{align}
\frac{\g {\gp}}{\g^{2} + {\gp}^2} (\mD^{2} - \mD'^{2})\times  A_0 Z_0
\;,
\end{align}
i.e.\ the $Z$-boson and photon are {\em not} mass eigenstates in the heat bath.
The actual resummed mass eigenvalues read
\begin{align}
m^2_{Z',\rmi{res.}} &= \frac{1}{8} \bigg(
    (\g^{2} + {\gp}^2 )\phi^2 + 4(\mD^{2} + \mD'^{2})
    \nn \\ &
    \hphantom{=\frac{1}{8} \bigg(}
    + \sqrt{\g^{4} \phi^4
    + 2 \g^{2} \phi^2 \big({\gp}^2 \phi^2 + 4(\mD^{2} - \mD'^{2}) \big)
    + \big({\gp}^2 \phi^2 - 4(\mD^{2} - \mD'^{2}) \big)^2
    }\,\bigg)
\;, \\
m^2_{A', \rmi{res.}} &= \frac{1}{8} \bigg(
    (\g^{2} + {\gp}^2 )\phi^2 + 4(\mD^{2} + \mD'^{2})
    \nn \\ &
    \hphantom{=\frac{1}{8} \bigg(}
    - \sqrt{\g^{4}\phi^4
    + 2\g^{2}\phi^2 \big({\gp}^{2} \phi^2 + 4(\mD^{2} - \mD'^{2}) \big)
    + \big({\gp}^{2}\phi^2 - 4(\mD^{2} - \mD'^{2}) \big)^2
    }\,\bigg)
\;, 
\end{align}
where the rotation angle to define eigenstates $Z', A'$ depends now on $\mD^{2}$ and $\mD'^{2}$ and exhibits a more complicated form compared to the usual vacuum Weinberg angle. 
Note that if these mass eigenvalues are linearised with respect to the thermal contributions $\mD^{2}$ and $\mD'^{2}$, one gets exactly the above resummed $Z$-boson and photon masses.
In fact this difference is numerically small, which suggests that one could still treat the $Z$-boson and the photon as mass eigenstates.
Also note that in the above $\Pi_{Z,W,\gamma}$ are for the longitudinal modes (perturbative Debye masses for the transverse modes vanish).

For the scalar doublet, thermal screening affects both its mass and self-coupling, with contributions
\begin{align}
\label{eq:higgs-thermal-mass}
\Pi_\phi(T) &=
    \frac{T^2}{12} \Big(6\lambda + \frac{3}{4}(3\g^{2} + {\gp}^{2}) + 3 \gY^{2} \Big) + \frac{1}{4} T^4 c_6
    \;, \\
\Gamma_\lambda(T) &= T^2 c_6
    \;,
\end{align}
and the resummed parameters become
\begin{align}
\mu^2_{h,\rmi{res.}} &= \mh^2 + \Pi^{ }_\phi(T)
\;, \\
\lambda_{\rmi{res.}} &= \lambda + \Gamma_\lambda(T)
\;.
\end{align}
The above $c_6$ contributions are those given by the flower diagrams in Eq.~\eqref{eq:flower}. Resummed scalar mass eigenvalues are then obtained by substituting the mass parameter $\mh^2$ and self-coupling $\lambda$ for their resummed values in $m^2_\phi$ and
$m^2_\chi$.

\subsection{One-loop thermal effective potential}
\label{appendix:V:1loop}

This appendix composes the thermal effective potential for the SMEFT in four dimensions and implements the leading ring resummations, utilising the mass eigenvalues and thermal corrections of Appendix~\ref{appendix:4Dmasses}.
As a reminder, we fix the \MSbar-parameters on the $Z$-pole as explained in Appendix~\ref{appendix:MSbar-relations} and evolve them using the renormalisation group equations (RGEs) in Appendix~\ref{appendix:zero_temp_matching}.

The one-loop correction to the effective potential can be computed as a sum over all one-particle irreducible diagrams with a single loop and zero external momenta.
This standard computation (see Ref.~\cite{Sher:1988mj} for an illuminating derivation) results in the master sum-integral $J_{\rmii{1-loop}}$ of Eq.~\eqref{eq:J}.
Using Eqs.~\eqref{eq:m:phi}--\eqref{eq:m:top}, the full (unresummed) one-loop correction reads
\begin{align}
V_{\rmii{1-loop}}(\phi,T,\bmu) &=
    J_{\rmii{1-loop}}(m_\phi)
    + 2 J_{\rmii{1-loop}}(m^+_2)
    + 2 J_{\rmii{1-loop}}(m^-_2) \nn \\ &
    + J_{\rmii{1-loop}}(m^+_1)
    + J_{\rmii{1-loop}}(m^-_1)
    - 4\Nc J_{\rmii{1-loop}}(m_t) \nn \\ &
    + (D-1)\Big(
    2 J_{\rmii{1-loop}}(m_W)
    + J_{\rmii{1-loop}}(m_Z)
    + J_{\rmii{1-loop}}(m_\gamma) \Big)
    \;.
\end{align} 
This can be divided into a zero-temperature Coleman-Weinberg piece and a thermal piece as in Eq.~\eqref{eq:J:split}, and can be evaluated by using
\begin{align}
\label{eq:J:cw}
J_{\rmii{CW}}(m)
\equiv \frac{1}{2} \Big( \frac{\bmu^{2}e^\gammaE}{4\pi} \Big)^\epsilon \int \frac{{\rm d}p^D}{(2\pi)^D} \ln(p^2+m^2) =
-\frac{1}{2} \Big( \frac{\bmu^{2}e^\gammaE}{4\pi} \Big)^\epsilon \frac{[m^2]^\frac{D}{2}}{(4\pi)^{\frac{D}{2}}} \frac{\Gamma(-\frac{D}{2})}{\Gamma(1)}
\;,
\end{align}
with $D=4-2\epsilon$.
The divergent $1/\epsilon$ terms are cancelled by the counterterm part
\begin{align}
V_{\rmii{CT}} =
      \frac{1}{2} \phi^2\,\delta\mh^{2}
    + \frac{1}{4} \phi^4\,\delta\lambda 
    + \frac{1}{8} \phi^6\,\delta c_6
    \;.
\end{align}
The thermal functions from Eq.~\eqref{eq:J:split} are expanded at high-$T$ in $d=3-2\epsilon$ with finite parts
\begin{align}
\label{eq:J:b}
J_{\T,b}(z)T^{-4} &=
    - \frac{\pi^2}{90}
    + \frac{1}{24} z^2
    - \frac{1}{12\pi} z^3
    - \frac{1}{4(4\pi)^2} z^4 \ln\Big(\frac{z^2}{a_b}\Big)
    + \mc{O}(z^6)
    \;, \\
\label{eq:J:f}
J_{\T,f}(z)T^{-4} &=
    \frac{7}{8}\frac{\pi^2}{90}
    - \frac{1}{48} z^2
    - \frac{1}{4(4\pi)^2} z^4 \ln\Big(\frac{z^2}{a_f}\Big)
    + \mc{O}(z^6)
    \;,
\end{align}
where
$z=m/T$,
$a_b = (4\pi)^2 \exp(\frac{3}{2}-2\gammaE)$ and
$a_f = 16 a_b$.
Here $z$-independent terms do not contribute to the dynamics of the phase transition and can be dropped.

With these tools -- together with the daisy prescription for resummation as explained in Sec.~\ref{sec:overview_4d} -- one can write down the familiar result as a sum of a temperature dependent and a temperature independent, Coleman-Weinberg piece 
\begin{equation}
V_{\rmii{1-loop}} =
      V_{\rmii{CW}}
    + V_{\T}
    + V_{\rmii{daisy}}
\;,
\end{equation}
where both terms in the one-loop correction rely on the field dependent masses. Explicitly, this is
\begin{align}
\label{eq:V:CW}
    V_{\rmii{CW}} &=
    - 12\frac{ m_t^4}{64 \pi^2} \left( \ln\left( \frac{m_t^2}{\bmu^2} \right)- \frac{3}{2} \right) 
    + 3\!\!\sum_{i\in\{W,Z\}} \frac{m_{i}^4}{64 \pi^2} \left( \ln\left( \frac{m_{i}^2}{\bmu^2} \right) - \frac{5}{6} \right)
    \nn\\ &
    + \!\!\sum_{i\in\{\rmi{scalars}\}} \frac{m_{i}^4}{64\pi^2} \left( \ln\left( \frac{m_{i}^2}{\bmu^2} \right) - \frac{3}{2} \right)
    \;,
\end{align}
where ${\rm scalars}$ includes the
Goldstone modes,
gauge bosons,
Higgs degrees of freedom, 
and
\begin{align}
\label{eq:VT}
V_{\T} &=
    - 12
    J_{\T,f} \left( \frac{m_{t}^2}{T^2} \right)
    + 3\!\!\sum_{i\in\{W,Z\}}
    J_{\T,b} \left( \frac{m_{i}^2}{T^2} \right)
    +\sum_{i\in\{\rmi{scalars}\}}
    J_{\T,b} \left( \frac{m_{i}^2}{T^2} \right)
    \;. 
\end{align}
The thermal part is accompanied with daisy terms for bosonic fields
\begin{equation}
\label{eq:arnold-espinosa}
    V_{\rmii{daisy}} =
        \sum_{\rmi{scalars},W,Z,\gamma}
        - \frac{T}{12 \pi} \left( [m^2_{\rmi{res.}} ]^\frac{3}{2}-m^3 \right)
    \;,
\end{equation}
where $m^2_{\rmi{res.}}$ denotes the mass eigenvalue with resummed parameters.
Note that the daisy contribution of the photon is non-zero due to thermal screening, but this does not give any field-dependent contribution. 
The gauge dependence manifests itself through the field dependent mass eigenvalues $m$.
These we give in Appendix~\ref{appendix:4Dmasses} together with the resummed mass eigenvalues. 

Finally, an alternative form for the one-loop part of the effective potential is provided by separation of the master sum-integral into the soft and hard parts following Eq.~\eqref{eq:I:split:s:h}, and upgrading the mass eigenvalues in the soft part to their resummed versions. The soft parts can be evaluated as
$T J_{\rmii{soft}}(m_{\rmii{res.}})$ (and noting that in the reduced dimension the overall factor for the gauge parts is $d-1$ and not $D-1$) and
the hard parts in the high-$T$ expansion are
\begin{align}
J^{b/f}_{\rmii{hard}}(m) \simeq
  \frac{1}{2} m^2 I^{4b/f}_1
- \frac{1}{4} m^4 I^{4b/f}_2 
+ \frac{1}{6} m^6 I^{4b/f}_3
+ \mc{O}(m^8/T^4)
\;,
\end{align}
where we have dropped a mass-independent piece.
The one-loop master sum-integrals
\begin{align}
\label{eq:I:4bf}
I^{4b/f}_{\alpha} \equiv \Tint{P/\{P \} }' \frac{1}{[P^2]^\alpha}
\;,
\end{align}
are expressed for example in Refs.~\cite{Gorda:2018hvi,Schicho:2020xaf}.

\subsubsection*{Correlators from the effective potential}

We have carried out a crosscheck of our results, utilising the fact that the effective potential is the generator of renormalised 1PI correlators of the Higgs field.
We utilised two different methods to compute $\hat{\Gamma}_{\phi^\dagger \phi}$,
$\hat{\Gamma}_{(\phi^\dagger \phi)^2}$, and
$\hat{\Gamma}_{(\phi^\dagger \phi)^3}$.
On the one hand, we computed them by taking derivatives of the effective potential with respect to the external scalar field.
On the other hand, we computed the same correlators directly in terms of Feynman diagrams.
For this we utilised computer algebra tools to automate the diagrammatic calculation, as discussed in Sec.~\ref{appendix:matching}.
We emphasise that this is a strong crosscheck, since for example the six-point correlator in the diagrammatic calculation is a sum of $\mc{O}(10^2)$ different diagrams, each with permutated external legs.%
\footnote{
    For fun: In total the six-point correlator contains 2185 diagrams including permutated external legs.
} 

When the scalar field is shifted by the background field $\phi/\sqrt 2$, one has at one-loop order
\begin{align}
V_{\rmii{eff}} &=
    \frac{1}{2} \mh^2 \phi^2
    + \frac{1}{4} \lambda \phi^4
    + \frac{1}{8} c_6 \phi^6
    + V_{\rmii{1-loop}}
    \;, 
\end{align}
where the one-loop piece expands in powers of the background field, also utilising the high-$T$ expansion, as
\begin{align}
V_{\rmii{1-loop}} &\simeq
    \frac{1}{2} \delta \mh^2 v^2
    + \frac{1}{4} \delta \lambda v^4
    + \frac{1}{8} \delta c_6 v^6
    + V_1
    \nn \\ &=
    \frac{1}{2} \Big( \delta \mh^2 - \Gamma_{\phi^\dagger \phi}  \Big) v^2
    + \frac{1}{4} \Big( \delta \lambda - \frac{1}{2} \Gamma_{(\phi^\dagger \phi)^2}  \Big) v^4
    + \frac{1}{8} \Big( \delta c_6 -\frac{1}{6} \Gamma_{(\phi^\dagger \phi)^3} \Big) v^6
    + \mc{O}(v^8)
    \nn \\ &=
    - \frac{1}{2} \hat{\Gamma}_{\phi^\dagger \phi} v^2
    - \frac{1}{8} \hat{\Gamma}_{(\phi^\dagger \phi)^{2}} v^4
    - \frac{1}{48}\hat{\Gamma}_{(\phi^\dagger \phi)^{3}} v^6
    + \mc{O}(v^8)
    \;,
\end{align}
in renormalised perturbation theory, where the hat denotes the renormalised correlation function.
The diagrammatic one-loop piece $V_1$ can be computed as (note that for simplicity we include only hard contributions -- and drop $\mc{O}(v^0)$ terms)
\begin{align}
V_1 &= 2d J^{b}_{\rmii{hard}}(m_W)
    + d J^{b}_{\rmii{hard}}(m_Z)
    + J^{b}_{\rmii{hard}}(m_\phi)
    + J^{b}_{\rmii{hard}}(m_{\rmii{2}^+})
    + J^{b}_{\rmii{hard}}(m_{\rmii{2}^-})
    \nn \\ &
    + J^{b}_{\rmii{hard}}(m_{\rmii{3}^+})
    + J^{b}_{\rmii{hard}}(m_{\rmii{3}^-})
    + J^{b}_{\rmii{hard}}(m_{\rmii{1}^+})
    + J^{b}_{\rmii{hard}}(m_{\rmii{1}^-})
    - 4 \Nc J^{f}_{\rmii{hard}}(m_t)
    \;,
\end{align}
where $d=3-2\epsilon$.
The $J^{b/f}_{\rmii{hard}}$-functions
read
\begin{align}
\label{eq:hard-expansion-bosonic}
J^{b}_{\rmii{hard}}(m) 
&\simeq
    + m^2 \frac{T^2}{24}
    - \frac{m^4}{4(4\pi)^2}\Big(\frac{1}{\epsilon} + L_b \Big)
    + \frac{\zeta(3)}{3 (4\pi)^4}\frac{m^6}{T^2}
    \nn \\ &
    - \frac{\zeta(5)}{2(4\pi)^6}\frac{m^8}{T^4}
    + \frac{\zeta(7)}{(4\pi)^8}\frac{m^{10}}{T^6}
    + \mathcal{O}\Bigl(\frac{m^{12}}{T^8}\Bigr)
    \;, \\
\label{eq:hard-expansion-fermionic}
J^{f}_{\rmii{hard}}(m) 
&\simeq
    - m^2 \frac{T^2}{48}
    - \frac{m^4}{4(4\pi)^2}\Big(\frac{1}{\epsilon} + L_f \Big)
    + \frac{7 \zeta(3)}{3 (4\pi)^4}\frac{m^6}{T^2}
    \nn \\ &
    - \frac{31 \zeta(5)}{2 (4\pi)^6}\frac{m^8}{T^4}
    + \frac{127 \zeta(7)}{(4\pi)^8}\frac{m^{10}}{T^6}
    + \mathcal{O}\Bigl(\frac{m^{12}}{T^8}\Bigr)
    \;,
\end{align}
with $\zeta(n)$ the Riemann zeta function and employing
the shorthand notation 
\begin{align}
\label{eq:Lb}
    L_b &\equiv2\ln\Big(\frac{\bmu}{T}\Big)-2[\ln(4\pi)-\gammaE]
    \;, \\
\label{eq:Lf}
    L_f &\equiv L_b + 4\ln2
    \;.
\end{align}
We have included terms with $\zeta(5)$ and $\zeta(7)$ that are used in computing some leading (1-loop) corrections to scalar 8- and 10-point correlators.
The correlators, $\hat{\Gamma}$, can then be extracted from the coefficients of the expansion in the background field.
One obtains the same pure scalar correlators as in
Appendix.~\ref{appendix:matching:correlators}.
As emphasised above, this is a strong crosscheck of the automated diagrammatic calculation outlined in the following section.

\section{Dimensionally reduced SMEFT in three dimensions}
\label{appendix:smeft:3d}

The construction of a dimensionally-reduced effective theory requires the computation of various correlation functions in the high-temperature, unbroken phase. For the Standard Model and its simplest extensions (e.g.\ 2HDM~\cite{Losada:1998at,Andersen:1998br,Gorda:2018hvi} or $Z_2$-symmetric real-triplet extension~\cite{Niemi:2018asa}), it is possible to perform these computations mostly by hand with little computer assistance. However, for more complicated BSM theories -- or in order to reach higher orders of perturbation theory in simpler models -- automation is inevitable.
In the past, dimensional reduction in electroweak theories has been performed in
Landau gauge since the computational effort related with a general covariant gauge is usually immense.
Landau gauge greatly reduces the effort, as propagators (and corresponding master sum-integrals) simplify and the number of contributing diagrams is reduced immensely. 
However, ideally we would carry out all computations in a general gauge, to explicitly test gauge invariance.
In this work, using tools presented in Ref.~\cite{Schicho:2020xaf}, we have fully automated the dimensional reduction of the SM in a general covariant gauge, and furthermore we have included effects of
the extra SMEFT dimension-six operator~\eqref{eq:O6}.
Using the general covariant gauge throughout the computation, shows explicitly that the matching relations -- for the super-renormalisable part of the EFT (cf.\ discussion in Appendix.~\ref{appendix:matching}) -- are gauge invariant.
This justifies previous computations conducted in Landau gauge~\cite{Kajantie:1995dw}.

Before diving into the details of our computation -- a combination of manual work and automation -- we give a pedagogic introduction to the different steps that call for automation in a dimensional reduction computation. For this, we use the example of the 2-point correlator or self-energy of the ${\rm SU}(2)$
gauge boson, with an explicit diagram-by-diagram calculation.

\subsection{Dimensional reduction for beginners: electroweak Debye mass}
\label{appendix:dr_for_beginners}

This section gives an explicit tutorial for a typical dimensional reduction calculation, by computing the ${\rm SU}(2)$ gauge boson self-energy at one-loop order in the unbroken phase, or the
$A^a_\mu A^b_\nu$-correlator.
The results for this correlator can then be used to obtain the Debye mass of the temporal component, as well as the field normalisations for both temporal and spatial fields.
Note that a classic reference for these computations is Ref.~\cite{Kajantie:1995dw}.
Here, we follow Ref.~\cite{Brauner:2016fla} in which Appendix~C.1 gives a diagram-by-diagram result for this correlator in Landau gauge, and all required master sum-integrals are listed in Appendix~B and Feynman rules in Appendix~A.
Note that the computation of the
$A^a_\mu A^b_\nu$-correlator resembles the calculation the of corresponding correlator in high-$T$ QCD which yields the thermal gluon mass.
This computation is found in Sec.~5.4 of the textbook Ref.~\cite{Laine:2016hma} using Feynman gauge.  

The outline
of a typical dimensional reduction computation is the following: (see Sec.~3 of Ref.~\cite{Schicho:2020xaf})
\bi
\item[{\bf Step 1}:]
    Choose a model Lagrangian.
\item[{\bf Step 2}:]
    Derive corresponding Feynman rules in the unbroken phase.
    Note that the computation of correlators is done in the symmetric -- unbroken phase of the theory -- in the gauge eigenstate basis where all gauge bosons and fermions are massless.
\item[{\bf Step 3}:]
    Generate all diagrams for each required correlator, and use Feynman rules to compose expressions for each individual diagram and compute related symmetry factors.
\item[{\bf Step 4}:]
    Perform all algebra contractions of Lorentz, isospin, Dirac etc. indices and manipulate sum-integrals to express all diagrams in terms of a basis of master sum-integrals.
\item[{\bf Step 5}:]
    Evaluate the basis of required master sum-integrals.
\item[{\bf Step 6}:]
    Match finite parts of the correlators to solve for the desired quantities of the 3d EFT.
\ei

For the ${\rm SU}(2)$ gauge boson self-energy, there is no new contribution from the dimension-six coupling of the SMEFT at one-loop order.
Therefore the computation is the same as in the SM.

\subsubsection*{Step 1}

The relevant Lagrangian of {\bf step 1} is of the form
\begin{align}
\label{eq:L:SU2}
\mathcal{L} \simeq
      \frac{1}{4} G^a_{\mu\nu} G^a_{\mu\nu}
    + \frac{1}{2\xi_{2}} (\partial_\mu A^a_\mu)^2
    + \sum_{i}\bar{\psi}_i \bsl{D} \psi_i
    + \partial_\mu \bar{\eta}^a D_\mu \eta^a
    + (D_\mu \phi)^\dagger (D_\mu \phi)
    \;,
\end{align}
where definitions and the exact form of the gauge field strength tensor, fermion structures, covariant derivatives, definitions of group indices etc.\ are found in Sec.~2.1 of Ref.~\cite{Brauner:2016fla}.
Here, we upgrade the computation to a general covariant gauge instead of using Landau gauge.
This Lagrangian suffices for the computation of the ${\rm SU}(2)$ gauge boson self-energy outlined here, but of course for the full dimensional reduction, the photon field, corresponding ghost field and Higgs potential should be added.

\subsubsection*{Step 2}

For {\bf step 2}, the relevant Feynman rules are listed in Appendix~A of Ref.~\cite{Brauner:2016fla}, with the replacement of the ${\rm SU}(2)$ gauge field propagator in a general covariant gauge,
\begin{align}
  D^{ab}_{\mu\nu}(P) \equiv  \delta_{ab}    
    \frac{1}{P^2}\Big(\delta_{\mu\nu}
  - (1-\xi_2) \frac{P_\mu P_\nu}{P^2} \Big)
  \;.
\end{align}
In addition, we use the following short-hand notations:
\begin{align}
\label{eq:SU2:vtx}
    V[
        A^{a}_{\mu}(K)\,
        A^{b}_{\nu}(P)\,
        A^{c}_{\lambda}(Q) ] &\equiv
        -i g \Delta_{abc}^{\mu\nu\lambda}(K,P,Q)
    \;,\\
    V[
        A^{a}_{\mu}\,
        A^{b}_{\nu}\,
        A^{c}_{\kappa}\, 
        A^{d}_{\lambda} ] &\equiv
        \g^{2}\Delta_{abcd}^{\mu\nu\kappa\lambda}
    \;,
\end{align}
for gauge field cubic and quartic self-interaction vertices.
The Lorentz indices and adjoint isospin structures are those of Eq.(A.3) and Eq.(A.5) of Ref.~\cite{Brauner:2016fla}, respectively.

For the generation of these rules one can resort to any
preferred Feynman rule generating software (e.g.
{\verb FeynRules }~\cite{Alloul:2013bka}).
However, since the computation is conducted in the unbroken phase,
the isospin structure of the fields is comparatively simple,
allowing the economical possibility of formulating Feynman rules for the full fields, rather than for their individual components.
This significantly reduces the number of different diagrams, but results in a non-trivial structure of the isopsin indices for some vertices (cf.~\eqref{eq:SU2:vtx}).
Our in-house software generates the Feynman rules starting from the model Lagrangian.
After going over to momentum space and symmetrising over fields, the final rules are fully symmetric in group indices and momenta.

\subsubsection*{Step 3}

Moving forward to {\bf step 3}, the ${\rm SU}(2)$ gauge field 2-point correlator,
$A^a_\mu(K) A^b_\nu(-K)$,
is evaluated at one-loop level with corresponding diagrams:
\begin{align}
\label{eq:su2:2pt:1l}
    \TopSi(\Lgeii,1) = 
    &
    \TopoST(\Lgeii,\Aglii)
    + \TopoST(\Lgeii,\Acsi)
    + \TopoSB(\Lgeii,\Aglii,\Aglii)   
    + \TopoSB(\Lgeii,\Ahgi,\Ahgi)
    + \TopoSB(\Lgeii,\Acsi,\Acsi)
    + \TopoSB(\Lgeii,\Auq,\Auq)
    \;.
\end{align}
For illustration, we evaluate the pure gauge bubble and ring diagrams that explicitly depend on the gauge parameter $\xi_2$, as well as the fermionic diagrams.
The ghost and scalar diagrams do not depend on the gauge parameters $\xi_i$ so results for them can be read from the Landau gauge calculation in Appendix~C.1 of Ref.~\cite{Brauner:2016fla}.
In renormalised perturbation theory, the UV-divergence is cancelled by a tree-level counterterm interaction diagram. 

The external momentum $K$ can be set to be purely spatial, i.e.\ $K=(0,\vec{k})$, and soft $K\sim g T$.
The softness of external momenta allows a series expansion in $K$, and its $K_{0}$-piece gives a contribution to the thermal mass, while its quadratic piece contributes to field normalisations between 4d and 3d fields of EFT; see Ref.~\cite{Kajantie:1995dw,Vepsalainen:2007ji}.
Note that since the heat bath breaks Lorentz invariance only in the temporal direction -- leaving spatial Lorentz (or rotational) symmetry intact -- only the temporal part of the fields can obtain a thermal mass.
However, thermal screening does effect the spatial fields in ways that do not break rotational symmetry, affecting their couplings and field normalisations when the hard scale is integrated out.
The following only includes contributions from the hard scale, as at one-loop level zero-mode contributions trivially cancel against 3d contributions when correlators are matched.

The pure gauge bubble diagram reads (with 4-momentum $P$ in the loop)
\begin{align}
\TopoST(\Lgeii,\Aglii)=
    \underbrace{\vphantom{\Tint{P}}
        \tfrac{1}{2}}_{s}\times\quad
    \g^{2} \quad\times
    \underbrace{
        \Tint{P}'}_{\substack{\text{integration over}\\\text{hard scale}}} 
    \underbrace{\vphantom{\Tint{P}}
        \Delta_{abcd}^{\mu\nu\kappa\lambda}
        D^{cd}_{\kappa \lambda}(P)}_{\text{contract indices}}
\;,
\end{align}
where $s$ is the symmetry factor of the diagram.

From a computational point of view,
the required diagrams of each correlator are generated using e.g.\
{\verb qgraf }~\cite{Nogueira:1991ex}.
All momenta are shifted onto a canonical momentum basis already at the diagram level.
Only then Feynman rules are inserted to compose expressions for each individual diagram.

\subsubsection*{Step 4}

For {\bf step 4}, while contraction over adjoint isospin indices is trivial, contraction over Lorentz indices is laborious already at one-loop level.
For adjoint isospin
$\delta_{aa} = N_{2}^2 - 1 = 3$,
where $N_{2} = 2$ is the fundamental isospin dimension.
For Lorentz indices
$\delta_{\mu\mu} = D = d+1$ in $D$ spacetime dimensions.
The pure gauge bubble reads
\begin{align}
    2\g^{2}\delta_{ab} \; \Big(
        (2-D-\xi_2) \delta_{\mu\nu} \Tint{P}' \frac{1}{P^2} +  (\xi_2-1)\Tint{P}' \frac{P_\mu P_\nu}{P^4} 
    \Big)\;.
\end{align}
Next, we separate the result into
temporal $\mu=\nu=0$ and
spatial $\mu=r$, $\nu=s$ ($r,s = 1,\dots,3$) parts  -- note that cross-terms vanish due to odd sum-integrations -- and manipulate sum-integrals in order to express everything in terms of one-loop {\em master sum-integrals}
$I^{4b}_{\alpha,\beta} \equiv \Tintip{P} \frac{P_{0}^{\beta}}{[P^2]^\alpha}$. 
In the case at hand, we need the relation
$
I^{4b}_{\alpha+1,\beta+2} = \left(1-\frac{d}{2\alpha}\right) I^{4b}_{\alpha,\beta}
$.
For the calculation at hand, we find all relevant reduction relations in Appendix~B of Ref.~\cite{Brauner:2016fla}.
In order to utilise these, we additionally scalarise integrals by e.g.\
$\Tintip{P} \frac{p_r p_s}{P^4} = \frac{\delta_{rs}}{d}\;\Tintip{P} \frac{p^2}{P^4}$
which follows from 3d Lorentz invariance, and use trivial manipulations such as
$p^2 = P^2 - P_{0}^{2}$.
Note that $P_0$ is always non-zero, as we integrate over non-zero modes only.
Eventually, we arrive at an intermediate result for the pure gauge bubble
\begin{align}
    \g^{2} \delta_{ab} I^{4b}_1
    \begin{cases}
        -d (1+\xi_2) & (\mu=\nu=0) \\
        -(-1+2d+\xi_2) & (\mu = r, \nu=s)
    \end{cases}
    \;.
\end{align}
Then, turning to the fermionic diagram, we find
\begin{align}
    \TopoSB(\Lgeii,\Auq,\Auq) &=
    \underbrace{\vphantom{\Tint{P}}
        1}_{s} \times
    \underbrace{\Nf(1+\Nc)}_{\text{\# of left-handed fermions}} \times\quad
    (-\frac{i}{2} \g)^2
    \underbrace{(\tau_a)^{im} (\tau_b)^{nj} \delta_{ij} \delta_{mn}}_{\text{contract fund. isospin}}
    \nn\\ & \times
    \underbrace{\vphantom{\frac{1}{P^2}}
        (-1)}_{\text{fermion loop}}
    \Tint{ \{ P \} }
    \underbrace{
        \frac{1}{P^2} \frac{1}{(P+K)^2} \text{Tr} \Big[ \mathcal P_{L} i \bsl{P} \gamma_\nu \mathcal P_{L} i (\bsl{P} + \bsl{K}) \gamma_\mu  \Big]  }_{\text{evaluate Dirac trace and contract Lorentz indices}}
    \;,
\end{align}
which becomes
\begin{align}
    -\frac{1}{2} \g^{2} \Nf (1+\Nc) \delta_{ab} \Tint{ \{ P \} } \frac{1}{P^2} \frac{1}{(P+K)^2} 2 \Big( 4 P_\mu (P_\nu + K_\nu) - \delta_{\mu\nu} P \cdot (P+Q) \Big)
    \;.
\end{align}
Due to the momentum being soft, the propagator can be $K$-expanded up to quadratic order as 
\begin{align}
    \frac{1}{(P+K)^2} \simeq \frac{1}{P^2} - 2 \frac{(P \cdot K)}{P^4} + 4\frac{(P \cdot K)^2}{P^6} - \frac{K^2}{P^4}
    \;.
\end{align}
Therefore, the fermion ring diagram reads, up to quadratic order in soft momentum $K$
\begin{align}
    &-\frac{1}{2} \g^{2} \Nf (1+\Nc) \delta_{ab} \nn \\
    &\times \; \Tint{ \{ P \} }
    \begin{cases}
        - 2 \frac{1}{P^2}
        + 4 \frac{P^2_0}{P^4}
        + 2 \frac{K^2}{P^4}
        - 4 \frac{K^2 P^2_0}{P^6}
        - 4 \frac{(P \cdot K)^2}{P^6}
        + 16 \frac{ P^2_0 (P \cdot K)^2}{P^8}
        & (\mu=\nu=0) \\
        - 2 \delta_{rs} \frac{1}{P^2}
        + 4 \frac{P_r P_s}{P^4}
        + 2 \delta_{rs} \frac{K^2}{P^4}
        - 4 \delta_{rs} \frac{(P \cdot K)^2}{P^6}
        - 8 P_r K_s \frac{(P \cdot K)}{P^6}
    \\ \hphantom{-2 \delta_{rs} \frac{1}{P^2}}
        - 4 K^2 \frac{P_r P_s}{P^6}
        + 16 \frac{ P_r P_s (P \cdot K)^2}{P^8}
        & (\mu = r, \nu=s)
    \end{cases}
    \;.
\end{align}
Again, all these sum-integrals can be expressed in terms of
$I^{4b}_1$ and $I^{4b}_2$ after scalarisation of integrals such as
$\Tinti{ \{ P \} } \frac{P_{r}P_{s} (P\cdot K)^2}{P^8} =
\frac{K^2 \delta_{rs} +2 K_r K_s}{(d+2)d}
\Tinti{ \{ P \} } \frac{p^4}{P^8}$ and by using the recursion relations from Appendix~B of Ref.~\cite{Brauner:2016fla}.
Eventually, this simplifies to
\begin{align}
    & \g^{2} \Nf (1+\Nc) \delta_{ab} \nn \\
    &\times \;
    \begin{cases}
        (2^{2-d}-1) (d-1) I^{4b}_1
        + K^2 \frac{1}{6}(2^{4-d}-1) (d-1) I^{4b}_2
        & (\mu=\nu=0) \\
        (-\frac{1}{3})(2^{4-d}-1) I^{4b}_2 (K^{2}\delta_{rs} - K_{r}K_{s})
        & (\mu = r, \nu=s)
    \end{cases}
    \;.
\end{align}
Here, the spatial part is individually transverse and without a $K_0$ contribution, as spatial fields do not generate a thermal mass.
Next, the pure gauge ring diagram reads (we denote here $Q=P+K$)
\begin{align}
\TopoSB(\Lgeii,\Aglii,\Aglii) =
    \underbrace{\vphantom{\Delta_{adc}^{\mu\sigma\lambda}}
        1}_{s} \times (-i\g)^2 \; \Tint{P}'
    \underbrace{
        \Delta_{adc}^{\mu\sigma\lambda}(K,P,-Q)   
        \Delta_{bef}^{\nu\alpha\rho}(-K,-P,Q)
        D^{cf}_{\lambda\rho}(Q)
        D^{de}_{\sigma\alpha}(P)}_{\text{contract indices}}
    \;.
\end{align}
Performing contractions and expanding in soft momentum yields multiple terms, but nothing qualitatively new compared to previous diagrams that we have already mastered.
An intermediate result in terms of familiar master integrals reads
\begin{align}
    & \g^{2} \delta_{ab} \nn \\
    &\times \;
    \begin{cases}
        d (4-2d+\xi_2) I^{4b}_1 + K^2 \frac{1}{6} \Big(16 -3d +2 d^2 -6(d-2)\xi_2 \Big) I^{4b}_2
        & (\mu=\nu=0) \\
        \delta_{rs} (2d+\xi_2) I^{4b}_1
        + \frac{1}{6} \Big(
            K^{2}\delta_{rs} (31-2d)
            - 2 K_{r}K_{s} (17-d)
        & \\ \hphantom{ \delta_{rs} (2d+\xi_2) I^{4b}_1 + \frac{1}{6} \Big(}
            + 6\xi_2 (K^{2}\delta_{rs} - K_r K_s)  \Big) I^{4b}_2 
        & (\mu = r, \nu=s)
    \end{cases}
    \;.
\end{align}
In its spatial part,
the $\xi_{2}$-independent piece is not individually transverse,
but together with the corresponding ghost diagram, the sum of diagrams is rendered transverse. Also the momentum independent part of the full spatial correlator vanishes. On the other hand,
the $\xi_{2}$-dependent part is transverse, because ghosts do not contribute to that. Both ghost and scalar diagrams can be evaluated in a manner similar to previous diagrams, and since they do not posses a $\xi_{2}$-dependence, we can read their Landau gauge results from Appendix~C.1 of Ref.~\cite{Brauner:2016fla}.
Finally, the counterterm diagram contributes by
\begin{align}
& - \delta Z_A\, \delta_{ab}
    \begin{cases}
        K^2 & (\mu=\nu=0) \\
        K^2 \delta_{rs} - K_r K_s & (\mu = r, \nu=s)
    \end{cases}
    \;,
\end{align}
where we {\em define}
$\delta Z_A = \frac{\g^{2}}{(4\pi)^2}\frac{1}{\epsilon} (\frac{27}{6} +\frac{4}{3}\Nf - \xi_2)$
so that our correlators become UV-finite in dimensional regularisation.
We emphasise, that counterterms within a finite-$T$ computation, are recycled from zero temperature, since the UV structure of the hard mode contributions remains unaltered by high temperature. This means that the dimensional reduction computation can obtain all required counterterms and eventually $\beta$-functions along with the construction of the high-$T$ EFT.

All algebraic manipulations encompassed in this step are handled using our favorite kitchen knife
{\verb FORM }~\cite{Ruijl:2017dtg}.
This includes contractions of Lorentz,
group (colour, isospin), and
Dirac indices.
Additionally we
manipulate sum-integrals to express all diagrams in terms of a basis of master sum-integrals,
shift onto different integral sectors, and
employ integration-by-parts methods~\cite{Nishimura:2012ee}
using a standard Laporta reduction~\cite{Laporta:2001dd}.
Their algorithmic implementation is documented in Sec.~3.4 of~\cite{Schicho:2020xaf}.

\subsubsection*{Step 5}

Proceeding, {\bf step 5} evaluates the basis of master sum-integrals in dimensional regularisation.
Due to the hierarchy of scales, these sum-integrals can be expanded in powers of $m/T$, and hence evaluated as massless sum-integrals, leading to significant simplifications.

At the order that we work, NLO in powers of $g^2$, only one-loop sum-integrals are needed to match the fields and couplings.
Therein, computations are straightforward and give results in terms of Gamma- and Zeta-functions,
see for example Appendix~B of Ref.~\cite{Brauner:2016fla}.

Two-loop sum-integrals are required for NLO matching of the masses of scalars and temporal gauge fields. 
At two-loop order there is only one master topology, the sunset diagram.
A direct evaluations of the sunset topology sum-integrals can be found in Refs.~\cite{Parwani:1991gq,Arnold:1994eb,Laine:2017hdk,Ekstedt:2020qyp}.
We also recommend an illuminating Ref.~\cite{Osterman:2019xx}.
However, our streamlined use of IBP relations~\cite{Nishimura:2012ee,Ghisoiu:2012yk,Schicho:2020xaf}, reduces all massless two-loop sum-integrals to products of one-loop sum-integrals.
No two-loop masters are needed.

Going beyond NLO dimensional reduction requires higher loop sum-integrals.
In this case the evaluation of master integrals becomes the most non-trivial part in the dimensional reduction pipeline.
Automation of such computations is still in its early stages.
For relevant computations of master sum-integrals at three-loop and higher loop orders, see Refs.~\cite{Arnold:1994eb, Arnold:1994ps,Braaten:1995jr,Andersen:2000zn,Vuorinen:2003fs,Gynther:2007bw,Andersen:2008bz,Andersen:2009ct,Schroder:2012hm, Moeller:2012da,Ghisoiu:2012kn,Ghisoiu:2012yk}.
Further, see Ref.~\cite{Gynther:2005dj}, for a three-loop computation of the pressure in the electroweak theory utilising the master integrals of the above references.

\subsubsection*{Step 6}

At last, \textbf{step 6} retrieves the final form of the correlators.
After the $\epsilon$-expansion, $1/\epsilon$-poles have been cancelled by the counterterm contribution and
\begin{align}
\hat{\Pi}_{A_{0}^{a}A_{0}^{b}} &=
    \g^{2} T^2 \Big(\frac{5}{6} + \frac{1}{3}\Nf \Big)
    \;, \\  
\hat{\Pi}'_{A_{0}^{a}A_{0}^{b}} &=
    \frac{\g^{2}}{(4\pi)^2} \bigg(3
        + \frac{4}{3}\Nf(L_f-1)
        + \Big(\xi_2 - \frac{25}{6} \Big) L_b
        - 2 \xi_2 \bigg)
    \;, \\ 
\hat{\Pi}'_{A_{r}^{a}A_{s}^{b}} &=
    \frac{\g^{2}}{(4\pi)^2} \bigg( -\frac{2}{3}
    + \frac{4}{3} \Nf L_f
    + \Big(\xi_2 - \frac{25}{6} \Big) L_b \bigg)
    \;,
\end{align}
where schematically
$\Pi' \equiv \frac{{\rm d}}{{\rm d}K^2} \Pi$.
Here it is crucial that $\Pi'$-parts encode both the explicit $\xi_2$-dependence and RG-scale dependence.
It is indeed these dependencies that cancel contributions from other correlators related to the matching of corresponding EFT parameters, namely the
3d gauge coupling $\g_{3}^{2}$ and
portal coupling $h_1$ between $A^a_0$ and 3d Higgs.
We will show this cancellation explicitly in Eqs.~\eqref{eq:matching-first}--\eqref{eq:matching-last}.
Finally, let us inspect the actual matching formula for the electroweak Debye mass. The matching of correlators can be achieved by equating effective Lagrangians in both 4d and 3d theories
\begin{align}
    \frac{1}{2} \Big(
    \underbrace{
        \Pi^\rmii{soft}_{A_{0}^{a}A_{0}^{b}}}_{\text{0-modes}}
    + \underbrace{
        \Pi^\rmii{soft/hard}_{A_{0}^{a}A_{0}^{b}}}_{\text{mixed modes}}
    + \underbrace{
          \hat{\Pi}^\rmii{1-loop}_{A_{0}^{a}A_{0}^{b}}
        + \hat{\Pi}^\rmii{2-loop}_{A_{0}^{a}A_{0}^{b}}
    }_{\text{non-zero modes}} \Big) \Big(A_{0}^{a}A_{0}^{b}\Big)_{\rmii{4d}} \equiv 
    T \frac{1}{2} \Big( \mD^{2} + \Pi^\rmii{3d}_{A_{0}^{a}A_{0}^{b}} \Big)
    \Big(A_{0}^{a}A_{0}^{b}\Big)_{\rmii{3d}}
    \;.
\end{align}
In this schematic matching example, we have in fact included both correlators at 2-loop level. At this level, one has to resum the zero-mode of the temporal gauge field by its one-loop thermal mass, and introduce a corresponding
{\em resummation counterterm interaction} to the Lagrangian.
As a result, the soft/hard mixing contribution of the correlator in 4d vanishes identically, and the 2-loop soft contribution of zero-modes matches exactly the loop corrections to the 3d correlator, i.e.\ cancelling soft and 3d parts in the above equation.
By accounting for the relation between 3d and 4d fields
\begin{align}
    \Big(A_{0}^{a}A_{0}^{b}\Big)_{\rmii{3d}} =
    \frac{1}{T} \Big(1 + \hat{\Pi}'_{A_{0}^{a}A_{0}^{b}} \Big)
    \Big(A_{0}^{a}A_{0}^{b}\Big)_{\rmii{4d}}
    \;,
\end{align}
one can finally solve
\begin{align}
    \mD^{2} =
    \underbrace{
    \underbrace{\hat{\Pi}^\rmii{1-loop}_{A_{0}^{a}A_{0}^{b}}}_{\mathcal{O}(\g^{2}(\bmu))}
    + \underbrace{
        \hat{\Pi}^\rmii{2-loop}_{A_{0}^{a}A_{0}^{b}}
        - \hat{\Pi}^\rmii{1-loop}_{A_{0}^{a}A_{0}^{b}} \hat{\Pi}'_{A_{0}^{a}A_{0}^{b}}}_{\mathcal{O}(\g^{4}), \ln(\bmu), \xi_2 \; \text{cancels}} }_{\bmu \; \text{cancels}}
    \;.
\end{align}
Here we have highlighted how gauge dependence drops out between 2-loop hard contributions and the 1-loop field normalisation contribution, and how the final 3d parameter is (in a power counting sense) independent of the RG-scale $\bmu$, as the running of the LO piece is cancelled by logarithms of the NLO piece.
However, an important comment is necessary here: for the electroweak phase transition the NLO piece of $\mD^{2}$ is not needed, as it contributes to the Higgs effective potential (or free energy) at $\mc{O}(g^5)$ and hence is of higher-order than we work.%
\footnote{
    For the SM, the free energy (or pressure) has been calculated to $\mc{O}(g^5)$ in Ref.~\cite{Gynther:2005dj}.
}
The transition is driven by the Higgs field at the ultrasoft scale, and in the fact temporal $A^a_0$ field can be integrated out in the second step of dimensional reduction going from the soft to the ultrasoft scale.
Therefore, only the LO piece of $\mD^{2}$ is needed and reads
\begin{align}
\label{eq:mD:1l}
    \mD^{2} = \hat{\Pi}_{A_{0}^{a}A_{0}^{b}} &=
    \frac{\g^{2} T^2}{12}\Big(
        \frac{9 N_{2}}{2}
        + 1
        + \Nf(1 + \Nc) \Big)
    =
    \g^{2} T^2 \Big(\frac{5}{6} + \frac{1}{3}\Nf \Big)
    \;.
\end{align}
The explicit parameters of
fundamental isospin dimension $N_2 = 2$ and 
fundamental colour dimension $\Nc = 3$ are set to their integer values henceforth.
Note that the result~\eqref{eq:mD:1l} is not RG-invariant at $\mc{O}(g^4T^2)$ due to the absence of the NLO contribution, but again this only contributes to the Higgs effective potential at order $\mathcal{O}(g^5)$.
As a final remark, we note that for studies of the EWPT although 2-loop level matching for $\mD^{2}$ is unnecessary, 2-loop level matching is important for the Higgs mass parameter, as it contributes to the Higgs effective potential at $\mc{O}(g^4)$.
Indeed later this appendix encounters a similar matching relation for the Higgs thermal mass, with similar qualitative features of the cancellation of the gauge parameter and RG-scale. 

As a final illustration, we show all the diagrams contributing to the Higgs thermal mass at one-loop level originating from the $(\phi^{\dagger}\phi)$-self-energy
\begin{align}
\label{eq:sc1:2pt:1l}
    \TopSi(\Lsci,1) =
    &
    \TopoST(\Lsci,\Agli)
    + \TopoST(\Lsci,\Aglii)
    + \TopoST(\Lsci,\Acsi)
    + \TopoSB(\Lsci,\Agli,\Asci)
    + \TopoSB(\Lsci,\Aglii,\Asci)
    + \TopoSB(\Lsci,\Aqu,\Aquu)
    \;.
\end{align}
We highly recommend interested readers to embark on this computation themselves starting from the above diagrams.
At leading order in the high-$T$ expansion, this correlator reads
\begin{align}
    \hat{\Pi}_{\phi^\dagger \phi} &= -T^2 \Big(
        \frac{1}{16}(3\g^{2} + {\gp}^{2})
        + \frac{1}{4} \gY^{2}
        + \frac{1}{2} \lambda \Big) 
\end{align}
and will result in a familiar one-loop thermal correction to the Higgs mass parameter, cf.\ Eq.~\eqref{eq:higgs-thermal-mass}.
The two-loop result for this correlator -- with NLO mass correction -- will be presented in Eq.~\eqref{eq:higgs-2loop-correlator}.

Next, we proceed from this pedagogic computation of thermal masses to the full dimensional reduction and construction of the 3d EFT that includes thermal screening effects to all EFT parameters.

\subsection{Results for dimensional reduction of SMEFT} 
\label{appendix:matching}

Dimensional reduction is performed by matching the infrared parts of correlators of the full 4d theory with those of the effective 3d theory. 
For leading order dimensional reduction, the 3d mass is computed at 1-loop and couplings at tree-level, i.e.\ couplings are only scaled by temperature. In terms of our power counting 
($\gp\sim g$,
$\gY\sim\g$,
$\lambda\sim\g^{2}$ and
$c_6\sim\g^{4}$)
{\em viz.} Eq.~\eqref{eq:g_scalings},
this corresponds to
$\mathcal{O}(g^2)$ accuracy for parameters at the soft scale after the first step of dimensional reduction, and
$\mathcal{O}(g^3)$ at the ultrasoft scale after the second step of dimensional reduction. 
At next-to-leading order,
i.e.\ $\mathcal{O}(g^4)$ accuracy at the soft scale,
the 3d mass is computed to 2-loop accuracy and couplings to 1-loop. 

Effective theory parameters are expected to be independent of the choice of gauge fixing parameters~\cite{Farakos:1994kx}.
This is true at least up to $\mc{O}(g^4)$.
In order to prove this, we perform dimensional reduction in a general covariant gauge and show below explicitly how the gauge parameters cancel.
Additionally, 3d parameters are independent of the 4d RG scale in terms of our power counting. Any leftover scale dependence is of higher order than $\mathcal{O}(g^4)$, and if numerically important, signals a bad convergence of perturbation theory. The cancellation of gauge parameters and RG-scale provide a very powerful cross-check for the validity of the calculation.

The 3d effective theory which we match to is (note the Euclidean metric as this is a purely spatial theory),
\begin{align}
\label{eq:3d:soft:action}
S^{\rmii{soft}}_{\rmii{3d}} = \int {\rm d}^3 x \biggl[
      \frac{1}{4}G_{rs}^{a}G_{rs}^{a}
    + \frac{1}{4}F_{rs}^{ }F_{rs}^{ }
    &+\frac{1}{2}(D_{r}^{ }A_{0}^{a})^2
    + \frac{1}{2}(\partial_{r}^{ }B_{0}^{ })^2
    + \frac{1}{2}(D_{r}^{ }C_{0}^{\alpha})^2
    \nn \\ &
    + (D_{r}\phi^{ })^\dagger (D_{r}\phi^{ })
    + V^{\rmii{soft}}_{\rmii{3d}}\biggr]
    \;, 
\end{align}
where
$G^{a}_{rs} =
\partial^{ }_{r} A^a_s -
\partial^{ }_{s} A^a_r + \g^{ }_{3} \epsilon^{abc}A_{r}^b A_{s}^c$,
$F_{rs} = \partial_{r}B_{s} - \partial_{s}B_{r}$ and
$D^{ }_{r}\phi^{ } = (\partial^{ }_{r} - i\g^{ }_{3} \tau^a A_{r}^a/2 - i\gp_{3} B^{ }_{r}/2)\phi^{ }$.
The $\tau^a$ are the Pauli matrices.
For simplicity we do not use a different notation for 3d and 4d fields,
following the convention of Ref.~\cite{Kajantie:1995dw}.
The scalar potential in the soft scale 3d theory reads%
\footnote{
    Several temporal gluon interaction terms are negligable. Since the Higgs field does not have colour, these neglected effects are formally of higher order in the final effective scalar theory at the ultrasoft scale.
}
\begin{align}
\label{eq:3d:soft:V}
V^{\rmii{soft}}_{\rmii{3d}} &=
    \mu_{h,3}^{2} \phi^\dagger\phi
    + \lambda^{ }_{3} (\phi^\dagger\phi)^2
    \nn \\ &
    + c^{ }_{6,3} (\phi^\dagger\phi)^3
    + c^{ }_{8,3} (\phi^\dagger\phi)^4
    + c^{ }_{10,3} (\phi^\dagger\phi)^5
    \nn \\ &
    + \frac{1}{2}\mD^2\,A^a_0A^a_0 
    + \frac{1}{2}\mD'^2\,B_0^2
    + \frac{1}{2}\mD''^2\,C^{\alpha}_0 C^{\alpha}_0 
    \nn \\ &
    + \frac{1}{4}\kappa_1 (A^a_0A^a_0)^2
    + \frac{1}{4}\kappa_2 B_0^4
    + \frac{1}{4}\kappa_3 A^a_0A^a_0B_0^2
    \nn \\ &
    + h^{ }_{1} \phi^\dagger\phi A^a_0A^a_0
    + h^{ }_{2} \phi^\dagger\phi B_0^2
    \nn \\ &
    + h^{ }_{3} B_0\phi^\dagger\vec{A}_0\cdot{\bm \tau}\phi
    + h^{ }_{4} \phi^\dagger\phi C^\alpha_0 C^\alpha_0
    \;,
\end{align}
and together with 3d gauge couplings $\g^{ }_{3}$ and $\gp_{3}$, our task in dimensional reduction is to find these parameters in terms of 4d parameters and the temperature.
Here
$\g_{3}^{2}$,
${\gp_{3}}^{2}$, and
$\lambda^{ }_{3}$
have dimensions of [GeV] and all the fields
have dimensions of [GeV]$^{1/2}$.
We regularise the theory in the \MSbar-scheme. 
In the following we utilise the class of covariant gauges,
\begin{equation}
\label{eq:gauge_fixing}
S^{\rmii{soft}}_\xi = \int {\rm d}^3 x \biggl[
    \frac{1}{2\xi_{3,2}}(\partial_{r}A_{r}^a)^2
    + \partial_{r}\bar{\eta}^a \partial_{r}\eta^a 
    + \g_{3}\epsilon^{abc}\partial_{r}\bar{\eta}^a A_{r}^b \eta^c
    + \frac{1}{2\xi_{3,1}}(\partial_{r}B_{r})^2
    + \partial_{r}\bar{\eta}\partial_{r}\eta \biggr]
    \;,
\end{equation}
and, in Secs.~\ref{appendix:Veff3d} and \ref{appendix:3d_thermodynamics} we adopt the Landau gauge,
$\xi_{3,1},\xi_{3,2} \to 0_+$
(taking this limit explicitly avoids certain IR divergences~\cite{Laine:1994bf,Laine:1994zq}).

At the ultrasoft scale the temporal scalars $A_0, B_0$ and $C_0$ are heavy and are integrated out.
The remaining ultrasoft EFT parameters are differentiated from the soft EFT parameters with a bar, i.e.\
$\bar{g}_{3}^{ }$,
$\bar{g}'_{3}$,
$\bar{\mu}_{h,3}^{2}$,
$\bar{\lambda}_{3}^{ }$,
$\bar{c}_{6,3}^{ }$,
$\bar{c}_{8,3}^{ }$, and
$\bar{c}_{10,3}^{ }$.
Although the ultrasoft theory differs only by these aforementioned changes, for clarity we show the ultrasoft action in full:
\begin{equation}
\label{eq:3d:ultrasoft:action}
S^{\rmii{ultrasoft}}_{\rmii{3d}} = \int {\rm d}^3 x \biggl[
      \frac{1}{4}G_{rs}^{a}G_{rs}^{a}
    + \frac{1}{4}F_{rs}^{ }F_{rs}^{ }
    + (D_{r}\phi^{ })^\dagger (D_{r}\phi^{ })
    + V^{\rmii{ultrasoft}}_{\rmii{3d}}\biggr]
    \;. 
\end{equation}
The implicit gauge couplings are
$\bar{g}^{ }_{3}$ for the ${\rm SU}(2)$ and
$\bar{g}'_{3}$ for the ${\rm U}(1)$ sectors.
The ultrasoft scalar potential reads
\begin{align}
\label{eq:3d:ultrasoft:V}
V^{\rmii{ultrasoft}}_{\rmii{3d}} &=
      \bar{\mu}_{h,3}^{2} \phi^\dagger\phi
    + \bar{\lambda}^{ }_{3} (\phi^\dagger\phi)^2
    \nn \\ &
    + \bar{c}^{ }_{6,3} (\phi^\dagger\phi)^3
    + \bar{c}^{ }_{8,3} (\phi^\dagger\phi)^4
    + \bar{c}^{ }_{10,3} (\phi^\dagger\phi)^5
    \;,
\end{align}
and the gauge fixing corresponds to the soft scale in Eq.~\eqref{eq:gauge_fixing}.

The dimensionally-reduced operator basis which we use is complete
at dimension [GeV$^2$], but is incomplete
at dimension [GeV$^3$], at which it contains only a single interaction operator, $c_{6,3}(\hsq)^3$.
This is because this operator appears also in the 4d SMEFT, and hence $c_{6,3}$ receives a tree-level contribution of order $\mc{O}(g^4)$, whereas all other coefficients of dimension [GeV$^3$] operators start at one-loop level, at order $\mc{O}(g^6)$.
Some examples of missing operators
at dimension [GeV$^3$] in the dimensionally-reduced SMEFT
are the following gauge invariant Higgs kinetic terms
(we 
use a Warsaw-like basis, cf.\ Ref.~\cite{Buchmuller:1985jz,Grzadkowski:2010es,deBlas:2014mba})
\begin{equation}
\label{eq:O:higher}
\mc{O}_{\phi\Box} \sim
    \bigl(\phi^\dagger \phi\bigr)\Box
    \bigl(\phi^\dagger \phi\bigr)
    \;,\quad
\mc{O}_{\phi D} \sim
    \bigl(\phi^\dagger D_{r}\phi\bigr)^*
    \bigl(\phi^\dagger D_{r}\phi\bigr)
    \;,
\end{equation}
with
$\Box = \partial_{r}\partial_{r}$.
There are also multiple dimension [GeV$^3$] operators of the Higgs coupling to temporal scalars at the soft scale, but these are also of order $\mc{O}(g^6)$.

\subsubsection{Results for correlators}
\label{appendix:matching:correlators}

At NLO in dimensional reduction, higher dimensional operators with $c_6, c_8$ and $c_{10}$ coefficients only affect pure Higgs correlators, while all other correlators are unaffected and therefore are already found in the literature. However, in order to demonstrate the cancellation of gauge parameters in parameter matching, we list all required correlators below. 

As explained in the previous section, we use an automated, computer algorithm computation of all correlators in the unbroken phase: diagram generation and computer algebra can be used to automate
Lorentz contractions,
Dirac traces and
IBP-methods to reduce sum-integral structures to a basis or set of master integrals. 
This calculation is implemented via
{\verb qgraf } for diagram generation and
{\verb FORM } for algebraic manipulations and IBP reduction.
Note that at NLO in dimensional reduction the scalar propagator can be treated as massless within two-loop order terms.
This demonstrates the power of IBP reduction, since all two-loop integrals reduce to products of one-loop master sum-integrals.

We denote
four-point correlators at zero external momentum by $\Gamma$, and
two-point correlators by
$\Pi$ for the part evaluated at zero external momentum and
$\Pi'$ for the coefficient of the term which is quadratic in external momenta.%
\footnote{
    Correlators can be expanded in external momenta $Q=(0,\vec q)$ when $Q \sim gT$ is soft.
}
For the spatial gauge fields this is the transverse part of the correlator.
For all correlators we only list contributions from the hard modes, as the soft contribution cancels with corresponding 3d contribution in the matching, as Sec.~\ref{sec:matching-relations} demonstrates. Note that all soft contributions at one-loop are finite, and do not require renormalisation. At two-loop, the scalar correlator requires resummation and soft contributions require renormalisation.   

All correlators are computed in renormalised perturbation theory, and are therefore finite, which we denote by hats in $\hat{\Pi}$ and $\hat{\Gamma}$ (note that we omit writing corresponding isospin, colour and Lorentz index structures explicitly).
An illustrative example of renormalisation is
the mixed scalar-gauge
$(\phi^{\dagger}\phi A A)$-correlator
\begin{align}
    \hat{\Gamma}_{\phi^\dagger \phi A^a_\mu A^b_\nu} &=
    -\bigg(
    \Gamma^{ }_{\phi^\dagger \phi A^a_\mu A^b_\nu}
    - \g\delta\g
    - \frac{1}{2}\g^{2} \Big(
        \delta Z^{ }_\phi + \delta Z^{ }_A
    \Big) \bigg)
    \;,
\end{align}
where the unhatted $\Gamma$ sums all contributing Feynman diagrams, excluding counterterms.%
\footnote{
    Note that correlator itself is negative of sum of Feynman diagrams.
}
Similar relations hold for other correlators, and emerge naturally in renormalised perturbation theory where counterterms are treated as interactions.

We start from one-loop contributions quadratic in the soft external momentum, and these read ($L_{b/f}$ are defined in Eqs.~\eqref{eq:Lb}--\eqref{eq:Lf} and
$\Nf=3$ is the number of fermion families)
\begin{align}
\hat{\Pi}'_{A_{0}^{a}A_{0}^{b}} &=
    \frac{\g^{2}}{(4\pi)^2} \bigg(3 + \frac{4}{3}\Nf(L_f-1) + \Big(\xi_2 - \frac{25}{6} \Big) L_b - 2 \xi_2 \bigg)
    \;, \\ 
\hat{\Pi}'_{A_{r}^{a}A_{s}^{b}} &=
    \frac{\g^{2}}{(4\pi)^2} \bigg( -\frac{2}{3} + \frac{4}{3} \Nf L_f + \Big(\xi_2 - \frac{25}{6} \Big) L_b \bigg)
    \;, \\ 
\hat{\Pi}'_{B_{0}B_{0}} &= 
    \frac{{\gp}^{2}}{(4\pi)^2} \bigg(\frac{1}{3} + \frac{20}{9}\Nf(L_f-1) + \frac{1}{6}L_b \bigg)
    \;, \\ 
\hat{\Pi}'_{B_{r}B_{s}} &=
    \frac{{\gp}^{2}}{(4\pi)^2} \bigg(\frac{20}{9}\Nf L_f + \frac{1}{6}L_b \bigg)
    \,, \\
\label{eq:phi:2pt:1l:1d}
\hat{\Pi}'_{\phi^\dagger \phi} &=
    \frac{1}{(4\pi)^2} \bigg( -\frac{L_b}{4} \Big( 3(3-\xi_2)\g^{2} + (3-\xi_1){\gp}^{2} \Big) + 3 L_f \gY^{2}  \bigg) 
    \;.
\end{align}
Again, due to the heat bath breaking the 4d Lorentz invariance, contributions to spatial and temporal gauge fields differ. Crucially, these quadratic momentum contributions to the SU(2) gauge field and to the Higgs explicitly depend on the gauge, and are required to render the final EFT parameters gauge independent in Sec.~\ref{sec:matching-relations}.
These contributions are evaluated at one-loop at
$\mathcal{O}(g^2)$ order -- which reaches the desired $\mathcal{O}(g^4)$ accuracy for the EFT parameters -- and at this order there is no $c_{6,8,10}$-dependence from higher dimensional operators. 

Turning to contributions at zero external momentum, the pure Higgs correlators read
\begin{align}
\hat{\Gamma}_{(\phi^\dagger \phi)^2} &=
    - 2 c_6 T^2
    - \frac{1}{(4\pi)^2} \bigg( -\frac{1}{4} \Big( 3\g^{4} + {\gp}^{4} + 2 \g^{2} {\gp}^{2} \Big) - 6 \gY^{4} L_f
    \nn \\ &
    + \Big[ \frac{3}{8} \Big( 3\g^{4} + {\gp}^{4} + 2 \g^{2} {\gp}^{2} \Big) + 24 \lambda^2 + 24 c_6 \mh^2 - \lambda \Big( 3 \g^{2} \xi_2 +  {\gp}^{2} \xi_1\Big) \Big] L_b \bigg)
    \nn \\ &
    + \Big( -\frac{5}{3} c_{8} T^4 -\frac{25}{18} c_{10} T^6 \Big)_{\rmii{flowers}}
    \;, \\ 
\label{eq:phi:3}
\hat{\Gamma}_{(\phi^\dagger \phi)^3} &=
    -\frac{\zeta(3)}{128 \pi^4 T^2} \bigg( -\frac{3}{8} \Big( 3\g^{6} + 3\g^{4} {\gp}^{2} + 3 \g^{2} {\gp}^{4} + {\gp}^{6} \Big)  - 240 \lambda^3
    \nn \\ &
    + 84 \gY^{6} + 6 \lambda^2 \Big(3 \g^{2} \xi_2 + {\gp}^{2} \xi_1 \Big)  \bigg) + \frac{1}{(4\pi)^2} L_b \bigg[ c_6 \bigg(324 \lambda - \frac{9}{2} \Big( 3 \g^{2} \xi_2 + {\gp}^{2} \xi_1 \Big) \bigg) + 120 c_8 \mh^2 \bigg]
    \nn \\ &
    - 10 c_8 T^2
    + \Big( -\frac{25}{2} c_{10} T^4 \Big)_{\rmii{flowers}}
    \;, \\ 
\label{eq:phi:4}
\hat{\Gamma}_{(\phi^\dagger \phi)^4} &=
    \frac{24}{(4\pi)^2} L_b \bigg(63 c^2_6 + 30 c_{10}\mh^2 + c_8 \Big(96 \lambda - 3\g^{2} \xi_2 - {\gp}^{2} \xi_1 \Big) \bigg)
    - 60 c_{10} T^2
    + \hat{\Gamma}^{\zeta}_{(\phi^\dagger \phi)^4}
    \;, \\
\label{eq:phi:5}
\hat{\Gamma}_{(\phi^\dagger \phi)^5} &=
    \frac{30}{(4\pi)^2} L_b \bigg(912 c_6 c_8 +  5 c_{10} \Big(120 \lambda - 3\g^{2} \xi_2 - {\gp}^{2} \xi_1 \Big) \bigg)
    + \hat{\Gamma}^{\zeta}_{(\phi^\dagger \phi)^5}
\;.
\end{align}
Here all contributions arise at one-loop order, {\em except} the $c_{8,10}$ pieces denoted with {\em flowers}.
These appear at two- and three-loop orders: these pieces are the leading contributions from $c_{8,10}$ and are included as they are associated with the bubble integral in Eq.~\eqref{eq:hard-loop}.
Note that for dimension-8 and -10 correlators, $\zeta$-terms of higher order in the high-$T$ expansion of one-loop integrals (Eqs.~\eqref{eq:hard-expansion-bosonic} and \eqref{eq:hard-expansion-fermionic}) are collected in $\hat{\Gamma}^{\zeta}_{(\phi^\dagger \phi)^{4}}$ and
$\hat{\Gamma}^{\zeta}_{(\phi^\dagger \phi)^{5}}$.
We do not write them down here explicitly due to the lengthiness of the expressions, but we do include their contributions to the final matching relations below in Eqs.~\eqref{eq:zetas-8} and \eqref{eq:zetas-10}.

Other correlators are independent of $c_{6,8,10}$-couplings, and read
\begin{align}
\hat{\Pi}_{A_{0}^{a}A_{0}^{b}} &=
    \g^{2} T^2 \Big(\frac{5}{6} + \frac{1}{3}\Nf \Big)
    \;, \\  
\hat{\Pi}_{B_{0}B_{0}} &=
    {\gp}^{2} T^2 \Big(\frac{1}{6} + \frac{5}{9}\Nf \Big)
    \;, \\ 
\hat{\Pi}_{C_{0}^\alpha C_{0}^\beta} &=
    \gs^{2} T^2 \Big(1 + \frac{1}{3}\Nf \Big)
    \;, \\[3mm]  
\hat{\Gamma}_{\phi^\dagger \phi A^a_r A^b_s} &=
    - \frac{\g^{2}}{(4\pi)^2} \bigg( \frac{3}{8} \Big(-\g^{2} + {\gp}^{2} \Big) L_b - \frac{3}{2} \gY^{2} L_f - \frac{1}{8} \Big( 7 \g^{2} \xi_2 +  {\gp}^{2} \xi_1  \Big) L_b \bigg)
    \;, \\ 
\hat{\Gamma}_{\phi^\dagger \phi B_r B_s} &=
    - \frac{{\gp}^{2}}{(4\pi)^2} \bigg( \frac{1}{8} \Big(3 \g^{2} + {\gp}^{2} \Big) L_b - \frac{3}{2} \gY^{2} L_f - \frac{1}{8} \Big( 3 \g^{2} \xi_2 +  {\gp}^{2} \xi_1  \Big) L_b \bigg)
    \;, \\ 
\hat{\Gamma}_{\phi^\dagger \phi A_{0}^{a}A_{0}^{b}} &=
    - \frac{\g^{2}}{(4\pi)^2} \bigg( -\frac{1}{4} \Big(23 \g^{2} + {\gp}^{2} \Big) + \frac{3}{8} \Big(-3\g^{2} + {\gp}^{2} \Big) L_b
    \nn \\ &
    \hphantom{=-\frac{\g^{2}}{(4\pi)^2} \bigg(}
    - \frac{3}{2} (L_f-2) \gY^{2} - 6 \lambda - \frac{1}{8} \xi_1 {\gp}^{2} L_b + \Big(1- \frac{7}{8}L_b \Big) \xi_2 \g^{2}  \bigg)
    \;, \\
\hat{\Gamma}_{\phi^\dagger \phi B_0 B_0} &=
    - \frac{{\gp}^{2}}{(4\pi)^2} \bigg( -\frac{1}{4} \Big(3 \g^{2} + {\gp}^{2} \Big) + \frac{3}{8} \Big(3\g^{2} + {\gp}^{2} \Big) L_b
    \nn \\ &
    \hphantom{=-\frac{{\gp}^{2}}{(4\pi)^2} \bigg(}
    + \frac{1}{6} (34 -9 L_f) \gY^{2} - 6 \lambda - \frac{1}{8} L_b \Big(\xi_1 {\gp}^{2} + 3 \xi_2 \g^{2} \Big)\bigg)
    \;, \\
\hat{\Gamma}_{\phi^\dagger \phi A^a_0 B_0} &=
    - \frac{\g{\gp}}{(4\pi)^2} \bigg( -\frac{1}{4} \Big(\g^{2} + {\gp}^{2} \Big) + \frac{3}{8} \Big(\g^{2} + {\gp}^{2} \Big) L_b
    \nn \\ &
    \hphantom{=-\frac{\g{\gp}}{(4\pi)^2} \bigg(}
    - \frac{1}{2} (2 + 3 L_f) \gY^{2} - 2 \lambda - \frac{1}{8} \xi_1 {\gp}^{2} L_b + \frac{1}{2}\Big(1-\frac{5}{4}L_b \Big) \xi_2 \g^{2}\bigg)
    \;, \\
\hat{\Gamma}_{\phi^\dagger \phi C^\alpha_0 C^\beta_0} &=
    - \frac{1}{(4\pi)^2} \Big(4 \gs^2 \gY^{2} \Big)
    \;, \\ 
\hat{\Gamma}_{A^a_0 A^b_0 A^c_0 A^d_0} &=
    - \frac{\g^{4}}{(4\pi)^2} \frac{2}{3} \Big(-17 + 4 \Nf \Big)
    \;, \\ 
\hat{\Gamma}_{B_0 B_0 B_0 B_0} &=
    - \frac{{\gp}^{4}}{(4\pi)^2} \Big(-2 + \frac{380}{81} \Nf \Big)
    \;, \\ 
\hat{\Gamma}_{A^a_0 A^b_0 B_0 B_0} &=
    - \frac{\g^{2} {\gp}^{2}}{(4\pi)^2} \Big(-2 + \frac{8}{3} \Nf \Big)
    \;. 
\end{align}

\begin{figure}[t]
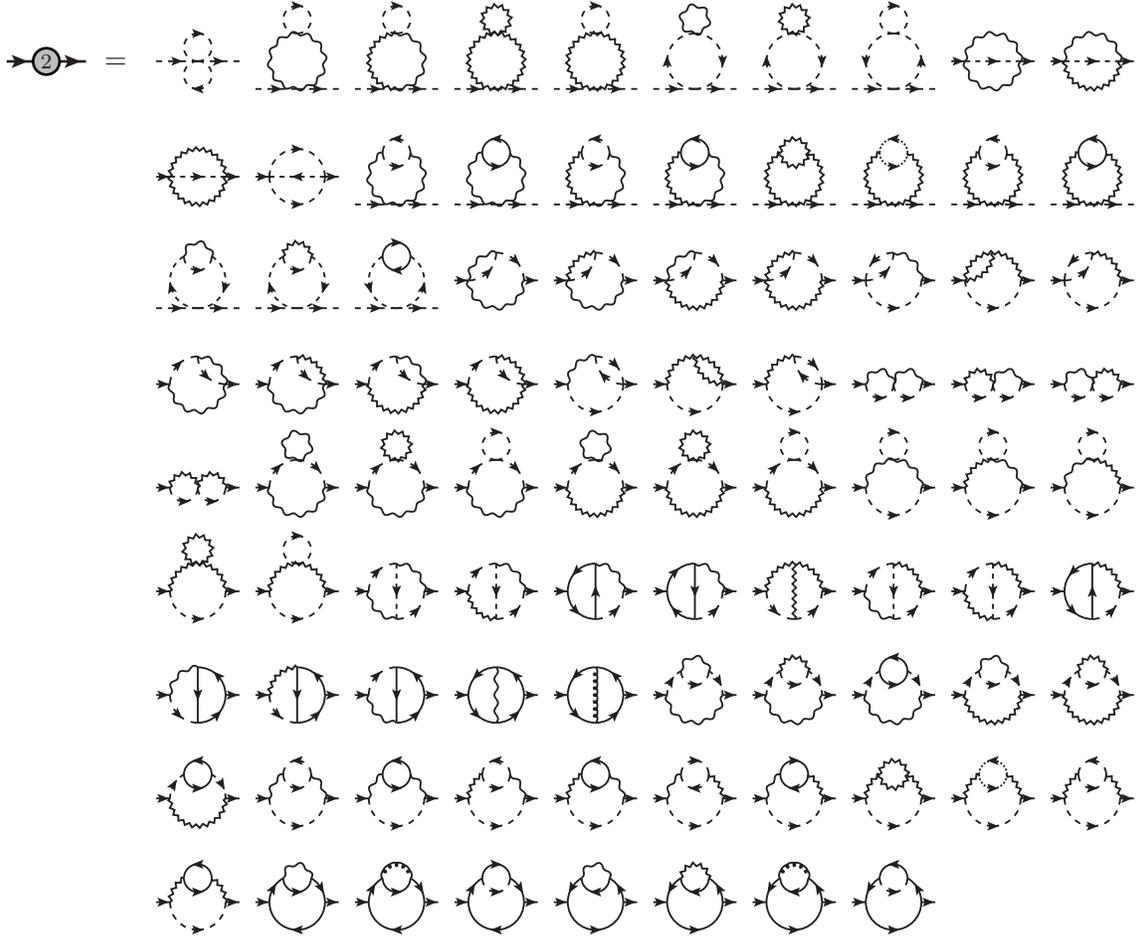

    \centering
\begin{eqnarray*}
 \TopSi(\Lsci,2) =
 &&
 \ToptSTTT(\Lsci,\Acsi,\Acsi)
 \ToptSTT(\Lsci,\Agli,\Agli,\Acsi)
 \ToptSTT(\Lsci,\Agli,\Aglii,\Acsi)
 \ToptSTT(\Lsci,\Aglii,\Aglii,\Aglii)
 \ToptSTT(\Lsci,\Aglii,\Aglii,\Acsi)
 \ToptSTT(\Lsci,\Acsi,\Acsi,\Agli)
 \ToptSTT(\Lsci,\Acsi,\Acsi,\Aglii)
 \ToptSTT(\Lsci,\Asci,\Asci,\Acsi)
 \ToptSS(\Lsci,\Agli,\Agli,\Lcsi)
 \ToptSS(\Lsci,\Agli,\Aglii,\Lcsi)
 \\[5mm]&&
 \ToptSS(\Lsci,\Aglii,\Aglii,\Lcsi)
 \ToptSS(\Lsci,\Acsi,\Asci,\Lsci)
 \ToptSTB(\Lsci,\Agli,\Agli,\Asci,\Asci)
 \ToptSTB(\Lsci,\Agli,\Agli,\Aqu,\Aqu)
 \ToptSTB(\Lsci,\Agli,\Aglii,\Asci,\Asci)
 \ToptSTB(\Lsci,\Agli,\Aglii,\Aqu,\Aqu)
 \ToptSTB(\Lsci,\Aglii,\Aglii,\Aglii,\Aglii)
 \ToptSTB(\Lsci,\Aglii,\Aglii,\Aghi,\Aghi)
 \ToptSTB(\Lsci,\Aglii,\Aglii,\Asci,\Asci)
 \ToptSTB(\Lsci,\Aglii,\Aglii,\Aqu,\Aqu)
 \\[5mm]&&
 \ToptSTB(\Lsci,\Acsi,\Acsi,\Agli,\Asci)
 \ToptSTB(\Lsci,\Acsi,\Acsi,\Aglii,\Asci)
 \ToptSTB(\Lsci,\Asci,\Asci,\Auq,\Auqu)
 \ToptSal(\Lsci,\Acsi,\Agli,\Agli,\Asci)
 \ToptSal(\Lsci,\Acsi,\Aglii,\Agli,\Asci)
 \ToptSal(\Lsci,\Acsi,\Agli,\Aglii,\Asci)
 \ToptSal(\Lsci,\Acsi,\Aglii,\Aglii,\Asci)
 \ToptSal(\Lsci,\Agli,\Asci,\Asci,\Asci)
 \ToptSal(\Lsci,\Aglii,\Aglii,\Asci,\Aglii)
 \ToptSal(\Lsci,\Aglii,\Asci,\Asci,\Asci)
  \\[5mm]&&
 \ToptSar(\Lsci,\Agli,\Acsi,\Agli,\Asci)
 \ToptSar(\Lsci,\Aglii,\Acsi,\Agli,\Asci)
 \ToptSar(\Lsci,\Agli,\Acsi,\Aglii,\Asci)
 \ToptSar(\Lsci,\Aglii,\Acsi,\Aglii,\Asci)
 \ToptSar(\Lsci,\Acsi,\Agli,\Asci,\Acsi)
 \ToptSar(\Lsci,\Aglii,\Aglii,\Asci,\Aglii)
 \ToptSar(\Lsci,\Acsi,\Aglii,\Asci,\Acsi)
 \ToptSE(\Lsci,\Agli,\Agli,\Asci,\Asci)
 \ToptSE(\Lsci,\Agli,\Aglii,\Asci,\Asci)
 \ToptSE(\Lsci,\Aglii,\Agli,\Asci,\Asci)
 \\[5mm]&&
 \ToptSE(\Lsci,\Aglii,\Aglii,\Asci,\Asci)
 \ToptSBT(\Lsci,\Acsi,\Acsi,\Agli,\Agli)
 \ToptSBT(\Lsci,\Acsi,\Acsi,\Agli,\Aglii)
 \ToptSBT(\Lsci,\Acsi,\Acsi,\Agli,\Acsi)
 \ToptSBT(\Lsci,\Acsi,\Acsi,\Aglii,\Agli)
 \ToptSBT(\Lsci,\Acsi,\Acsi,\Aglii,\Aglii)
 \ToptSBT(\Lsci,\Acsi,\Acsi,\Aglii,\Acsi)
 \ToptSBT(\Lsci,\Agli,\Agli,\Asci,\Acsi)
 \ToptSBT(\Lsci,\Agli,\Aglii,\Asci,\Acsi)
 \ToptSBT(\Lsci,\Aglii,\Agli,\Asci,\Acsi)
 \\[5mm]&&
 \ToptSBT(\Lsci,\Aglii,\Aglii,\Asci,\Aglii)
 \ToptSBT(\Lsci,\Aglii,\Aglii,\Asci,\Acsi)
 \ToptSM(\Lsci,\Agli,\Acsi,\Agli,\Asci,\Lsci)
 \ToptSM(\Lsci,\Agli,\Acsi,\Aglii,\Asci,\Lsci)
 \ToptSM(\Lsci,\Agli,\Aqu,\Aquu,\Asci,\Luq)
 \ToptSM(\Lsci,\Agli,\Auqu,\Auq,\Asci,\Lquu)
 \ToptSM(\Lsci,\Aglii,\Aglii,\Asci,\Asci,\Lglii)
 \ToptSM(\Lsci,\Aglii,\Acsi,\Agli,\Asci,\Lsci)
 \ToptSM(\Lsci,\Aglii,\Acsi,\Aglii,\Asci,\Lsci)
 \ToptSM(\Lsci,\Aglii,\Aqu,\Aquu,\Asci,\Luq)
 \\[5mm]&&
 \ToptSM(\Lsci,\Aqu,\Agli,\Asci,\Aquu,\Lqu)
 \ToptSM(\Lsci,\Aqu,\Aglii,\Asci,\Aquu,\Lqu)
 \ToptSM(\Lsci,\Aqu,\Acsi,\Agli,\Aquu,\Lquu)
 \ToptSM(\Lsci,\Aqu,\Aqu,\Aquu,\Aquu,\Lgli)
 \ToptSM(\Lsci,\Aqu,\Aqu,\Aquu,\Aquu,\Lgliii)
 \ToptSBB(\Lsci,\Acsi,\Acsi,\Agli,\Agli,\Asci)
 \ToptSBB(\Lsci,\Acsi,\Acsi,\Agli,\Aglii,\Asci)
 \ToptSBB(\Lsci,\Acsi,\Acsi,\Agli,\Aqu,\Aquu)
 \ToptSBB(\Lsci,\Acsi,\Acsi,\Aglii,\Agli,\Asci)
 \ToptSBB(\Lsci,\Acsi,\Acsi,\Aglii,\Aglii,\Asci)
 \\[5mm]&&
 \ToptSBB(\Lsci,\Acsi,\Acsi,\Aglii,\Aqu,\Aquu)
 \ToptSBB(\Lsci,\Agli,\Agli,\Asci,\Asci,\Asci)
 \ToptSBB(\Lsci,\Agli,\Agli,\Asci,\Aqu,\Aqu)
 \ToptSBB(\Lsci,\Agli,\Aglii,\Asci,\Asci,\Asci)
 \ToptSBB(\Lsci,\Agli,\Aglii,\Asci,\Aqu,\Aqu)
 \ToptSBB(\Lsci,\Aglii,\Agli,\Asci,\Acsi,\Acsi)
 \ToptSBB(\Lsci,\Aglii,\Agli,\Asci,\Auq,\Auq)
 \ToptSBB(\Lsci,\Aglii,\Aglii,\Asci,\Aglii,\Aglii)
 \ToptSBB(\Lsci,\Aglii,\Aglii,\Asci,\Aghi,\Aghi)
 \ToptSBB(\Lsci,\Aglii,\Aglii,\Asci,\Asci,\Asci)
 \\[5mm]&&
 \ToptSBB(\Lsci,\Aglii,\Aglii,\Asci,\Aqu,\Aqu)
 \ToptSBB(\Lsci,\Auqu,\Auqu,\Auq,\Agli,\Aquu)
 \ToptSBB(\Lsci,\Auqu,\Auqu,\Auq,\Agliii,\Aquu)
 \ToptSBB(\Lsci,\Auqu,\Auqu,\Auq,\Auq,\Asci)
 \ToptSBB(\Lsci,\Aqu,\Aqu,\Aquu,\Agli,\Auq)
 \ToptSBB(\Lsci,\Aqu,\Aqu,\Aquu,\Aglii,\Auq)
 \ToptSBB(\Lsci,\Aqu,\Aqu,\Aquu,\Agliii,\Auq)
 \ToptSBB(\Lsci,\Aqu,\Aqu,\Aquu,\Aquu,\Asci)
\end{eqnarray*}
\caption{
    Diagrams contributing to the Higgs two-point correlator at two-loop level.
    Scalar fields are represented by arrowed dashed lines,
    fermions as solid arrowed lines,
    ${\rm U}(1)$ gauge fields as wiggly lines,
    ${\rm SU}(2)$ gauge fields as zig-zag lines,
    ${\rm SU}(3)$ gauge fields as curly lines, and
    corresponding ghosts as arrowed dotted lines.
    }
\label{fig:sc1:2pt:2l}
\end{figure}

The computation of the Higgs two-point correlator at two-loop level is the most complicated part of dimensional reduction. To illustrate this -- and the power of our automated computation -- we  depict all the occurring diagrams in Fig.~\ref{fig:sc1:2pt:2l}.
Note that as this correlator is computed at zero external momentum, in Landau gauge many of these diagrams vanish trivially, significantly reducing the computational effort -- and allowing even a manual, non-automated computation such as the one performed in Ref.~\cite{Gorda:2018hvi} in case of the Two-Higgs-Doublet Model. 

By denoting
\begin{align}
c \equiv 
    \frac{1}{2}\bigg(\ln\Big(\frac{8\pi}{9}\Big) + \frac{\zeta'(2)}{\zeta(2)} - 2 \gammaE \bigg)
    \;,
\end{align}
where $\gammaE$ is the Euler-Mascheroni constant,
we obtain the result
\begin{align}
\label{eq:higgs-2loop-correlator}
\hat{\Pi}_{\phi^\dagger \phi} &=
    \bigg(
        - T^2 \Big( \frac{1}{16}(3\g^{2} + {\gp}^{2}) + \frac{1}{4} \gY^{2} + \frac{1}{2} \lambda \Big)
        \nn \\ &
        \hphantom{=\bigg(}
        + \frac{\mh^{2}}{(4\pi)^2} L_b \Big( 6 \lambda - \frac{1}{4} (3 \xi_2 \g^{2} + \xi_1 {\gp}^{2}) \Big)
        + \hat{\Gamma}^{\zeta}_{(\phi^\dagger \phi)}
    \bigg)_{\rmii{1-loop}}
    \nn \\ &
    + \bigg( -\frac{5}{36}T^6 c_8 -\frac{25}{288}T^8 c_{10} \bigg)_{\rmii{flowers}}
    \nn \\ &
    + \bigg\{ -\frac{1}{4}T^4 c_6 -\frac{T^2}{(4\pi)^2} \bigg[ \frac{167}{96}\g^{4} + \frac{1}{288} {\gp}^{4} - \frac{3}{16}\g^{2} {\gp}^{2} + \frac{1}{4}\lambda_2(3\g^{2}+{\gp}^2)
    \nn \\ &
    \hphantom{+\bigg\{}
    + L_{b}\Big( \frac{41}{64}(\g^{4} - {\gp}^{4}) - \frac{15}{32}\g^{2}{\gp}^{2} + \frac{1}{8}\lambda \Big(3(3+\xi_2)\g^{2}+(3+\xi_1){\gp}^{2} \Big) - 6 \lambda^2
    \nn \\ &
    \hphantom{+\bigg\{ + L_{b}\Big(}
    + \frac{1}{64}(3\g^{2} + {\gp}^{2})(3 \xi_2 \g^{2} + \xi_1 {\gp}^{2}) \Big)
    \nn \\ &
    \hphantom{+\bigg\{}
    + \Big( c + \ln\Big(\frac{3T}{\bmu_{3}}\Big) \Big)\Big( \frac{81}{16}\g^{4} +  3\lambda ( 3\g^{2} + {\gp}^{2})  - 12\lambda^2 -\frac{7}{16} {\gp}^{4} - \frac{15}{8}\g^{2} {\gp}^{2} \Big)
    \nn \\ &
    \hphantom{+\bigg\{}
    - \gY^{2} \Big(\frac{3}{16}\g^{2} + \frac{11}{48}{\gp}^{2} + 2\gs^{2} \Big)
    + \frac{9\g^{4} + 5 {\gp}^4}{108}\Nf
    \nn \\ &
    \hphantom{+\bigg\{}
    + L_{f}\Big(
        \gY^{2} \Big(\frac{9}{16}\g^{2} + \frac{17}{48}{\gp}^{2} + 2 \gs^{2} - 3 \lambda + \frac{1}{16}(3 \xi_2 \g^{2} + \xi_1 {\gp}^{2}) \Big)
        \nn \\ &
        \hphantom{+\bigg\{ + L_{f}\Big(} 
        + \frac{9}{8}\gY^{4}
        - (\frac{1}{4}\g^{4} + \frac{5}{36}{\gp}^{4})\Nf
    \Big)
    \nn \\ &
    \hphantom{+\bigg\{}
    + \ln(2)\Big(
        \gY^{2}\Big(\frac{7}{72}\g^{2} - \frac{3}{8}{\gp}^{2} + \frac{8}{3} \gs^{2} + 9 \lambda -\frac{1}{4}(3 \xi_2 \g^{2} + \xi_1 {\gp}^{2} ) \Big)
        \nn \\ &
        \hphantom{+\bigg\{ + L_{f}\Big(}
        - \frac{3}{2}\gY^{4}
        + (\frac{3}{2}\g^{4} + \frac{5}{6}{\gp}^{4} ) \Nf \Big)  \bigg] \bigg\}_{\rmii{2-loop}}
    \;.
\end{align}
Again, the leading $c_{8,10}$ pieces are denoted with {\em flowers} and they appear at three- and four-loop orders.
In order to obtain this expression, resummation is required 
to cancel the IR sensitive mixed hard/soft modes, that are non-analytic in the mass parameter $\mh^2$.
This resummation can be performed by adding and subtracting one-loop corrections to the mass parameter, in which case terms with plus signs contribute to the mass in the scalar propagator, while terms with minus signs are treated as (counterterm-like) interactions. For details of this procedure, see Refs.~\cite{Gorda:2018hvi,Kajantie:1995dw}.
However, a shortcut to circumvent this procedure is provided by using IBP reduction to evaluate sum-integrals:
since in dimensional reduction at NLO two-loop mass effects are of higher order, all two-loop sum-integrals can be treated as massless.
Therefore, all mixed soft/hard contributions vanish trivially in dimensional regularisation, and non-analytic IR structures in need of resummation never appear. 

Additionally, renormalisation of $\Pi_{\phi^\dagger \phi}$ at two-loop level is more complicated than for other correlators at one-loop. In fact, if one includes only hard contributions to this correlator, a divergence proportional to $T^2$ remains. This leftover divergence is cancelled by the divergence in the soft part of this correlator, and on the 3d theory side it corresponds exactly to the 3d mass counterterm, cf.\ Eq.~\eqref{eq:3d-mass-ct}.
What remains after cancellation of this divergence is the logarithm of $\bmu_{3}$ visible in Eq.~\eqref{eq:higgs-2loop-correlator}.

\subsubsection{Parameter matching and gauge invariance}
\label{sec:matching-relations}

The matching of the parameters can be performed by equating effective vertices in both 4d and 3d theories. For this, the general relation between generic 4d and 3d fields $\psi$ reads
\begin{align}
\label{eq:dr_field_matching}
(\psi^2)_{\rmii{3d}} = \frac{1}{T} (\psi^2)_{\rmii{4d}} \Big(1 + \hat{\Pi}'_{\psi^2} \Big)
\;.
\end{align}
This relation can be derived from the condition that the physical, screened masses of the fields agree between the 4d and 3d theories~\cite{Kajantie:1995dw}.
For an illustrative example of matching, let us consider the scalar quartic coupling.
The renormalised
$(\phi^\dagger_{i}\phi^{ }_{j}\phi^\dagger_{k} \phi^{ }_{l})$-correlators read (here we write isospin structure
$
\Delta_{ijkl}\equiv
\delta_{ij}\delta_{kl} +
\delta_{il}\delta_{jk}
$
explicitly)
\begin{align}
\underbrace{
    T \Big( - 2\lambda_3 + \Gamma^{\rmii{3d}}_{(\phi^\dagger \phi)^2} \Big) 
    \Delta_{ijkl}}_{{\rm 3d}} \equiv
\underbrace{
    \Big( - 2\lambda + \Gamma^{\rmii{soft}}_{(\phi^\dagger \phi)^2} + \hat{\Gamma}_{(\phi^\dagger \phi)^2} \Big)
    \Delta_{ijkl}}_{{\rm 4d}}
    \;.
\end{align}
The contribution from soft modes equals the 1-loop 3d contribution, i.e.\
$
\Gamma^{\rmii{soft}}_{(\phi^\dagger \phi)^2} =
\Gamma^{\rmii{3d}}_{(\phi^\dagger \phi)^2}
$
so these contributions cancel each other.
Effective vertices are then
\begin{align}
T \lambda_3 (\phi^\dagger \phi)^2_\rmii{3d} = \Big(\lambda -\frac{1}{2} \hat{\Gamma}_{(\phi^\dagger \phi)^2} \Big)  (\phi^\dagger \phi)^2_\rmii{4d}
\end{align}
which leads to the following NLO matching relation
\begin{align}
\lambda_3 = T \Big(\lambda -\frac{1}{2} \hat{\Gamma}_{(\phi^\dagger \phi)^2} - 2 \lambda \hat{\Pi}'_{\phi^\dagger \phi}  \Big)
\;,
\end{align}
once the relation between the 4d and 3d fields is taken into account.
In a similar manner, we obtain a complete list of the other matching relations in terms of other $n$-point correlators:
\begin{align}
\label{eq:matching-first}
\mu_{h,3}^2 &=
    \mh^2
    - \Big(
      \hat{\Pi}^{\rmii{1-loop}}_{\phi^\dagger \phi}
    + \hat{\Pi}^{\rmii{2-loop}}_{\phi^\dagger \phi} \Big)
    - \Big( \mh^2 - \hat{\Pi}^{\rmii{1-loop}}_{\phi^\dagger \phi} \Big) \hat{\Pi}'_{\phi^\dagger \phi}
    \;, \\
c_{6,3} &=
    T^2 \Big( c_6
        - 3 c_6 \hat{\Pi}'_{\phi^\dagger \phi}
        - \frac{1}{6} \hat{\Gamma}_{(\phi^\dagger \phi)^3} 
        \Big)
    \;, \\
c_{8,3} &=
    T^3 \Big( c_8 -4 c_8  \hat{\Pi}'_{\phi^\dagger \phi}  - \frac{1}{24} \hat{\Gamma}_{(\phi^\dagger \phi)^4}  \Big)
    \;, \\
c_{10,3} &=
    T^4  \Big( c_{10} -5 c_{10}  \hat{\Pi}'_{\phi^\dagger \phi}  - \frac{1}{120} \hat{\Gamma}_{(\phi^\dagger \phi)^5}  \Big)
    \;, \\
\g_{3}^2 &=
    T \bigg[ \g^{2} \bigg(1 - \Big( \hat{\Pi}'_{\phi^\dagger \phi} + \hat{\Pi}'_{A_{r}^{a}A_{s}^{b}}  \Big) \bigg) + 2 \hat{\Gamma}_{\phi^\dagger \phi A_{r}^{a}A_{s}^{b}}\bigg]
    \;, \\
{\gp_{3}}^2 &=
    T \bigg[ \g^{2} \bigg(1 - \Big( \hat{\Pi}'_{\phi^\dagger \phi} + \hat{\Pi}'_{B_r B_s}  \Big) \bigg) + 2 \hat{\Gamma}_{\phi^\dagger \phi B_r B_s}\bigg]
    \;,
\end{align}
\begin{align}
h_{1} &=
    - \frac{1}{2} T \bigg[ -\frac{1}{2}\g^{2} \bigg(1 - \Big( \hat{\Pi}'_{\phi^\dagger \phi} + \hat{\Pi}'_{A_{0}^{a}A_{0}^{b}}  \Big) \bigg) + \hat{\Gamma}_{\phi^\dagger \phi A_{0}^{a}A_{0}^{b}}    \bigg]
    \;, \\
h_{2} &=
    - \frac{1}{2} T \bigg[ -\frac{1}{2}{\gp}^{2} \bigg(1 - \Big( \hat{\Pi}'_{\phi^\dagger \phi} + \hat{\Pi}'_{B_0 B_0}  \Big) \bigg) + \hat{\Gamma}_{\phi^\dagger \phi B_0 B_0} \bigg]
    \;, \\
h_{3} &=
    - T \bigg[ -\frac{1}{2}\g{\gp} \bigg(1 - \Big( \hat{\Pi}'_{\phi^\dagger \phi} + \frac{1}{2} \hat{\Pi}'_{A_{0}^{a}A_{0}^{b}} + \frac{1}{2} \hat{\Pi}'_{B_0 B_0}  \Big) \bigg) + \hat{\Gamma}_{\phi^\dagger \phi A_{0}^{a} B_0}\bigg]
    \;, \\
h_{4} &=
    - \frac{1}{2} T \; \hat{\Gamma}_{\phi^\dagger \phi C^\alpha_0 C^\beta_0}
    \;, \\
\kappa_{1} &=
    - \frac{1}{2} T \; \hat{\Gamma}_{A^a_0 A^b_0 A^c_0 A^d_0}
    \;, \\
\kappa_{2} &=
    - \frac{1}{6} T \; \hat{\Gamma}_{B_0 B_0 B_0 B_0}
    \;, \\
\kappa_{3} &=
    - T \; \hat{\Gamma}_{A^a_0 A^b_0 B_0 B_0} 
    \;, \\
\mD^{2} &= \hat{\Pi}_{A_{0}^{a}A_{0}^{b}}
    \;, \\
\mD'^{2} &= \hat{\Pi}_{B_{0}B_{0}}
    \;, \\
\mD''^{2} &= \hat{\Pi}_{C_{0}^\alpha C_{0}^\beta}
    \;. 
\label{eq:matching-last}
\end{align}
The key feature of these formulas is that all 3d parameters are {\em gauge independent} up to $\mc{O}(g^4)$: at one-loop level Debye masses and quartic temporal scalar self-interactions are immediately gauge independent since the correponding correlators are (field normalisation contributes to these parameters only at NNLO).
In the other parameters we observe an explicit cancellation of gauge parameters between
correlators $\hat{\Pi},\hat{\Gamma}$ and
field normalisations $\hat{\Pi}'$,
when we insert the corresponding expressions.

However, there is a notable exception, the higher order corrections we include for $c_{6,3}$, $c_{8,3}$ and $c_{10,3}$ are gauge dependent, at orders $\mc{O}(g^6)$, $\mc{O}(g^8)$ and $\mc{O}(g^{10})$ respectively.
For e.g.\ for $c_{6,3}$
the corresponding correlator
$\hat{\Gamma}_{(\phi^\dagger \phi)^3}$ in Eq.~\eqref{eq:phi:3} is gauge-dependent at subleading order, as the gauge-dependent parts proportional to $\zeta(3)$ are not cancelled by the field normalisation piece.
A possible explanation of this leftover gauge-dependence is that the operator basis is incomplete.
We have not included higher dimensional kinetic scalar operators of the schematic form $(\phi^\dagger D \phi)^2$ where $D$ represents covariant derivative.
Our guess is, that when a complete operator basis is used, one can perform a field redefinition into a basis that is manifestly gauge invariant.
However, we do not tackle this problem further here, but leave this topic for future research. 

The final results for the 3d parameters at the soft scale read
\begin{align}
\lambda_{3} &=
    T\Big(\lambda(\bmu) + \frac{1}{(4\pi)^2}\bigg[\frac{1}{8}\Big(3\g^{4} + {\gp}^{4} +2 \g^{2} {\gp}^2 \Big)+ 3 L_f \Big(\gY^{4} - 2\lambda \gY^{2} \Big)
    \nn \\ &
    - L_b \bigg(\frac{3}{16}\Big(3\g^{4} + {\gp}^{4} + 2 \g^{2} {\gp}^{2} \Big) -\frac{3}{2}\Big(3\g^{2}+{\gp}^{2} -8 \lambda \Big) \lambda \bigg) \bigg]\Big)
    \nn \\ &
    + T^3 c_6
    - \frac{\mh^2}{(4\pi)^2} 12 T c_{6}L_{b}
    + \frac{5}{6} c_8 T^5
    + \frac{25}{36} c_{10} T^7
    \;, \\
\label{eq:c63}
c_{6,3} &=
    T^2 c_6(\bmu) \bigg (1 + \frac{1}{(4\pi)^2} \bigg[ \Big( -54 \lambda + \frac{9}{4} (3 \g^{2} + {\gp}^{2}) \Big) L_b - 9 \gY^{2} L_f \bigg] \bigg)
    \nn \\ &
    - \frac{\zeta(3)}{768 \pi^4} \bigg( -\frac{3}{8} \Big( 3 \g^{6} + 3 \g^{4} {\gp}^{2} + 3 \g^{2} {\gp}^{4} + {\gp}^{6} \Big)  - 240 \lambda^3
    \nn \\ &
    + 84 \gY^{6}
    + \underbrace{6 \lambda^2 \Big(3 \g^{2} \xi_2 + {\gp}^{2} \xi_1 \Big)}_{\xi \text{-dependence at } \mathcal{O}(g^6) }\bigg)
    + c_8 T^2 \Big( \frac{5}{3} T^2 - \frac{\mh^2}{(4\pi)^2} 20 L_b  \Big)
    + \frac{25}{12} c_{10} T^6
    \;, \\
\label{eq:c83}
c_{8,3} &=
    T^3 \bigg( c_8(\bmu)  + \frac{1}{(4\pi)^2} \bigg[ \Big[ \Big(-96 \lambda + 3(3 \g^{2} + {\gp}^{2})\Big) c_8 -63 c^2_6 - 30 \mh^2 c_{10} \Big] L_b - 12 \gY^{2} c_8 L_f \bigg] \bigg)
    \nn \\ &
    + \frac{5}{2} T^5 c_{10}
    + c_{8,3,\zeta}
    \;, \\
\label{eq:c103}
c_{10,3} &=
    T^4 \bigg( c_{10}(\bmu)  + \frac{1}{(4\pi)^2} \bigg[ \Big[ \Big( -150 \lambda + \frac{15}{4} (3 \g^{2} + {\gp}^{2}) \Big)c_{10} -228 c_6 c_8 \Big] L_b - 15 \gY^{2} c_{10} L_f \bigg] \bigg)
    \nn \\ &
    + c_{10,3,\zeta}
    \;, \\
\g_{3}^2 &=
    \g^{2}(\bmu)T\bigg[1 +\frac{\g^{2}}{(4\pi)^2}\bigg(\frac{43}{6}L_b+\frac{2}{3}-\frac{4\Nf}{3}L_f\bigg)\bigg]
    \;,\\
{\gp_{3}}^{2} &=
    {\gp}^{2}(\bmu)T\bigg[1 +\frac{{\gp}^{2}}{(4\pi)^2}\bigg(-\frac{1}{6}L_b-\frac{20\Nf}{9}L_f\bigg)\bigg]
    \;,\\
h_{1} &=
    \frac{\g^{2}(\bmu)T}{4}\bigg(1+\frac{1}{(4\pi)^2}\bigg\{\bigg[\frac{43}{6}L_b+\frac{17}{2}-\frac{4\Nf}{3}(L_f-1)\bigg]\g^{2}+\frac{{\gp}^{2}}{2}-6\gY^{2}+12\lambda\bigg\} \bigg)
    \;,\\
h_{2} &=
    \frac{{\gp}^{2}(\bmu)T}{4}\bigg(1 +\frac{1}{(4\pi)^2}\bigg\{\frac{3\g^{2}}{2}
    - \bigg[\frac{(L_b-1)}{6}  + \frac{20\Nf(L_f-1)}{9}\bigg]{\gp}^{2}- \frac{34}{3} \gY^{2}+12\lambda\bigg\} \bigg)
    \;,\\
h_{3} &=
    \frac{\g(\bmu){\gp}(\bmu)T}{2}\bigg\{1+\frac{1}{(4\pi)^2}\bigg[-\g^{2}+ \frac{1}{3}{\gp}^{2}+L_b\bigg(\frac{43}{12}\g^{2} -\frac{1}{12}{\gp}^{2}\bigg)
    \nn \\ &
    \hphantom{=\frac{\g(\bmu){\gp}(\bmu)T}{2}\bigg\{1}
    - \Nf(L_f-1)\bigg(\frac{2}{3}\g^{2}+\frac{10}{9}{\gp}^{2}\bigg)+4\lambda+ 2\gY^{2}\bigg] \bigg\}
    \;, \\
h_{4} &= -T\frac{1}{(4\pi)^2} 2 \gs^{2} \gY^{2}
    \;, 
\end{align}
\begin{align}
\kappa_{1}&= T\frac{\g^{4}}{(4\pi)^2} \bigg(\frac{17-4\Nf}{3}\bigg)
    \;,\\
\kappa_{2} &= T\frac{{\gp}^{4}}{(4\pi)^2} \bigg(\frac{1}{3}-\frac{380}{81} \Nf\bigg)
    \;,\\
\kappa_{3} &= T\frac{\g^{2} {\gp}^{2}}{(4\pi)^2}\bigg(2-\frac{8}{3}\Nf\bigg)
    \;, \\
\mD^{2} &=
    \g^{2}(\bmu) T^2\bigg(\frac{5}{6}+\frac{\Nf}{3}\bigg)
    \;,\\
\mD'^{2} &=
    {\gp}^{2}(\bmu)  T^2\bigg(\frac{1}{6}+\frac{5\Nf}{9}\bigg)
    \;,\\
\mD''^{2} &=
    \gs^{2} T^2 \bigg(1+\frac{\Nf}{3}\bigg)
    \;, \\
\mu_{h,3}^2 &=
    \mu^2_{h}(\bmu)
    + \frac{T^2}{16}\Big(3\g^{2}(\bmu) + {\gp}^{2}(\bmu) + 4 \gY^{2}(\bmu) + 8 \lambda(\bmu) \Big)
    \nn \\ &
    + \frac{1}{4} T^4 c_6
    + \frac{5}{36}T^6 c_8
    + \frac{25}{288}T^8 c_{10}
    \nn \\ &
    + \frac{1}{(4\pi)^2} \bigg\{ \mh^2 \bigg[ \Big(\frac{3}{4}(3\g^{2} + {\gp}^{2}) - 6 \lambda \Big)L_b - 3 \gY^{2} L_f \bigg]
    \nn \\ &
    + T^2 \bigg[ \frac{167}{96}\g^{4} + \frac{1}{288}{\gp}^{4} - \frac{3}{16}\g^{2} {\gp}^{2} + \frac{1}{4}\lambda (3\g^{2}+{\gp}^{2})
    \nn \\ &
    + L_b \Big( \frac{17}{16}\g^{4} - \frac{5}{48}{\gp}^{4} - \frac{3}{16}\g^{2}{\gp}^{2} + \frac{3}{4}\lambda(3\g^{2}+{\gp}^{2}) - 6 \lambda^2 \Big)
    \nn \\ &
    + \Big( c + \ln\left(\frac{3T}{\bmu_{3}}\right) \Big)\Big(  \frac{81}{16}\g^{4} +  3\lambda ( 3\g^{2} + {\gp}^{2})  - 12\lambda^2 -\frac{7}{16} {\gp}^{4} - \frac{15}{8}\g^{2} {\gp}^{2} \Big)
    \nn \\ &
    - \gY^{2} \Big(\frac{3}{16}\g^{2} + \frac{11}{48}{\gp}^{2} + 2 \gs^{2} \Big)
    + \Big(\frac{1}{12}\g^{4} + \frac{5}{108}{\gp}^{4}\Big)\Nf
    \nn \\ &
    + L_f \Big( \gY^{2} \Big(\frac{9}{16}\g^{2} + \frac{17}{48}{\gp}^{2} + 2 \gs^{2} - 3 \lambda \Big) +\frac{3}{8}\gY^{4} - \Big(\frac{1}{4}\g^{4} + \frac{5}{36}{\gp}^{4}\Big) \Nf \Big)
    \nn \\ &
    + \ln(2) \Big( \gY^{2} \Big(-\frac{21}{8}\g^{2} - \frac{47}{72}{\gp}^{2} + \frac{8}{3} \gs^{2} + 9 \lambda \Big) -\frac{3}{2}\gY^{4} + \Big(\frac{3}{2}\g^{4} + \frac{5}{6}{\gp}^{4}\Big) \Nf \Big) \bigg] \bigg\}
    \;.
\end{align}
These formulas emphasise that the LO pieces run in terms of 4d RG-scale $\bmu$. By applying corresponding $\beta$-functions, one can observe that this running cancels the explicit logarithmic scale dependence of the $L_{b/f}$-terms. 
However, there remains a scale dependence which is formally of $\mc{O}(g^6)$ for the parameters of the Higgs and spatial gauge bosons, as discussed in Sec.~\ref{sec:scale_dependence}.
There is also a scale dependence at $\mc{O}(g^4T^2)$ for the Debye masses $\mD^{2}$, $\mD'^{2}$ and $\mD''^{2}$.
To cancel this scale dependence, the Debye masses should be evaluated at two-loop order.
However, this scale dependence only contributes to the Higgs effective potential at $\mc{O}(g^5)$, and a full $\mc{O}(g^5)$ calculation goes beyond the scope of this paper;
see for example Refs.~\cite{Zhai:1995ac,Ghiglieri:2020dpq}.
In practice this leftover scale dependence is numerically insignificant since the running of the gauge couplings is
small.

The above Eqs.~\eqref{eq:c83} and \eqref{eq:c103}, employ the following shorthand notation for terms proportional to $\zeta$-functions
\begin{align}
\label{eq:zetas-8}
c_{8,3,\zeta} &\equiv
    \frac{1}{T} \bigg( \frac{\zeta(3)}{(4\pi)^4} \bigg[ -6 T^2 c_6 \lambda \Big(-96 \lambda + 3 g^2 \xi_2 + {\gp}^2 \xi_1  \Big)  \bigg]
    \nn \\ &
    + \frac{\zeta(5)}{(4\pi)^6} \bigg[ -\frac{3}{32} \Big( 3\g^{8} + 4\g^6 {\gp}^2 + 6\g^4 {\gp}^4 + 4\g^2{\gp}^6 + {\gp}^8 \Big) + 186 \gY^8
    \nn \\ &
    - 672 \lambda^4 + 8 \lambda^3 {\gp}^2 \xi_1 - \lambda^2 {\gp}^4 \xi^2_1 + 2 \lambda^2 \g^2 \xi_2 (12 \lambda-{\gp}^2 \xi_1 ) - 3 \lambda^2 \g^{4} \xi^2_2 \bigg] \bigg)
    \;, \\
\label{eq:zetas-10}
c_{10,3,\zeta} &\equiv
    \frac{1}{T^2} \bigg(
    \frac{\zeta(3)}{(4\pi)^4}  \bigg[ -T^4 \Big( \frac{9}{2} c^2_6 (-312 \lambda + 3\g^2 \xi_2 + {\gp}^2 \xi_1 ) + 8 c_8 \lambda (-132 \lambda + 3\g^2 \xi_2 + {\gp}^2 \xi_1 )  \Big) \bigg]
    \nn \\ &
    + \frac{\zeta(5)}{(4\pi)^6}  \bigg[ -3 c_6 \lambda T^2 \Big(2208 \lambda^2 + {\gp}^4 \xi_1 + 2 g^2 {\gp}^2 \xi_1 \xi_2 + 3 g^4 \xi_2 - 12 \lambda (3 g^2 \xi_2 + {\gp}^2 \xi_1) \Big)  \bigg]
    \nn \\ &
    + \frac{\zeta(7)}{(4\pi)^8}  \bigg[ \frac{1}{32} \Big( 3 {\gp}^{10} + 15\g^2 {\gp}^8 + 30\g^4 {\gp}^6 + 30\g^6 {\gp}^4 + 15\g^8 {\gp}^2 + 9\g^{10} - 48768 \gY^{10}
    \nn \\ &
    + 251904 \lambda^5 -1280 \lambda^4 {\gp}^2 \xi_1 + 320 \lambda^3 {\gp}^4 \xi^2_1 + 640 \lambda^3 \g^2 \xi_2 (-6 \lambda + {\gp}^2 \xi_1) + 960 \lambda^3 \g^4 \xi^2_2   \Big) \bigg] \bigg)
    \;.
\end{align}
These contributions -- with leftover gauge dependence -- are formally $\mathcal{O}(g^8)$ and $\mathcal{O}(g^{10})$ respectively, and we include them as they contribute at leading (one-loop) order to the respective 3d parameters.

Finally, when the soft temporal scalar scalars are integrated out in the second step of dimensional reduction (cf.\ Ref.~\cite{Kajantie:1995dw}), the action of the final ultrasoft scale EFT is given in Eq.~\eqref{eq:3d:ultrasoft:action}.
The parameters of this ultrasoft EFT read
\begin{align}
\bar{g}^2_3 &= \g_{3}^{2} \Big( 1 - \frac{\g_{3}^{2}}{6 (4\pi) \mD} \Big)
\;, \\
\bar{g}'^2_3 &= {\gp_{3}}^2
\;, \\
\bar{\mu}^2_{h,3} &= \mu^2_{h,3}
    - \frac{1}{4\pi}\Big(3 h_{1}\mD +  h_{2}\mD' + 8 h_{4}\mD'' \Big)
    \nn \\ &
    \hphantom{= \mu^2_{h,3}}
    + \frac{1}{(4\pi)^2} \bigg( 3\g_{3}^{2}h_{1} - 3 h_{1}^2 - h_{2}^2 - \frac{3}{2} h_{3}^{2}
    \nn \\ &
    \hphantom{= \mu^2_{h,3}+ \frac{1}{(4\pi)^2} \bigg(}
    + \Big(-\frac{3}{4}\g_{3}^{4} + 12\g_{3}^{2}h_{1} \Big) \ln\Big(\frac{\bmu_{3}}{2\mD} \Big) 
    - 6 h_{1}^{2} \ln\Big(\frac{\bmu_{3}}{2\mD} \Big)
    \nn \\ &
    \hphantom{= \mu^2_{h,3}+ \frac{1}{(4\pi)^2} \bigg(}
    - 2 h_{2}^{2} \ln\Big(\frac{\bmu_{3}}{2\mD'} \Big)
    - 3 h_{3}^{2} \ln\Big(\frac{\bmu_{3}}{\mD+\mD'} \Big)
    \bigg)
\;, \\
\bar{\lambda}_{3} =& \lambda_{3} - \frac{1}{2(4\pi)}\Big(
      \frac{3 h_{1}^{2}}{\mD}
    + \frac{h_{2}^{2}}{\mD'}
    + \frac{h_{3}^{2}}{\mD+\mD'}
    \Big)
    \;.
\end{align}
In this step of dimensional reduction, the two-loop matching of mass parameters requires resummation (by adding and subtracting the one-loop contribution to ultrasoft mass). Consequently the soft Higgs mass parameter does not appear inside the two-loop piece -- despite mixed diagrams involving both Higgs and soft temporal scalars -- since the effect of the Higgs mass is formally of higher order.
We point out that this technical detail was overlooked in Ref.~\cite{Gorda:2018hvi} in Eq.~(3.45).  

For higher dimensional operator couplings, for simplicity we neglect contributions from $B_0$ and $C_0$ fields, as they are numerically insignificant.
In fact, already corrections from $A^a_0$ are heavily suppressed.
At the ultrasoft scale we have
\begin{align}
\bar{c}_{6,3} &=
    c_{6,3} + \frac{1}{2(4\pi)}\frac{h_{1}^{3}}{\mD^3}
\;, \\
\bar{c}_{8,3} &=
    c_{8,3} - \frac{1}{8(4\pi)}\frac{h_{1}^{4}}{\mD^5}
\;, \\
\bar{c}_{10,3} &=
    c_{10,3} + \frac{1}{32(4\pi)}\frac{h_{1}^{5}}{\mD^7}
\;.
\end{align} 
These formulas complete our construction of the 3d EFTs that we use to incorporate higher order resummations in our perturbative analysis of the phase transition in the SMEFT.

\subsection{The 3d perturbative expansion parameter}
\label{appendix:3d_perturbation_theory}

For the case of the 3d EFT considered in this paper (cf.~Eq.~\eqref{eq:3d:ultrasoft:action}), we show that the $\hbar$-expansion for the phase transition equals an expansion in powers of $\sqrt{\bar{c}_{6,3}}$.
Near the critical temperature two minima are separated by a barrier, and are of similar heights. For the tree-level Higgs potential to show this structure, all three terms must be of the same order.
From this one can derive that the broken minimum scales as $\phi\propto \sqrt{-\bar{\lambda}_3/\bar{c}_{6,3}}$.
To expand around the broken phase or around the critical bubble, we shift the Higgs by a background field, $\phi\to\left(0,\Phi/\sqrt{2}\right)+\phi$, a saddle point of the tree-level scalar action,
\begin{align}
S_0[\Phi] &= \int {\rm d}^3 x \left( \frac{1}{2}\partial_{r}\Phi\partial_{r} \Phi + V(\Phi)\right)
\;, \\
\frac{\delta S_0}{\delta \Phi} &= 0
\;.
\end{align}
Expanding around this background and scaling
\begin{align}
\label{eq:scalings}
\Phi &\to \sqrt{\frac{-\bar{\lambda}_3}{\bar{c}_{6,3}}} \Phi
\;, \qquad
\phi \to \frac{\bar{g}_3}{\bar{c}_6^{1/4}} \phi
\;,  \qquad
x^{r} \to \frac{\bar{c}_{6,3}^{1/2}}{\bar{g}_3^2} x^{r}
\;, \nn \\
A_{r} &\to \frac{\bar{g}_3}{\bar{c}_{6,3}^{1/4}} A_{r}
\;, \qquad
B_{r} \to \frac{\bar{g}_3}{\bar{c}_{6,3}^{1/4}} B_{r}
\;, \qquad
\eta \to \frac{\bar{g}_3}{\bar{c}_{6,3}^{1/4}} \eta
\;,
\end{align}
the action takes the form 
\begin{equation}
\label{eq:action_scaling}
S = \frac{1}{\sqrt{\bar{c}_{6,3}}} S_0[\Phi]
+ \sum_{n=2}^6\frac{1}{n!}\bar{c}_{6,3}^{(n-2)/4}\frac{\delta^n S}{\delta\varphi_{\alpha_1}\cdots\delta\varphi_{\alpha_n}}\bigg|_{\Phi}\varphi_{\alpha_1}\cdots\varphi_{\alpha_n}
\;,
\end{equation}
where $\{\varphi_\alpha\}=\{\phi,A_{r},B_{r},\eta,\bar{\eta}\}$ runs over all of the fields, and all factors of $\bar{c}_{6,3}$ have been made explicit. This shows that the effective loop expansion parameter is
proportional to $\bar{c}_{6,3}^{1/2}$. This perhaps surprising observation, that $\bar{c}_{6,3}$ controls the loop expansion in the coupled gauge-Higgs theory (even e.g. the $(A_{r})^4$ interaction!), is a special property of the truncated 3d theory we are considering, and would not hold in a more general truncation including, for example, higher order derivative terms.

As regards the $\bar{g}_3^2$, $\bar{g}_3^{'2}$ and $\bar{\lambda}_3$ dependence of physical quantities computed in the 3d loop expansion, by scaling out overall powers of $\bar{g}_3^2$ to fix dimensions, one can see that only the ratios $\bar{\lambda}_3/\bar{g}_3^2$ and $\bar{g}_3^{'2}/\bar{g}_3^2$ may arise to modify the expansion parameter. In our analysis of the SMEFT, $\bar{c}_{6,3}$ is naturally the smallest parameter, and hence the 3d loop expansion will generally converge well, at least around the broken phase.

The structure of Eq.~\eqref{eq:action_scaling} is not modified in the symmetric phase, meaning that $\bar{c}_{6,3}^{1/2}$ acts as the loop counting parameter there too. However, in the symmetric phase the tree-level mass of the 3d gauge bosons is zero, which leads to IR divergences, and consequently nonperturbativity. A fuller discussion of this is given in Sec.~\ref{sec:nonperturbativity}.

\subsection{Two-loop thermal effective potential} 
\label{appendix:Veff3d}

For the ${\rm SU}(2)$ gauge theory, with a Higgs field in the fundamental representation, the effective potential has been calculated to two-loop order in Refs.~\cite{Farakos:1994kx,Laine:1994bf,Kripfganz:1995jx}. Ref.~\cite{Farakos:1994kx} also generalises this to include the ${\rm U}(1)$-hypercharge. For the SMEFT, the $(\hsq)^3$ term introduces corrections to the Feynman rules.
However, there are no new connected vacuum diagrams to be added, as we now show.
Using the usual topological identities for connected graphs (see for example the chapter on divergences and regularisation in Ref.~\cite{ZinnJustin:2002ru}), one can derive the following equation for
the number of loops, $L$, of a vacuum graph containing
$N_6$ 6-point,
$N_4$ 4-point, and
$N_3$ 3-point vertices,
\beq
L = 1 + 2N_6 + N_4 + \frac{1}{2}N_3
\;.
\eeq
This shows that there are no connected vacuum diagrams containing 6-point vertices (i.e.\ with $N_6>0$) below $L=3$. Hence, working to two-loop order we need only keep track of the corrections to the mass and lower-point vertex rules in the computation of the effective potential.

The tree-level 3d potential at the ultrasoft scale, after inserting the Higgs background field $\phi=(0,v_3)/\sqrt{2}$ 
reads
\begin{align}
V_{3(0)} &=
    \frac{1}{2} \bar{\mu}_{h,3}^2 v_3^2
    + \frac{1}{4} \bar{\lambda}_3 v_3^4
    + \frac{1}{8} \bar{c}_{6,3} v_3^6
    \nn \\ &
    + \frac{1}{2} \delta\bar{\mu}_{h,3}^2 v_3^2
    + \frac{1}{4} \delta \bar{\lambda}_3 v_3^4
    + \frac{1}{8} \delta \bar{c}_{6,3} v_3^6
    + \delta V
    \;,
\end{align}
where $\delta V$ is the vacuum counterterm, and all counterterms arise at two-loop order. Due to the presence of the dimension-6 operator, the 3d EFT is not super-renormalisable, and the 2-loop counterterms are not exact, as in the pure SM 3d EFT.

Note that as we utilise the gauge-invariant $\hbar$-expansion for our 3d computations, we are free to fix a gauge in calculating the 3d effective potential.
In Landau gauge, the one-loop contribution is given by
\begin{align}
V_{3(1)} &=
    J_{\rmii{soft}}(m_{\phi,3})
    + 3 J_{\rmii{soft}}(m_{\chi,3})
    \nn \\  &
    + (d-1)\Big( 2 J_{\rmii{soft}}(m_{W,3}) + J_{\rmii{soft}}(m_{Z,3}) \Big)
\;, 
\end{align}
where  the master integral is given by Eq.~\eqref{eq:Jsoft} and
the mass eigenvalues, for the Higgs, Goldstones, $W$ and $Z$ bosons, are functions of the 3d parameters and the 3d background field $v_3$,
\begin{align}
m^2_{\phi,3} &=
    \bar{\mu}_{h,3}^2
    + 3 \bar{\lambda}_3 v^2_3
    + \frac{15}{4} \bar{c}_{6,3} v^4_3
\;, \\
m^2_{\chi,3} &=
    \bar{\mu}_{h,3}^2
    + \bar{\lambda}_3  v^2_3
    + \frac{3}{4} \bar{c}_{6,3} v^4_3
\;,\\
m^2_{W,3} &=
    \frac{1}{4}\bar{g}_3^2 v^2_3
\;,\\
m^2_{Z,3} &=
    \frac{1}{4} \Big( \bar{g}_3^2
    + \bar{g}_3^{\prime 2} \Big) v^2_3
\;.
\end{align}

\begin{figure}[t]
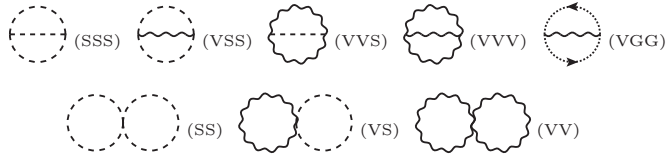

    \centering
    \begin{align*}
        &\ToptVS(\Axx,\Axx,\Lxx)_{\rmii{(SSS)}}\;
        \ToptVS(\Axx,\Axx,\Lgli)_{\rmii{(VSS)}}\;
        \ToptVS(\Agli,\Agli,\Lxx)_{\rmii{(VVS)}}\;
        \ToptVS(\Agli,\Agli,\Lgli)_{\rmii{(VVV)}}\;
        \ToptVS(\Aghi,\Aghi,\Lgli)_{\rmii{(VGG)}}
        \\[3mm] & \quad\quad
        \ToptVE(\Axx,\Axx)_{\rmii{(SS)}}\;
        \ToptVE(\Agli,\Axx)_{\rmii{(VS)}}\;
        \ToptVE(\Agli,\Agli)_{\rmii{(VV)}}
    \end{align*}
\caption{Diagram topologies contributing to the two-loop effective potential. Dashed lines denote scalars (S), wavy lines denote vector bosons (V) and dotted lines refer to ghost fields (G).
}
\label{fig:NP2}
\end{figure}

The two-loop contribution to the effective potential in the 3d EFT is straightforward to include, by following Refs.~\cite{Farakos:1994kx,Laine:1994bf,Niemi:2020hto}.
Since new contributions from $\bar{c}_{6,3}$ appear only through
the modified mass eigenvalues $m_{\chi,3}$ and $m_{\phi,3}$, and through pure scalar vertices, where the latter affects only the (SS) and (SSS) topology classes below.
\begin{align}
V_{3(2)} =  -\Big(
    \text{(SSS)}
    + \text{(VSS)}
    + \text{(VVS)}
    + \text{(VVV)}
    + \text{(VGG)}
    + \text{(SS)}
    + \text{(VS)}
    + \text{(VV)}
    \Big)\;,
\end{align} 
where different topology classes are illustrated in Fig.~\ref{fig:NP2} and their results read (again, in Landau gauge)
\begin{align}
{\rm (SSS)} &=
    \frac{3}{4} v^2_3 \Big(2 \bar{\lambda}_3 + 3 \bar{c}_{6,3} v_3^2 \Big)^2 \mc{D}_{\rmii{SSS}}(m_{\phi,3},m_{\chi,3},m_{\chi,3})
    \nn \\ &
    + \frac{3}{4} v^2_3 \Big( 2 \bar{\lambda}_3 + 5 \bar{c}_{6,3} v_3^2 \Big)^2 \mc{D}_{\rmii{SSS}}(m_{\phi,3},m_{\phi,3},m_{\phi,3})
    \;,\\
{\rm (VSS)} &=
    \frac{1}{4} \bar{g}_3^2 \mc{D}_{\rmii{VSS}}(m_{\chi,3},m_{\chi,3},m_{W,3})
    + \frac{1}{4} \bar{g}_3^2 \mc{D}_{\rmii{VSS}}(m_{\phi,3},m_{\chi,3},m_{W,3})
    \nn \\ &
    + \frac{1}{8} (\bar{g}_3^2 + \bar{g}_3^{\prime 2})  \mc{D}_{\rmii{VSS}}(m_{\phi,3},m_{\chi,3},m_{Z,3})
    \nn \\ &
    + \frac{1}{8} \frac{(\bar{g}_3^2 - \bar{g}_3^{\prime 2})^2}{\bar{g}_3^2 + \bar{g}_3^{\prime 2}} \mc{D}_{\rmii{VSS}}(m_{\chi,3},m_{\chi,3},m_{Z,3})
    + \frac{1}{2} \frac{\bar{g}_3^2 \bar{g}_3^{\prime 2}}{\bar{g}_3^2 + \bar{g}_3^{\prime 2}} \mc{D}_{\rmii{VSS}}(m_{\chi,3},m_{\chi,3},0)
    \;,\\
{\rm (VVS)} &=
    \frac{1}{8} \bar{g}_3^4 v_3^2  \mc{D}_{\rmii{VVS}}(m_{\phi,3},m_{W,3},m_{W,3})
    + \frac{1}{16} (\bar{g}_3^2 + \bar{g}_3^{\prime 2})^2 v_3^2  \mc{D}_{\rmii{VVS}}(m_{\phi,3},m_{Z,3},m_{Z,3})
    \nn \\ & + \frac{1}{4} \frac{\bar{g}_3^4 \bar{g}_3^{\prime 2} v_3^2}{\bar{g}_3^2 + \bar{g}_3^{\prime 2}} \mc{D}_{\rmii{VVS}}(m_{\chi,3},m_W,0) + \frac{1}{4} \frac{\bar{g}_3^2 \bar{g}_3^{\prime 4} v_3^2}{\bar{g}_3^2 + \bar{g}_3^{\prime 2}} \mc{D}_{\rmii{VVS}}(m_{\chi,3},m_{W,3},m_{Z,3})
    \;,\\
{\rm (VVV)} &=
    -\frac{1}{2} \frac{\bar{g}_3^4}{\bar{g}_3^2+\bar{g}_3^{\prime 2}} \mc{D}_{\rmii{VVV}}(m_{W,3},m_{W,3},m_{Z,3})
    -\frac{1}{2} \frac{\bar{g}_3^2 \bar{g}_3^{\prime 2}}{\bar{g}_3^2+\bar{g}_3^{\prime 2}} \mc{D}_{\rmii{VVV}}(m_{W,3},m_{W,3},0)
    \;,\\
{\rm (VGG)} &=
    -2 \bar{g}_3^2 \mc{D}_{\rmii{VGG}}(m_{W,3})
    -\frac{\bar{g}_3^4}{\bar{g}_3^2+\bar{g}_3^{\prime 2}} \mc{D}_{\rmii{VGG}}(m_{Z,3})
    \;,\\
{\rm (SS)} &=
    -\frac{15}{8} \Big(2\bar{\lambda}_3 + 3 \bar{c}_{6,3} v_3^2 \Big) \Big( I^3_1(m_{\chi,3}) \Big)^2 - \frac{3}{4} \Big( 9 \bar{c}_{6,3} v_3^2 +  2 \bar{\lambda}_3 \Big) I^3_1(m_{\chi,3}) I^3_1(m_{\phi,3})
    \nn \\ & 
    - \frac{3}{8} \Big( 2\bar{\lambda}_3 + 15 \bar{c}_{6,3} v_3^2 \Big) \Big( I^3_1(m_{\phi,3}) \Big)^2
    \;,\\
{\rm (VS)} &=
    - \frac{3}{4} (d-1) \bar{g}_3^2  I^3_1(m_{\chi,3}) I^3_1(m_{W,3})
    - \frac{1}{4} (d-1) \frac{(\bar{g}_3^2-\bar{g}_3^{\prime 2})^2}{\bar{g}_3^2+\bar{g}_3^{\prime 2}}  I^3_1(m_{\chi,3}) I^3_1(m_{Z,3})
    \\ &
    - \frac{1}{4} (d-1) \bar{g}_3^2  I^3_1(m_{\phi,3}) I^3_1(m_{W,3})
    - \frac{1}{8} (d-1) (\bar{g}_3^2 + \bar{g}_3^{\prime 2}) I^3_1(m_{\chi,3}) I^3_1(m_{Z,3})
    \nn \\ &
    - \frac{1}{8} (d-1) (\bar{g}_3^2 + \bar{g}_3^{\prime 2})  I^3_1(m_{\phi,3}) I^3_1(m_{Z,3})
    \;,\\
{\rm (VV)} &=
    - \frac{1}{2} \bar{g}_3^2 \mc{D}_{\rmii{VV}}(m_{W,3}, m_{W,3})
    - \frac{\bar{g}_3^4}{\bar{g}_3^2+\bar{g}_3^{\prime 2}}\mc{D}_{\rmii{VV}}(m_{W,3}, m_{Z,3})
    \;.
\end{align}
The above diagrams compose of several master integrals, that are defined and computed in Ref.~\cite{Niemi:2020hto}.
The counterterms required by renormalisation read
\begin{align}
\label{eq:3d-mass-ct}
    \delta\bar{\mu}^2_{h,3} &= -\frac{1}{\epsilon} \frac{1}{(4\pi)^2} \frac{1}{4} \Big( \frac{51}{16} \bar{g}^4_3 + 9 \bar{g}^2_3 \bar{\lambda}_3 - 12 \bar{\lambda}^2_3 - \frac{5}{16} \bar{g}'^4_3 - \frac{9}{8}  \bar{g}^2_3  \bar{g}'^2_3 + 3 \bar{g}'^2_3 \bar{\lambda}_3  \Big)
    \;, \\
    \delta \bar{\lambda}_3 &= -\frac{1}{\epsilon} \frac{1}{(4\pi)^2} \frac{3}{2} \bar{c}_{6,3} \Big( 3 \bar{g}^2_3 + \bar{g}'^2_3 - 16 \bar{\lambda}_3 \Big)
    \;, \\
    \delta \bar{c}_{6,3} &= \frac{1}{\epsilon} \frac{1}{(4\pi)^2} 51 \bar{c}^2_{6,3}
    \;, \\
    \delta V &= -\frac{1}{\epsilon} \frac{1}{(4\pi)^2} \frac{1}{4} \bar{\mu}^2_{h,3} \Big( 3 \bar{g}^2_3 + \bar{g}'^2_3 \Big)
    \;,
\end{align}
and these agree with Eqs.~(C.133)--(C.135) of Ref.~\cite{Gorda:2018hvi}, except that therein contributions of the U(1) gauge field have not been included.

\subsection{Computation of thermodynamics}
\label{appendix:3d_thermodynamics}

As Sec.~\ref{sec:recipe_3d} discusses, one vital advantage of dimensional reduction over daisy-resummation is the ability to perform consistent $\hbar$-expansions within the low-energy effective theory, at least in the case where there is a tree-level barrier. This ensures order-by-order gauge invariance, and further allows one to calculate the rate of bubble nucleation self-consistently, without double-counting degrees of freedom. In this subsection, we apply these ideas to the SMEFT.

One starts by solving everything at tree-level. In this case, and assuming $\bar{\lambda}_3<0$ (which is indeed the case for the entire range of $M$ we consider), the positive broken minimum is given by,
\beq
\label{eq:3d-tree-minimum}
v_{3,(0)} = \sqrt{\frac{2}{3}} \sqrt{\frac{-\bar{\lambda}_3}{\bar{c}_{6,3}} + \frac{1}{\bar{c}_{6,3}}\sqrt{\bar{\lambda}_3^2 - 3 \bar{c}_{6,3} \bar{\mu}_{h,3}^2}}
\;.
\eeq
The point at which this minimum is degenerate with the symmetric phase gives the critical mass. At tree-level this happens when the mass parameter is equal to,
\beq
\bar{\mu}^2_{h,3,c(0)} = \frac{\bar{\lambda}_3^2}{4 \bar{c}_{6,3}}
\;.
\eeq
One- and two-loop corrections to the broken minimum and to the critical mass are given in terms of the $\hbar$-expansion of the effective potential in
Eqs.~\eqref{eq:v_hbar} and \eqref{eq:mass_hbar}.
Explicitly to one-loop order, we find the critical mass,
\beq
\bar{\mu}^2_{h,3,c} =
    \frac{\bar{\lambda}_3^2}{4 \bar{c}_{6,3}}
    + \frac{\hbar}{3(4\pi)}\sqrt{\frac{-\bar{\lambda}_3}{\bar{c}_{6,3}}} \left[\bar{g}_3^3+\frac{1}{2}(\bar{g}_3^2+\bar{g}_3'^2)^{3/2}
    + (-\bar{\lambda}_3)^{3/2} \right]
   + \mc{O}(\hbar^2)
   \;,
\eeq
at which point the broken minimum of the potential is
\beq
v_{3,c} =
    \sqrt{-\frac{\bar{\lambda}_3}{\bar{c}_{6,3}}}
    + \frac{\hbar}{6(4\pi)}\frac{1}{(-\bar{\lambda}_3)}
    \left[\bar{g}_3^3+\frac{1}{2}\left(\bar{g}_3^2+\bar{g}_3'^2\right)^{3/2} + 25(-\bar{\lambda}_3)^{3/2}\right]
    + \mc{O}(\hbar^2)
    \;.
\eeq
Corrections at $\mc{O}(\hbar^2)$ are straightforward to construct, using
Eqs.~\eqref{eq:v_hbar} and \eqref{eq:mass_hbar} and the results of Appendix~\ref{appendix:Veff3d}, but the expressions are long, so we do not quote them explicitly here. The only subtlety, discussed below Eq.~\eqref{eq:v_hbar}, is the presence of infrared divergences, which must be regulated by taking the Goldstone boson mass to zero only at the end of the computation.

The change in the trace anomaly, which determines $\alpha$ in Eq.~\eqref{eq:alpha_definition}, is determined in terms of the $\hbar$-expansion for the effective potential, Eq.~\eqref{eq:V_hbar}, and its derivatives. As discussed in Ref.~\cite{Farakos:1994xh}, the derivatives of the effective potential can be interpreted in terms of condensates of the operators present in the Lagrangian.
For the SMEFT, we have that,
\begin{align} 
\frac{\Delta\Theta}{T} &=
    - \frac{3}{4}\Delta V_3
    + \frac{1}{4}\sum_{i}\frac{{\rm d}\kappa_i}{{\rm d}\ln T}\frac{\partial \Delta V_3}{\partial \kappa_i}
    \;,\\ &=
    - \frac{3}{4}\Delta V_3 
    + \frac{1}{4}\bigg(\frac{{\rm d} \bar{\mu}_{h,3}^2}{{\rm d}\ln T}\Delta \langle \hsq \rangle 
    + \frac{{\rm d} \bar{\lambda}_3}{{\rm d}\ln T}\Delta \langle (\hsq)^2 \rangle
    + \frac{{\rm d} \bar{c}_6}{{\rm d}\ln T}\Delta \langle (\hsq)^3 \rangle
    \nn \\ &
    \hphantom{=-\frac{3}{4}\Delta V_3}
    - \frac{1}{\bar{g}_3^2} \frac{{\rm d}\bar{g}_{3}^{2}}{{\rm d}\ln T}\Delta \langle \tfrac{1}{4}G_{rs}^{a} G_{rs}^{a} \rangle
    - \frac{1}{\bar{g}_3'^2} \frac{{\rm d}{\bar{g}_{3}}'^{2}}{{\rm d}\ln T}\Delta\langle \tfrac{1}{4}F_{rs}^{ }F_{rs}^{ }\rangle \bigg)
    \;,
\end{align}
where $\kappa_i$ runs over the parameters of the theory,
$\{\bar{g}_{3}^2, \bar{g}_{3}'^2, \bar{\mu}_{h,3}^{2},\bar{\lambda}^{ }_{3},\bar{c}^{ }_{6,3} \}$,
and $\Delta V$ is expanded in $\hbar$.
In this expression, the condensates of the 3d EFT are manifestly gauge-invariant, as is $\Delta V$ because it has been evaluated at its tree-level minimum.
The whole expression is therefore gauge-independent if the parameters of the 3d EFT are separately gauge-invariant.
As shown explicitly in Appendix~\ref{appendix:matching}, the parameters of the SMEFT ultrasoft EFT are gauge-invariant up to $\mc{O}(g^4)$, therefore this approach gives a gauge-invariant result up to this order.
We expect this to be true generically.
However, due to the incompleteness of the basis of higher-dimensional operators in our SMEFT, a numerically small $\mc{O}(g^6)$ gauge dependence of $c_{6,3}$ remains, which will inevitably affect $\Delta\Theta$.

To add some flavour of the computation, we explicitly express the 
five
condensates at one-loop order, evaluated at the critical mass,
\begin{align}
    \Delta \langle \hsq \rangle &= 
    - \frac{\bar{\lambda}_3}{2 \bar{c}_{6,3}}
    + \frac{\hbar}{6 (4 \pi) \sqrt{-\bar{\lambda}_3 \bar{c}_{6,3}}}
    \left[\bar{g}_3^3 + \frac{1}{2}\left(\bar{g}_3^2+\bar{g}_3'^2\right){}^{3/2} + 28 \left(-\bar{\lambda}_3\right)^{3/2}\right]
    + \mc{O}(\hbar^2)
    \;, \\
    \Delta \langle (\hsq)^2 \rangle &= 
    \frac{\bar{\lambda}_3^2}{4 \bar{c}_{6,3}^2}
    + \frac{\hbar\sqrt{-\bar{\lambda}_3}}{6 (4 \pi)  \bar{c}_{6,3}^{3/2}} 
    \left[\bar{g}_3^3 + \frac{1}{2}\left(\bar{g}_3^2+\bar{g}_3'^2\right){}^{3/2} + 16 \left(-\bar{\lambda}_3\right)^{3/2}\right]
    + \mc{O}(\hbar^2)
    \;, \\
    \Delta \langle (\hsq)^3 \rangle &= 
    - \frac{\bar{\lambda}_3^3}{8 \bar{c}_{6,3}^3}
    + \frac{\hbar\left(-\bar{\lambda}_3\right)^{3/2}}{8 (4 \pi)  \bar{c}_{6,3}^{5/2}} 
    \left[\bar{g}_3^3 + \frac{1}{2}\left(\bar{g}_3^2+\bar{g}_3'^2\right){}^{3/2} + 10 \left(-\bar{\lambda}_3\right)^{3/2}\right]
    + \mc{O}(\hbar^2)
    \;, \\
    - \frac{1}{\bar{g}_3'^2}\Delta \langle \tfrac{1}{4} F_{rs}F_{rs} \rangle &= -
    \frac{\hbar\left(-\bar{\lambda}_3 \right)^{3/2}}{8 (4 \pi) \bar{c}_{6,3}^{3/2}} \sqrt{ \bar{g}_3^2 + \bar{g}_3'^2}
    + \mc{O}(\hbar^2)
    \;, \\
    - \frac{1}{\bar{g}_3^2}\Delta \langle \tfrac{1}{4} G_{rs}G_{rs} \rangle &=
    - \frac{\hbar \left(-\bar{\lambda}_3\right){}^{3/2}}{4 (4 \pi)  \bar{c}_{6,3}^{3/2}} \left[\sqrt{\bar{g}_3^2}+\frac{1}{2} \sqrt{\bar{g}_3^2+\bar{g}_3'^2}\right]
    + \mc{O}(\hbar^2)
    \;.
\end{align}
We do not quote two-loop results explicitly due to their length, but they are straightforwardly constructed using the same procedures as above. Note that these condensates are quoted at the critical temperature. To calculate $\alpha$ they are needed at the percolation temperature $\Tp$, in which case there are square roots following from Eq.~\eqref{eq:3d-tree-minimum} which complicate the expressions, though procedurally there is no difference in the computation.

\section{Estimates for the nucleation prefactor}
\label{appendix:prefactor}

In a semiclassical evaluation, the bubble nucleation rate takes the following form,
\beq
    \Gamma = A \mbox{e}^{-S_{\bub}}
    \;,
\eeq
where $S_{\bub}$ is dimensionless and $A$ has mass dimension 4.
The nucleation prefactor, $A$, is discussed in Sec.~\ref{sec:nucleation_corrections}.
We do not calculate it explicitly in this paper, but instead give some rough estimates for it. The prefactor naturally splits into the product of two parts, a {\em dynamical} part $A_{\rmi{dyn}}$ and a {\em statistical} part $A_{\rmi{stat}}$%
\footnote{
    This terminology follows Ref.~\cite{langer1973hydrodynamic}.
},
as shown in Eq.~\eqref{eq:prefactor_split}.
The statistical part of the prefactor has mass dimension 3. Although it is difficult to calculate, the definition of this part is agreed upon in the literature. It is given by a ratio of functional determinants of the second-order fluctuations around the critical bubble and the symmetric phase,
\begin{align}
    A_{\rmi{stat}} = \frac{2}{V} {\rm Im} \sqrt{\frac{\det S_{,\alpha\beta}[\phi_f]}{\det S_{,\alpha\beta}[\phi_b]}}
    \;,
\end{align}
where $S$ is the action, $\alpha$ and $\beta$ run over the fields which fluctuate about the critical bubble, and we are using DeWitt notation~\cite{DeWitt:2003pm} for the functional derivatives.
In the 3d theory $\alpha$ and $\beta$ run over $\{\phi,A_{r}^a,B_{r},\eta^a,\bar{\eta}^a,\eta,\bar{\eta} \}$.
At this stage the ratio of determinants is formal, as there are zero and negative eigenvalues which must be dealt with separately.
The imaginary part arises due to the presence of a single negative eigenvalue, for which the corresponding integral must be carried out by analytic continuation~\cite{Langer:1967ax}.

The dynamical part of the prefactor arises for thermal, and not for vacuum, transitions and is essentially an inverse timescale for the critical bubble to grow,
\beq
A_{\rmi{dyn}} = \frac{\freq }{2\pi}
\;.
\eeq
This should depend both on the exponential growth rate of undamped linear perturbations to the bubble radius, and on the damping due to the thermal bath. However, its precise definition is not widely agreed upon (see e.g. Refs.~\cite{langer1973hydrodynamic,Affleck:1980ac,Linde:1980tt,Arnold:1987mh,Moore:2000jw,Berera:2019uyp}).

\subsection{Dynamical prefactor}
\label{appendix:dynamical_prefactor}

We consider two different estimates of the dynamical prefactor. The first is essentially parametric, and follows from the dominance of infrared gauge bosons in the time evolution of the critical bubble.
The second is a detailed formula which relies both upon on the thin wall approximation and a hydrodynamic approximation.

\tocless\subsubsection{Infrared gauge boson dominance}
\label{appendix:infrared_gauge_bosons}

Here we follow Refs.~\cite{Arnold:1987mh,Arnold:1996dy,Bodeker:1998hm,Moore:2000jw}. The exponential growth of the critical bubble is checked by the parametrically slower evolution of the infrared modes of the gauge bosons, as shown concretely in lattice simulations (see Fig.~11 in~\cite{Moore:2000jw}). Thus, the dynamical prefactor should be of order $g^4 T$, the inverse timescale for the evolution of the infrared modes of the gauge bosons~\cite{Arnold:1996dy}. This is the same reason why the sphaleron rate in the symmetric phase is $\mc{O}(g^{10}T^4)$%
\footnote{
    Or, more precisely, the sphaleron rate is $\mc{O}(g^{10}\ln(1/g)T^4)$. Here we will ignore the $\mc{O}(\ln(1/g))$ correction, which is not large for physical values of the weak coupling.
}.
More precisely, up to an $\mc{O}(1)$ multiplicative factor, the result is
\beq
A_{\rmi{dyn}} \sim
    \frac{\bar{g}_3^2}{(4\pi)}\frac{\bar{g}_3^2/(4\pi)}{\sigma_{\rm SU(2)}} \sim
    \frac{g^4 T}{(4\pi)^2}
    \;,
\eeq
where $\sigma_{\rm SU(2)}\approx 0.477\, T$ is the ``colour'' conductivity of the weak sector. The addition of powers of $4\pi$ here follows Ref.~\cite{Moore:2000jw}. We use this for our first estimate of the dynamical prefactor.

\tocless\subsubsection{Hydrodynamic approximation}
\label{appendix:prefactor_hydrodynamic}

Refs.~\cite{langer1973hydrodynamic,kawasaki1975growth,Csernai:1992tj,Carrington:1993ng} adopted a coarse-grained, hydrodynamic approach to calculating the nucleation prefactor. For the purposes of describing bubble nucleation, it was assumed that the dynamical degrees of freedom could be modelled by the macroscopic energy density, rather than working directly with the quantum fields. Further the thin-wall approximation was adopted.

With the proceeding caveats and approximations, the expression for the dynamical prefactor obtained is~\cite{Csernai:1992tj,Carrington:1993ng},
\beq
A_{\rmi{dyn}} = \frac{8\sigma \eta_S}{3\pi r_b^3(\Delta w)^2}
\;,
\eeq
where $\sigma$ is the surface tension, $\eta_S$ is the shear viscosity, $r_b$ is the bubble radius and $\Delta w$ is the change in the enthalpy, or latent heat, of the transition. Here the bubble radius can be consistently replaced with its thin-wall expression, $r_b=2\sigma/\Delta V$. We have used $\eta\approx 82.5 T^3$ following Ref.~\cite{Carrington:1993ng}. Note that in Refs.~\cite{Csernai:1992tj,Carrington:1993ng} the statistical prefactor was also estimated, however in this case the contribution of the IR geometric deformations of the bubble was wrongly dropped (see Sec.~\ref{appendix:prefactor_thin_wall}).

In the SMEFT in the 3d~approach, the remaining pieces of the expression are,
\begin{align}
\frac{\sigma}{T} &=
    \frac{\bar{\lambda}_3^2}{8 \bar{c}_{6,3}^{3/2}}
    \;,\qquad
    \frac{\Delta V}{T} = \frac{-\bar{\lambda}_3}{2 \bar{c}_{6,3}}\left(\frac{\bar{\lambda}_3^2}{4\bar{c}_{6,3}}-\bar{\mu}_{h,3}^2\right)
    \;,\\
\frac{\Delta w}{T} &=
    \left(\frac{-\bar{\lambda}_3}{2 \bar{c}_{6,3}}\right)\frac{{\rm d}\bar{\mu}_{h,3}^2}{{\rm d}\ln T}
    + \left(\frac{-\bar{\lambda}_3}{2 \bar{c}_{6,3}}\right)^2\frac{{\rm d}\bar{\lambda}_3}{{\rm d}\ln T}
    + \left(\frac{-\bar{\lambda}_3}{2 \bar{c}_{6,3}}\right)^3\frac{{\rm d}\bar{c}_{6,3}}{{\rm d}\ln T}
    \;.
\end{align}
In the 4d~approach, these expressions are replaced by expressions which are evaluated numerically. Again, one must {\em ad hoc} throw away the imaginary part of the effective potential.

\subsection{Statistical prefactor}
\label{appendix:statistical_prefactor}

The statistical prefactor has been computed numerically for the thermal phase transition of a real scalar field by several groups~\cite{Baacke:1993ne,Brahm:1993bm,Surig:1997ne,Baacke:2003uw,Dunne:2005rt}, and various approximation schemes were proposed in Refs.~\cite{Brahm:1993bm,Gleiser:1993hf,Baacke:2003uw}. For more complicated theories, such as the electroweak sector, related computations have been performed for the sphaleron rate~\cite{Arnold:1987mh,Carson:1990jm,Baacke:1993aj,Baacke:1994ix} and the Higgs vacuum decay rate~\cite{Andreassen:2017rzq,Chigusa:2017dux}. A complete calculation of the statistical prefactor for the bubble nucleation rate in the electroweak sector, or related BSM extensions, is made difficult by the radiatively induced nature of the transition (see Sec.~\ref{sec:nucleation_corrections}), and associated infrared divergences~\cite{Kripfganz:1994ha}. To our knowledge, there is no complete calculation in the literature of the statistical prefactor in the electroweak sector, or related BSM extensions. However approximations to it have been proposed in Refs.~\cite{Csernai:1992tj,Carrington:1993ng,Kripfganz:1994ha}.

The statistical prefactor can be further broken up into the contributions from the zero modes, the negative mode, and the positive modes,
\beq
A_{\rmi{stat}} =
    A_{\rmi{stat}}^{(0)}\,
    A_{\rmi{stat}}^{(-)}\,
    A_{\rmi{stat}}^{(+)}\;.
\eeq
The contributions from the zero modes, and the negative mode can be calculated using standard methods. It is $A_{\rmi{stat}}^{(+)}$ which is most difficult to calculate. The factor of one over the volume of space is included in the zero mode contribution.

\tocless\subsubsection{Zero modes}
\label{appendix:prefactor_zero_modes}

Some knowledge about the statistical prefactor can be gained simply by knowledge of the zero modes of the fluctuation spectrum about the critical bubble (see e.g.\ Ref.~\cite{Weinberg:1992ds}). In particular, we will be able to determine the dependence of the prefactor on $\bar{c}_{6,3}$, which is, in all cases, the smallest parameter in the 3d effective theory.

Zero modes arise due to global symmetries of the theory which are broken by the bubble configuration. Integration over the zero modes can be carried out with the method of collective coordinates~\cite{Vainshtein:1981wh} (see also Refs.~\cite{Arnold:1987mh,Carson:1990jm} for similar computations). There are 3 zero modes corresponding to the breaking of spatial translations. These are well known and integration over them results in the following volume and Jacobian factors,
\beq
\int \prod_{i=1}^3\frac{{\rm d}a_i}{\sqrt{2\pi}} = V \left(\frac{S}{2\pi}\right)^{3/2},
\eeq
where on the left hand side we have shown the form of the measure on these modes before the collective coordinate transformation and $V$ is the volume of space. There are also 3 zero modes due to the breaking of the global symmetry part of
${\rm SU}(2)\times {\rm U}(1) \to {\rm U}(1)$. Note that these global symmetries are not broken by our choice of general covariant (or Fermi) gauge, Eq.~\eqref{eq:gauge_fixing}, unlike in the case of $R_\xi$ gauges. Integration over these zero modes has been carried out in Appendix~C of Ref.~\cite{Buchmuller:1993bq}, and results in,
\beq
\int \prod_{i=4}^6\frac{{\rm d}a_i}{\sqrt{2\pi}} = \frac{\pi^2}{2} \left(\frac{2}{2\pi}\int {\rm d}^3x \Phi^\dagger_0(x)\Phi_0(x)\right)^{3/2},
\eeq
where the integral over ${\rm d}^3x$ is over all of space. Altogether, the zero modes contribute the factor,
\beq
\label{eq:zero_modes}
A_{\rmi{stat}}^{(0)} = \frac{1}{4\sqrt{2}\pi} S^{3/2}\left(\int {\rm d}^3x \Phi^\dagger_0(x)\Phi_0(x)\right)^{3/2}
\;.
\eeq
Extracting the zero modes leaves 6 eigenvalues in the ratio of functional determinants which are not matched up. These are positive eigenvalues of Higgs fluctuations about the symmetric phase.

\tocless\subsubsection{Nonzero modes: thick walls}
\label{appendix:prefactor_nonzero_modes}

For bubbles which are not in the thin wall regime~\cite{Coleman:1977py}, the only length scale entering the bounce solution is $1/\bar{\mu}_{h,3}$, which is in turn of order $1/\bar{\mu}_{h,3,c}\sim 2\sqrt{\bar{c}_{6,3}}/\bar{\lambda}_3$ when the supercooling is not parametrically large. Noting this, and scaling the operators in the ratio of determinants by $\bar{\mu}_{h,3}$, we arrive at
\begin{align}
    A_{\rmi{stat}}^{(-)}
    A_{\rmi{stat}}^{(+)} = \bar{\mu}_{h,3}^6\, \kappa\left(\frac{\bar{\lambda}_3}{\bar{g}_3^2},\frac{\bar{g}_3'^2}{\bar{g}_3^2}\right)
    \;,
\end{align}
where the function $\kappa$ is given by
\begin{align}
\kappa\left(\frac{\bar{\lambda}_3}{\bar{g}_3^2},\frac{\bar{g}_3'^2}{\bar{g}_3^2}\right) &= 2\mbox{Im}
    \sqrt{\frac{\det S_{,\alpha\beta}^f}{\det' S_{,\alpha\beta}^b}}
    \\
    &=2\mbox{Im}
    \sqrt{\frac{\det S_{,hh}^f}{\det' S_{,hh}^b}}
    \sqrt{\frac{\det S_{,WG}^f}{\det S_{,WG}^b}}
    \sqrt{\frac{\det S_{,ZG}^f}{\det S_{,ZG}^b}}
    \;.
\end{align}
Here the eigenvalues of the operators have all been scaled by $\bar{\mu}_{h,3}^2$ to be dimensionless and the dash on $\det '$ denotes that the six zero modes are removed. In the second line $S_{,WG}$ and $S_{,ZG}$ refer to the second derivative matrices in the subspaces spanned by the $W$ bosons and their Goldstone modes and the $Z$ boson and their Goldstone modes respectively. In going from the first to the second line we have used that, in the Landau gauge, neither the ghost propagators nor the photon propagator depend on the background Higgs field. Thus, only the physical Higgs particle, the $W$ and the $Z$ bosons and their respective Goldstone modes contribute to the statistical prefactor~\cite{Isidori:2001bm,Andreassen:2017rzq}.

Our first approximation to the statistical prefactor is simply $\kappa = 1$, which we call our thick wall approximation. It is accurate up to the multiplicative function, $\kappa$, which should be of $\mc{O}(1)$ if the ratios of couplings are themselves of $\mc{O}(1)$. In reality these coupling ratios lie in the region $\sim 0.1-0.4$, and hence one might expect corrections to the nucleation rate of a few orders of magnitude. Note that we have not included a $\bar{c}_{6,3}$ dependence for $\kappa$, as the $\bar{c}_{6,3}$ dependence of the prefactor is fixed by our assumption that $\bar{\mu}_{h,3}$ is parametrically of the same order as $\bar{\mu}_{h,3,c}$, i.e.\ that the supercooling is not parametrically large. At larger supercooling the $\bar{c}_{6,3}$ term becomes irrelevant to tunnelling, as the scalar field only tunnels to $\phi \sim \bar{\mu}_{h,3}/\sqrt{\bar{\lambda}_3}$, much short of the broken minimum, hence $\kappa$ has no $\bar{c}_{6,3}$ dependence in this case too.

\tocless\subsubsection{Nonzero modes: thin wall approximation}
\label{appendix:prefactor_thin_wall}

For very small supercooling from the critical temperature, the bubble radius, $r_b$, will be much larger than the thickness of the bubble wall, $r_w \sim 1/\bar{\mu}_{h,3}$. In this case one can make a thin wall approximation. The large hierarchy of scales, $r_w/r_b \ll 1$, leads to a multiplicative correction to the statistical prefactor unique to the thin wall limit. This is made up by low-energy geometric deformations of the bubble, with eigenvalues which scale as $\propto 1/r_b^2$~\cite{Guenther:1979td,Garriga:1993fh,Garriga:1994ut,Munster:1999hr}. For thermal transitions in the 3d effective theory it is
\beq
A_{\rmi{stat}}^{(-)} \approx \frac{r_b}{2\sqrt{2}}
\;, \qquad
A_{\rmi{stat}}^{(+)} \propto \left(\frac{r_b}{r_w}\right)^{-5/3}
\;,
\eeq
which when combined with the contribution of the zero modes in the thin wall approximation,
\beq
A_{\rmi{stat}}^{(0)} \approx \frac{4 \pi^2}{27} \left(\frac{\sigma}{T}\right)^{3/2} \left(\frac{r_b}{r_w}\right)^{15/2} \left(\frac{v_3^2}{\bar{\mu}_{h,3}}\right)^{3/2} \frac{1}{\bar{\mu}_{h,3}^2}
\;, 
\eeq
leads to,
\beq
\label{eq:statistical_prefactor_thin_wall}
A_{\rmi{stat}} \sim \left(\frac{\sigma}{T}\right)^{3/2} \left(\frac{r_b}{r_w}\right)^{41/6} \left(\frac{v_3^2}{\bar{\mu}_{h,3}}\right)^{3/2}
\;,
\eeq
where we dropped the $\mc{O}(1)$ constant, given the uncertainty in the contribution of the positive modes. 
Note that this result differs from Ref.~\cite{Csernai:1992tj} (and consequently Refs.~\cite{Carrington:1993ng,Caprini:2019egz}) for two reasons: first, we have included the additional zero modes from the breaking of custodial symmetry and second, we have included the parametric contribution of the positive low-lying modes, i.e.\ those considered in this Ref.~\cite{Guenther:1979td}. These both affect the statistical prefactor at leading order. Our result agrees with Refs.~\cite{Buchmuller:1992en,Buchmuller:1992rs}.

Eq.~\eqref{eq:statistical_prefactor_thin_wall} is applicable only after performing dimensional reduction. The reason is that only spatial deformations of the critical bubble are included in
$A_{\rmi{stat}}^{(+)}$.
A similar analysis is carried out included the deformations of the bubble in the compact thermal direction. This was carried out in Ref.~\cite{Garriga:1994ut} where it was shown that they contribute exponentially,
\beq
A_{\rmi{stat,4d}} \propto \exp\left( \frac{2}{3}\pi T\, \bar{\mu}_{h,{\rm res}}\, r_b^2 \right)
\;.
\eeq
This result suggests that, at least in the thin wall approximation, the nucleation prefactor gives a much larger correction to the nucleation rate if one does not perform dimensional reduction.

\tocless\subsubsection{Derivative expansion}
\label{appendix:prefactor_derivative}

Modes with wavelengths much shorter than the critical bubble allow for a derivative expansion of the fluctuation determinant.
In this case, the leading order approximation takes the background Higgs field to be locally constant. The wall of the critical bubble has a width of order the inverse Higgs mass. As such the prefactor of particles which are much heavier than the Higgs can be well approximated in the derivative expansion.

The magnetic gauge bosons can never be heavier than the Higgs in the symmetric phase, as they are massless, at least in perturbation theory. However, for $|\bar{\lambda}_3| \ll \bar{g}_3^2$,
the gauge bosons are much heavier than the Higgs in the broken phase. For sufficiently strong phase transitions, the broken phase contributions are expected to dominate over the symmetric phase ones as discussed in Sec.~\ref{sec:nonperturbativity}, and hence one might expect the derivative expansion to give a reasonable approximation for the contribution of the $W$ and $Z$ bosons to the nucleation prefactor, as in the following,
\beq
\sqrt{\frac{\det S_{,WG}^f}{\det S_{,WG}^b}} \sqrt{\frac{\det S_{,ZG}^f}{\det S_{,ZG}^b}}\approx \exp\left(\sum_{\alpha=W,Z,G}\frac{1}{12\pi}\int {\rm d}^3 x\ \left( \left(m^2_\alpha(\phi_b)\right)^{3/2} - \left(m^2_\alpha(\phi_f)\right)^{3/2} \right) \right),
\eeq
where the sum over $\alpha$ runs over the modes of the $W$, $Z$ and corresponding Goldstone bosons. This approach has been taken for the sphaleron rate in Refs.~\cite{Shaposhnikov:1987tw,Carson:1990jm,Baacke:1993aj,Baacke:1994ix}, and was shown to give a good approximation, at least for the case $|\bar{\lambda}_3| \lesssim \bar{g}_3^2$ (see in particular Fig.~1 in Ref.~\cite{Baacke:1994ix}).

On the other hand, the derivative expansion is never formally justified for the prefactor of fluctuations of the Higgs particle itself, or for lighter particles. If one were to apply the derivative expansion to the Higgs itself, the negative effective mass near the top of the potential barrier would result in a spurious nonzero imaginary part (not systematically related to the expected imaginary part from analytic continuation~\cite{Langer:1967ax}). Nevertheless, ploughing on and applying the derivative expansion to the Higgs particle fluctuation determinant, setting imaginary parts to zero by hand, one can derive a rough order-of-magnitude estimate of the statistical prefactor. The leading order contribution is taken to be,
\begin{align}
\mbox{Im}\sqrt{\frac{\det S_{,\alpha\beta}^f}{\det' S_{,\alpha\beta}^b}} &\approx
    (\text{zero modes})
    \exp\left(-\int {\rm d}^3 x\,
    \mbox{Re}\left( V_{(1)}(\phi_b) - V_{(1)}(\phi_f)\right) \right)
    \;, \\ &\approx
    \eqref{eq:zero_modes}\,\bar{\mu}_{h,3}^6\,
    \exp\left(\sum_{\alpha}\frac{1}{12\pi}\int {\rm d}^3 x\,
    \mbox{Re}\left( \left(m^2_\alpha(\phi_b)\right)^{3/2} - \left(m^2_\alpha(\phi_f)\right)^{3/2} \right) \right)
    \;,
\end{align}
where we have followed Ref.~\cite{Baacke:2003uw} in extracting, {\em ad hoc}, the zero-mode contribution, Eq.~\eqref{eq:zero_modes}, and have extracted powers of the mass $\bar{\mu}_{h,3}$ to make up the dimensions. Note the necessity of the introduction of the real part, justified on practical grounds. An analogous expression was shown in Ref.~\cite{Baacke:2003uw} to result in a good approximation of the statistical prefactor for a real scalar field in the case of thin-walled bubbles.

\clearpage
\bibliographystyle{JHEP}
\bibliography{references}
\end{document}